\documentclass[conference]{IEEEtran}
\IEEEoverridecommandlockouts
\usepackage{graphicx}
\usepackage{amsmath}
\usepackage{xcolor}
\usepackage{pdflscape}
\providecommand{\keywords}[1]{\textbf{\textit{Index terms---}} #1}
\usepackage{hyperref}
\usepackage{cite}
\usepackage{amsmath,amssymb,amsfonts}
\usepackage{algorithmic}
\usepackage{graphicx}
\usepackage{textcomp}
\usepackage{xcolor}
\usepackage{graphicx}
\usepackage{dcolumn}
\usepackage{bm}
\usepackage{cuted}
\usepackage{xcolor}
\usepackage[utf8]{inputenc}
\usepackage[T1]{fontenc}
\usepackage{lscape}
\usepackage{mathptmx}
\usepackage{amsmath}

\newcommand{\ket}[1]{\left|#1\right\rangle}
\newcommand{\bra}[1]{\left\langle#1\right|}

\def\BibTeX{{\rm B\kern-.05em{\sc i\kern-.025em b}\kern-.08em
    T\kern-.1667em\lower.7ex\hbox{E}\kern-.125emX}}
\usepackage{graphicx}
\usepackage{amsmath}

\begin{document}
\onecolumn
\title{Equivalence between classical epidemic model and non-dissipative and dissipative quantum tight-binding model }
\author{Krzysztof Pomorski$^{1,2,3}$ \\ \\

1: Cracow University of Technology \\ Faculty of Computer Science and Telecommunications \\ Department of Computer Science \\ \\
2: Quantum Hardware Systems ($www.quantumhardwaresystems.com$) \\ \\ E-mail: $kdvpomorski@gmail.com$ }
\maketitle

\begin{abstract}
The equivalence between classical epidemic model and non-dissipative and dissipative quantum tight-binding model is derived.  Classical epidemic model can reproduce the quantum entanglement emerging in the case of electrostatically coupled qubits described by von-Neumann entropy both in non-dissipative and dissipative case. The obtained results shows that quantum mechanical phenomena might be almost entirely simulated by classical statistical model. It includes the quantum like entanglement and superposition of states. Therefore coupled epidemic models expressed by classical systems in terms of classical physics can be the base for possible incorporation of quantum technologies and in particular for quantum like computation and quantum like communication. The classical density matrix is derived and described by the equation of motion in terms of anticommutator. Existence of Rabi like oscillations is pointed in classical epidemic model. Furthermore the existence of Aharonov-Bohm effect in quantum systems can also be
reproduced by the classical epidemic model. Every quantum system made from quantum dots and described by simplistic tight-binding model by use of position-based qubits can be effectively described by classical model with very specific structure of S matrix that has twice bigger size as it is the case of quantum matrix Hamiltonian. Obtained results partly question fundamental and unique character of quantum mechanics and are placing ontology of quantum mechanics much in the framework of classical statistical physics what can bring motivation for emergence of other fundamental theories bringing suggestion that quantum mechanical is only effective and phenomenological but not fundamental picture of reality.
\end{abstract}
\section{Introduction to classical epidemic model}
Epidemic model can model sickness propagation and various phenomena in sociology, physics and biology. The most basic form of epidemic model relies on co-dependence of two probabilities of occurrence of state 1 and 2 that can be identified at the state of being healthy and sick as it is being depicted in Fig.1. It is expressed in the compact way in the following way:
\begin{eqnarray}
(s_{11}(t)\ket{1}\bra{1}+s_{22}(t)\ket{2}\bra{2}+s_{12}(t)\ket{2}\bra{1}+s_{21}(t)\ket{1}\bra{2})(p_{1}(t)\ket{1}+p_{2}(t)\ket{2})=
\frac{d}{dt}(p_{1}\ket{1}+p_{2}\ket{2})
 = \nonumber \\
=\frac{d}{dt}
\begin{pmatrix}
p_{1}(t) \\
p_{2}(t) \\
\end{pmatrix}=
\begin{pmatrix}
s_{11}(t) & s_{12}(t) \\
s_{21}(t) & s_{22}(t) \\
\end{pmatrix}
\begin{pmatrix}
p_{1}(t) \\
p_{2}(t) \\
\end{pmatrix}=
\begin{pmatrix}
s_{11}(t) & s_{12}(t) \\
s_{21}(t) & s_{22}(t) \\
\end{pmatrix}|\psi_{classical}>= \nonumber \\
=\hat{S}_t(p_1(t)\ket{1}+p_2(t)\ket{2})=\frac{d}{dt}|\psi_{classical}>.
\end{eqnarray}
Quite naturally such system evolves in natural statistical environment before the measurement is done.
Once the measurement is done the statistical system state is changed
from undetermined and spanned by two probabilities into the case of being either with probability $p_1=1$ or $p_1=0$ so it corresponds to two projections:
\begin{eqnarray}
\hat{P}_{\rightarrow 1}=\ket{1}\bra{1}=
\begin{pmatrix}
1 & 0 \\
0 & 0 \\
\end{pmatrix},
\hat{P}_{\rightarrow 2}=\ket{2}\bra{2}=
\begin{pmatrix}
0 & 0 \\
0 & 1 \\
\end{pmatrix},\hat{P}_{\rightarrow 1}+\hat{P}_{\rightarrow 2}=\hat{I}=
\begin{pmatrix}
1 & 0 \\
0 & 1 \\
\end{pmatrix}
 \nonumber \\
\hat{P}_{\rightarrow 1}\ket{\psi}_{classical}= \ket{1}=\ket{\psi_1}_{after},
\begin{pmatrix}
0 & 0 \\
0 & 1 \\
\end{pmatrix},
\hat{P}_{\rightarrow 2}\ket{\psi}_{classical}= \ket{2}=\ket{\psi_2}_{after},
\end{eqnarray}
where occurrence of $\ket{\psi1}_{after}$ and $\ket{\psi2}_{after}$ occurs with probability $p_1(t_{measurement})$ and $p_2(t_{measurement})$.
\begin{figure}
\centering
\includegraphics[scale=0.9]{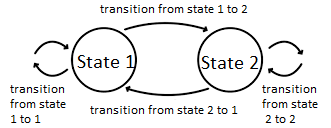}
\caption{Illustration of epidemic model referring to stochastic finite state machine being 2 level system with 2 distinguished states 1 and 2. 4 possible transitions are characterized by 4 time-dependent coefficients $s_{1 \rightarrow 1}(t)=s_{11}(t)$ ,$s_{1 \rightarrow 2}(t)=s_{12}(t)$, $s_{2 \rightarrow 1}(t)=s_{21}(t)$, $s_{2 \rightarrow 2}(t)=s_{22}(t)$.}
\end{figure}

We notice that matrix $\hat{S}$
\begin{eqnarray}
\hat{S}=
\begin{pmatrix}
s_{11}(t) & s_{12}(t) \\
s_{21}(t) & s_{22}(t) \\
\end{pmatrix}
\end{eqnarray}
has 2 eigenvalues
\begin{eqnarray}
  E_1(t) &=& \frac{1}{2} \left(-\sqrt{(s_{11}(t)-s_{22}(t))^2+4s_{12}(t)s_{21}(t)}+s_{11}(t)+s_{22}(t)\right) \\
  E_2(t) &=& \frac{1}{2} \left(+\sqrt{(s_{11}(t)-s_{22}(t))^2+4 s_{12}(t)s_{21}(t)}+s_{11}(t)+s_{22}(t)\right).
\end{eqnarray}
 and we have the corresponding classical eigenstates
\begin{eqnarray}
\ket{\psi_{E_1}}=\frac{2s_{21}}{2 s_{21}+(-\sqrt{(s_{11}-s_{22})^2+4s_{12}s_{21}}+s_{11}-s_{22})}
\begin{pmatrix}
\frac{-\sqrt{(s_{11}-s_{22})^2+4s_{12}s_{21}}+s_{11}-s_{22}}{2 s_{21}} \\ 1 
\end{pmatrix},   \\
\ket{\psi_{E_2}}=\frac{2s_{21}}{2s_{21}+(+\sqrt{(s_{11}-s_{22})^2+4 s_{12}s_{21}}+s_{11}-s_{22})}
\begin{pmatrix}
\frac{+\sqrt{(s_{11}-s_{22})^2+4 s_{12} s_{21}}+s_{11}-s_{22}}{2 s_{21}} \\
1
\end{pmatrix}.
\end{eqnarray}
We recognize that two states $\ket{\psi_{E_1}}$ and $\ket{\psi_{E_2}}$ are orthogonal, so $\bra{\psi_{E_1}}\ket{\psi_{E_2}}=\bra{\psi_{E_2}}\ket{\psi_{E_1}}=0$.
We also recognize that
\begin{eqnarray}
\bra{\psi_{E_1}}\ket{\psi_{E_1}}=[\frac{-\sqrt{(s_{11}-s_{22})^2+4s_{12}s_{21}}+s_{11}-s_{22}}{2 s_{21}+(-\sqrt{(s_{11}-s_{22})^2+4s_{12}s_{21}}+s_{11}-s_{22})}]^2+[\frac{2s_{21}}{2 s_{21}+(-\sqrt{(s_{11}-s_{22})^2+4s_{12}s_{21}}+s_{11}-s_{22})}]^2= \\
=1-[\frac{4s_{21}(-\sqrt{(s_{11}-s_{22})^2+4s_{12}s_{21}}+s_{11}-s_{22})}{(2 s_{21}+(-\sqrt{(s_{11}-s_{22})^2+4s_{12}s_{21}}+s_{11}-s_{22}))^2}]=n_{E_1}(t), \nonumber \\
\bra{\psi_{E_2}}\ket{\psi_{E_2}}=[\frac{+\sqrt{(s_{11}-s_{22})^2+4 s_{12} s_{21}}+s_{11}-s_{22}}{2s_{21}+(+\sqrt{(s_{11}-s_{22})^2+4 s_{12}s_{21}}+s_{11}-s_{22})}]^2+[\frac{2s_{21}}{2s_{21}+(+\sqrt{(s_{11}-s_{22})^2+4 s_{12}s_{21}}+s_{11}-s_{22})}]^2= \nonumber \\
=1-[\frac{4s_{21}(+\sqrt{(s_{11}-s_{22})^2+4 s_{12}s_{21}}+s_{11}-s_{22})}{(2s_{21}+(+\sqrt{(s_{11}-s_{22})^2+4 s_{12}s_{21}}+s_{11}-s_{22}))^2}]
=n_{E_2}(t).
\end{eqnarray}

It shall be underlined that necessary condition for identification the superposition of two classical eigenstates is expressed by $(-\sqrt{(s_{11}-s_{22})^2+4s_{12}s_{21}}+s_{11}-s_{22})>0$ and $(+\sqrt{(s_{11}-s_{22})^2+4 s_{12}s_{21}}+s_{11}-s_{22})>0$ what preimposes some constrains on
real value functions $s_{11}(t)$, $s_{12}$, $s_{21}$ and $s_{22}$.
The full classical state can be written as the superposition of two ensembles with probabilities $p_I$ and $p_II$ expressed by the following classical state
\begin{eqnarray}
\ket{\psi(t)}_{classical}=p_{I}(t)\ket{\psi_{E_1}}+p_{2}(t)\ket{\psi_{E_2}}= \nonumber \\
=p_{I}(t)\frac{2s_{21}}{2 s_{21}+(-\sqrt{(s_{11}-s_{22})^2+4s_{12}s_{21}}+s_{11}-s_{22})}
\begin{pmatrix}
\frac{-\sqrt{(s_{11}-s_{22})^2+4s_{12}s_{21}}+s_{11}-s_{22}}{2 s_{21}} \\ 1 
\end{pmatrix}+ \nonumber  \\
+p_{II}(t)\frac{2s_{21}}{2s_{21}+(+\sqrt{(s_{11}-s_{22})^2+4 s_{12}s_{21}}+s_{11}-s_{22})}
\begin{pmatrix}
\frac{+\sqrt{(s_{11}-s_{22})^2+4 s_{12} s_{21}}+s_{11}-s_{22}}{2 s_{21}} \\
1
\end{pmatrix}= \nonumber \\
=p_{I}(t)
\begin{pmatrix}
\frac{-\sqrt{(s_{11}-s_{22})^2+4s_{12}s_{21}}+s_{11}-s_{22}}{2 s_{21}+(-\sqrt{(s_{11}-s_{22})^2+4s_{12}s_{21}}+s_{11}-s_{22})} \\
\frac{2s_{21}}{2 s_{21}+(-\sqrt{(s_{11}-s_{22})^2+4s_{12}s_{21}}+s_{11}-s_{22})} 
\end{pmatrix}+
p_{II}(t)
\begin{pmatrix}
\frac{+\sqrt{(s_{11}-s_{22})^2+4 s_{12} s_{21}}+s_{11}-s_{22}}{2s_{21}+(+\sqrt{(s_{11}-s_{22})^2+4 s_{12}s_{21}}+s_{11}-s_{22})} \\
\frac{2s_{21}}{2s_{21}+(+\sqrt{(s_{11}-s_{22})^2+4 s_{12}s_{21}}+s_{11}-s_{22})}
\end{pmatrix}=
\nonumber  \\
=
\begin{pmatrix}
p_{I}(t)\frac{-\sqrt{(s_{11}-s_{22})^2+4s_{12}s_{21}}+s_{11}-s_{22}}{2 s_{21}+(-\sqrt{(s_{11}-s_{22})^2+4s_{12}s_{21}}+s_{11}-s_{22})}+p_{II}(t)\frac{+\sqrt{(s_{11}-s_{22})^2+4 s_{12} s_{21}}+s_{11}-s_{22}}{2s_{21}+(+\sqrt{(s_{11}-s_{22})^2+4 s_{12}s_{21}}+s_{11}-s_{22})} \\
p_{I}(t)\frac{2s_{21}}{2 s_{21}+(-\sqrt{(s_{11}-s_{22})^2+4s_{12}s_{21}}+s_{11}-s_{22})}+p_{II}(t)\frac{2s_{21}}{2s_{21}+(+\sqrt{(s_{11}-s_{22})^2+4 s_{12}s_{21}}+s_{11}-s_{22})} 
\end{pmatrix}=
\begin{pmatrix}
p_{1}(t) \\
p_{2}(t)
\end{pmatrix}=\ket{\psi(t)}_{classical}
.
\end{eqnarray}
We have superposition of states with 2 statistical ensembles occurring with probabilities $p_{I}(t)$ and $p_{II}(t)$ that are encoded in probabilities $p_1(t)$ and $p_2(t)$ that are directly observable. We can extract probabilities $p_{I}(t)$ and $p_{II}(t)$ from $\ket{\psi(t)}_{classical}$ in the following way
\begin{eqnarray}
p_{I}(t)=\frac{1}{n_{E_1}(t)}\bra{\psi_{E_1}} \ket{\psi(t)}_{classical}=\Bigg[1-\Bigg[\frac{4s_{21}(-\sqrt{(s_{11}-s_{22})^2+4s_{12}s_{21}}+s_{11}-s_{22})}{(2 s_{21}+(-\sqrt{(s_{11}-s_{22})^2+4s_{12}s_{21}}+s_{11}-s_{22}))^2}\Bigg]\Bigg]\bra{\psi_{E_1}} \ket{\psi(t)}_{classical}= \nonumber \\
=\Bigg[1-\Bigg[\frac{4s_{21}(-\sqrt{(s_{11}-s_{22})^2+4s_{12}s_{21}}+s_{11}-s_{22})}{(2 s_{21}+(-\sqrt{(s_{11}-s_{22})^2+4s_{12}s_{21}}+s_{11}-s_{22}))^2}\Bigg]\Bigg] \times \nonumber \\
\times
\begin{pmatrix}
\frac{-\sqrt{(s_{11}-s_{22})^2+4s_{12}s_{21}}+s_{11}-s_{22}}{2 s_{21}+(-\sqrt{(s_{11}-s_{22})^2+4s_{12}s_{21}}+s_{11}-s_{22})}, & \frac{2s_{21}}{2 s_{21}+(-\sqrt{(s_{11}-s_{22})^2+4s_{12}s_{21}}+s_{11}-s_{22})} 
\end{pmatrix}
\begin{pmatrix}
p_{1}(t), \\
p_{2}(t)
\end{pmatrix} 
\end{eqnarray}
and
\begin{eqnarray}
p_{II}(t)=\frac{1}{n_{E_2}(t)}\bra{\psi_{E_2}} \ket{\psi(t)}_{classical}=\Bigg[1-\Bigg[\frac{4s_{21}(+\sqrt{(s_{11}-s_{22})^2+4 s_{12}s_{21}}+s_{11}-s_{22})}{(2s_{21}+(+\sqrt{(s_{11}-s_{22})^2+4 s_{12}s_{21}}+s_{11}-s_{22}))^2}\Bigg]\Bigg]\bra{\psi_{E_2}} \ket{\psi(t)}_{classical}= \nonumber \\
=\Bigg[1-\Bigg[\frac{4s_{21}(+\sqrt{(s_{11}-s_{22})^2+4s_{12}s_{21}}+s_{11}-s_{22})}{(2 s_{21}+(+\sqrt{(s_{11}-s_{22})^2+4s_{12}s_{21}}+s_{11}-s_{22}))^2}\Bigg]\Bigg] \times \nonumber \\
\times
\begin{pmatrix}
\frac{+\sqrt{(s_{11}-s_{22})^2+4s_{12}s_{21}}+s_{11}-s_{22}}{2 s_{21}+(+\sqrt{(s_{11}-s_{22})^2+4s_{12}s_{21}}+s_{11}-s_{22})}, & \frac{2s_{21}}{2 s_{21}+(+\sqrt{(s_{11}-s_{22})^2+4s_{12}s_{21}}+s_{11}-s_{22})} 
\end{pmatrix}
\begin{pmatrix}
p_{1}(t), \\
p_{2}(t)
\end{pmatrix}= \nonumber \\ =
\Bigg[1-\Bigg[\frac{4s_{21}(+\sqrt{(s_{11}-s_{22})^2+4s_{12}s_{21}}+s_{11}-s_{22})}{(2 s_{21}+(+\sqrt{(s_{11}-s_{22})^2+4s_{12}s_{21}}+s_{11}-s_{22}))^2}\Bigg]\Bigg] \times \nonumber \\
\times
\begin{pmatrix}
\frac{+\sqrt{(s_{11}-s_{22})^2+4s_{12}s_{21}}+s_{11}-s_{22}}{2 s_{21}+(+\sqrt{(s_{11}-s_{22})^2+4s_{12}s_{21}}+s_{11}-s_{22})}p_1(t)+ \frac{2s_{21}}{2 s_{21}+(+\sqrt{(s_{11}-s_{22})^2+4s_{12}s_{21}}+s_{11}-s_{22})}p_2(t) 
\end{pmatrix}. 
\end{eqnarray}
Probabilities $p_{I}(t)$ and $p_{II}(t)$ will describe the occupancy of energy levels $E_1$ and $E_2$ in real time domain of epidemic simplistic model.
We have the same superposition of two eigenergies as in the case of quantum tight-binding model.
The same reasoning can be conducted for N-th state classical epidemic model expressed as
\begin{eqnarray}
(s_{11}(t)\ket{1}\bra{1}+s_{12}(t)\ket{2}\bra{1}+s_{13}(t)\ket{3}\bra{1}+.. + s_{1N}(t)\ket{N}\bra{1}+ \nonumber \\
+s_{21}(t)\ket{2}\bra{1}+s_{22}(t)\ket{2}\bra{2}+s_{23}(t)\ket{3}\bra{2}+.. + s_{2N}(t)\ket{N}\bra{2}+ \nonumber \\
.... + \nonumber \\
+s_{1N}(t)\ket{1}\bra{N}+s_{2N}(t)\ket{2}\bra{N}+s_{3N}(t)\ket{3}\bra{N}+.. + s_{NN}(t)\ket{N}\bra{N})
\nonumber \\
(p_{1}(t)\ket{1}+p_{2}(t)\ket{2}+..+p_{N}(t)\ket{N})=
\frac{d}{dt}(p_{1}\ket{1}+p_{2}\ket{2}+..+p_{N}\ket{N})
 = \nonumber \\
=\frac{d}{dt}
\begin{pmatrix}
p_{1}(t) \\
p_{2}(t) \\
.. \\
p_{N}(t) \\
\end{pmatrix}=
\begin{pmatrix}
s_{11}(t) & s_{12}(t) & .. & s_{1N}(t) \\
s_{21}(t) & s_{22}(t) & .. & s_{1N}(t  \\
.. \\
s_{N1}(t) & s_{N2}(t) & .. & s_{NN}(t  \\
\end{pmatrix}
\begin{pmatrix}
p_{1}(t) \\
p_{2}(t) \\
..
p_{N}(t) \\
\end{pmatrix}=
\begin{pmatrix}
s_{11}(t) & s_{12}(t) & .. & s_{1N}(t) \\
.. \\
s_{1N}(t) & s_{2N}(t)  & .. & s_{NN}(t) \\
\end{pmatrix}|\psi_{classical}>= \nonumber \\
=\hat{S}_t(p_1(t)\ket{1}+p_2(t)\ket{2}+..+p_N(t)\ket{N})=\frac{d}{dt}|\psi_{classical}>.
\end{eqnarray}
\section{Analytical solutions of simplistic classical epidemic model}
In principle we can also introduce weak measurement procedure that will be partly omitted in this work. In very real way if we have the population of N individuals possibly infected with COVID we can inspect $N_1$ individuals, where $N_1<N$ and we can introduce some corrections to $p_1(t_{measurement}^{-})\rightarrow \frac{1}{N}[(N-N_1)p_1(t_{measurement}^{-})+N_{1}p_1(t_{test})]=p_1(t_{measurement}^{+})$ and $p_2(t_{measurement}^{-})\rightarrow \frac{1}{N}[(N-N_1)p_2(t_{measurement}^{-})+N_{1}p_2(t_{test})]=p_2(t_{measurement}^{+})$, where $p_1(t_{test}),p_2(t_{test})$ are probabilities obtained by testing $N_1$ individuals what could correspond to weak measurement conducted on assemble of N individuals.  Let us consider the state of the system before measurement and its natural evolution
Such set of equations has two analytical solutions for probabilities $p_1(t)$ and $p_2(t)$ expressed as
\begin{eqnarray}
exp
\begin{pmatrix}
\int_{t0}^{t} s_{11}(t)dt' & \int_{t0}^{t} s_{11}(t)dt' \\
\int_{t0}^{t} s_{21}(t)dt' & \int_{t0}^{t} s_{22}(t)dt' \\
\end{pmatrix}
\begin{pmatrix}
p_{1}(t_0) \\
p_{2}(t_0) \\
\end{pmatrix}
=exp
\begin{pmatrix}
S_{11}(t,t_0) & S_{12}(t,t_0) \\
S_{21}(t,t_0) & S_{22}(t,t_0) \\
\end{pmatrix}
\begin{pmatrix}
p_{1}(t_0) \\
p_{2}(t_0) \\
\end{pmatrix}
= \nonumber \\
\begin{pmatrix}
U_{11}(t,t_0) & U_{12}(t,t_0) \\
U_{21}(t,t_0) & U_{22}(t,t_0) \\
\end{pmatrix}
\begin{pmatrix}
p_{1}(t_0) \\
p_{2}(t_0) \\
\end{pmatrix}
= \hat{U}(t,t_0)
\begin{pmatrix}
p_{1}(t_0) \\
p_{2}(t_0) \\
\end{pmatrix}
=
\begin{pmatrix}
p_{1}(t) \\
p_{2}(t) \\
\end{pmatrix}
\end{eqnarray}
with
\begin{eqnarray}
S_{11}(t,t_0)=\int_{t0}^{t} s_{11}(t)dt', S_{12}(t,t_0)=\int_{t0}^{t} s_{12}(t)dt', S_{22}(t,t_0)=\int_{t0}^{t} s_{22}(t)dt', S_{21}(t,t_0)=\int_{t0}^{t} s_{21}(t)dt',
\end{eqnarray}
and
\begin{eqnarray}
U_{1,1}(t,t_0)=e^{\frac{S_{11}(t,t_0)+S_{22}(t,t_0)}{2}}\Bigg[+\frac{(S_{11}(t,t_0)-S_{22}(t,t_0)) \sinh \left(\frac{1}{2} \sqrt{(S_{11}(t,t_0)-S_{22}(t,t_0))^2+4 S_{12}(t,t_0)
   S_{21}(t,t_0)}\right)}{\sqrt{(S_{11}(t,t_0)-S_{22}(t,t_0))^2+4 S_{12}(t,t_0) S_{21}(t,t_0)}}+\nonumber \\ +\cosh \left(\frac{1}{2} \sqrt{(S_{11}(t,t_0)-S_{22}(t,t_0))^2+4
   S_{12}(t,t_0)S_{21}(t,t_0)}\right)\Bigg]
\end{eqnarray}
\begin{eqnarray}
U_{2,2}(t,t_0)=e^{\frac{S_{11}(t,t_0)+S_{22}(t,t_0)}{2}} \Bigg[-\frac{(S_{11}(t,t_0)-S_{22}(t,t_0)) \sinh \left(\frac{1}{2} \sqrt{(S_{11}(t,t_0)-S_{22}(t,t_0))^2+4 S_{12}(t,t_0)
   S_{21}(t,t_0)}\right)}{\sqrt{(S_{11}(t,t_0)-S_{22}(t,t_0))^2+4 S_{12}(t,t_0)S_{21}(t,t_0)}}+ \nonumber \\ + \cosh \left(\frac{1}{2} \sqrt{(S_{11}(t,t_0)-S_{22}(t,t_0))^2+4
   S_{12}(t,t_0) S_{21}(t,t_0)}\right)\Bigg]
\end{eqnarray}
\begin{eqnarray}
U_{1,2}(t,t_0)=\frac{2 S_{12}(t,t_0) e^{\frac{S_{11}(t,t_0)+S_{22}(t,t_0)}{2}} \sinh \left(\frac{1}{2} \sqrt{(S_{11}(t,t_0)-S_{22}(t,t_0))^2+4 S_{12}(t,t_0)
   S_{21}(t,t_0)}\right)}{\sqrt{(S_{11}(t,t_0)-S_{22}(t,t_0))^2+4 S_{12}(t,t_0) S_{21}(t,t_0)}},
\end{eqnarray}
\begin{eqnarray}
U_{2,1}(t,t_0)=\frac{2 S_{21}(t,t_0) e^{\frac{S_{11}(t,t_0)+S_{22}(t,t_0)}{2}} \sinh \left(\frac{1}{2} \sqrt{(S_{11}(t,t_0)-S_{22}(t,t_0))^2+4S_{12}(t,t_0)
   S_{21}(t,t_0)}\right)}{\sqrt{(S_{11}(t,t_0)-S_{22}(t,t_0))^2+4S_{12}(t,t_0)S_{21}(t,t_0)}}
\end{eqnarray}
We obtain explicit formula for probabilities
\begin{eqnarray}
p_1(t)=e^{\frac{S_{11}(t,t_0)+S_{22}(t,t_0)}{2}}\Bigg[\Bigg[+\frac{(S_{11}(t,t_0)-S_{22}(t,t_0)) \sinh \left(\frac{1}{2} \sqrt{(S_{11}(t,t_0)-S_{22}(t,t_0))^2+4 S_{12}(t,t_0)
   S_{21}(t,t_0)}\right)}{\sqrt{(S_{11}(t,t_0)-S_{22}(t,t_0))^2+4 S_{12}(t,t_0) S_{21}(t,t_0)}}+\nonumber \\ +\cosh \left(\frac{1}{2} \sqrt{(S_{11}(t,t_0)-S_{22}(t,t_0))^2+4
   S_{12}(t,t_0)S_{21}(t,t_0)}\right)\Bigg]p_1(t_0)+ \nonumber \\ +\Bigg[\frac{2 S_{12}(t,t_0)\sinh \left(\frac{1}{2} \sqrt{(S_{11}(t,t_0)-S_{22}(t,t_0))^2+4 S_{12}(t,t_0)
   S_{21}(t,t_0)}\right)}{\sqrt{(S_{11}(t,t_0)-S_{22}(t,t_0))^2+4 S_{12}(t,t_0) S_{21}(t,t_0)}}\Bigg] p_2(t_0)\Bigg],
\end{eqnarray}
\begin{eqnarray}
p_2(t)=e^{\frac{S_{11}(t,t_0)+S_{22}(t,t_0)}{2}}\Bigg[\Bigg[\frac{2 S_{21}(t,t_0)\sinh \left(\frac{1}{2} \sqrt{(S_{11}(t,t_0)-S_{22}(t,t_0))^2+4S_{12}(t,t_0)
   S_{21}(t,t_0)}\right)}{\sqrt{(S_{11}(t,t_0)-S_{22}(t,t_0))^2+4S_{12}(t,t_0)S_{21}(t,t_0)}}\Bigg]p_1(t_0)+ \nonumber \\
   +\Bigg[-\frac{(S_{11}(t,t_0)-S_{22}(t,t_0)) \sinh \left(\frac{1}{2} \sqrt{(S_{11}(t,t_0)-S_{22}(t,t_0))^2+4 S_{12}(t,t_0)
   S_{21}(t,t_0)}\right)}{\sqrt{(S_{11}(t,t_0)-S_{22}(t,t_0))^2+4 S_{12}(t,t_0)S_{21}(t,t_0)}}+ \nonumber \\ + \cosh \left(\frac{1}{2} \sqrt{(S_{11}(t,t_0)-S_{22}(t,t_0))^2+4
   S_{12}(t,t_0) S_{21}(t,t_0)}\right)\Bigg]p_2(t_0)\Bigg].
\end{eqnarray}
It is useful to express the ratio of probabilities $p_1(t)$ and $p_2(t)$ in the following analytical way as
\begin{eqnarray}
r_{12}(t)=\frac{p_1(t)}{p_2(t)}=\nonumber \\
=\Bigg[[(S_{11}(t,t_0)-S_{22}(t,t_0))p_1(t_0)+2 S_{21}(t,t_0)p_2(t_0)] \tanh \left(\frac{1}{2} \sqrt{(S_{11}(t,t_0)-S_{22}(t,t_0))^2+4 S_{12}(t,t_0)
   S_{21}(t,t_0)}\right) +\nonumber \\+ [p_1(t_0)\sqrt{(S_{11}(t,t_0)-S_{22}(t,t_0))^2+4
   S_{12}(t,t_0)S_{21}(t,t_0)}]\Bigg]/ \nonumber \\ \Bigg[ -[(S_{11}(t,t_0)-S_{22}(t,t_0))p_2(t_0)+2 S_{21}(t,t_0)p_1(t_0)] \tanh \left(\frac{1}{2} \sqrt{(S_{11}(t,t_0)-S_{22}(t,t_0))^2+4 S_{12}(t,t_0)
   S_{21}(t,t_0)}\right) +\nonumber \\+ [p_2(t_0)\sqrt{(S_{11}(t,t_0)-S_{22}(t,t_0))^2+4
   S_{12}(t,t_0)S_{21}(t,t_0)}] \Bigg] \nonumber \\
\end{eqnarray}

\section{Equations of motion for classical epidemic model in projector representation in case of time independent matrix $S$}
Let us consider the equations of motion for the case of time-independent $s_{11}$, $s_{12}$, $s_{21}$ and $s_{22}$. In Dirac notation it can be written in the following way:
\begin{eqnarray}
(E_1\ket{\psi_{E_1}}\frac{1}{n_{E_1}}\bra{\psi_{E_1}}+E_2\ket{\psi_{E_2}}\frac{1}{n_{E_2}}\bra{\psi_{E_2}})(p_{I}\ket{\psi_{E_1}}+p_{II}\ket{\psi_{E_2}}).
=\frac{d}{dt}(p_{I}\ket{\psi_{E_1}}+p_{II}\ket{\psi_{E_2}})=((\frac{d}{dt}p_{I})\ket{\psi_{E_1}}+(\frac{d}{dt})p_{II}\ket{\psi_{E_2}}),
\end{eqnarray}
since $E_1$, $E_2$, $\ket{\psi_{E_1}}$ and $\ket{\psi_{E_2}}$ are time independent. By applying $\ket{\psi_{E1}}$ and $\ket{\psi_{E1}}$ on the left side we obtain
and using orthogonality relation between $\ket{\psi_{E_1}}$ and $\ket{\psi_{E_2}}$ we obtain set of equations
\begin{equation}
\frac{E_1}{n_{E_1}}p_{I}=\frac{d}{dt}p_{I}, \frac{E_2}{n_{E_2}}p_{II}=\frac{d}{dt}p_{II},
\end{equation}
that has the solutions
\begin{equation}
e^{\frac{E_1}{n_{E_1}}(t-t_0)}p_{I}(t_0)=p_{I}(t), e^{\frac{E_2}{n_{E_2}}(t-t_0)}p_{II}(t_0)=p_{II}(t).
\end{equation}
Sum of probabilities is not normalized. However physical significance between ratio $p_1(t)$ and $p_2$ that is expressed by ratio
\begin{equation}
r_{12}(t)=\frac{p_1(t)}{p_2(t)}=\frac{p_{I}(t_0)}{p_{II}(t_0)}exp((\frac{E_{1}}{n_{E_1}}-\frac{E_2}{n_{E_2}})(t-t_0)).
\end{equation}
It means that Rabi oscillations or more precisely change of occupancy among levels is naturally build in classical epidemic model.
Still superposition of two states is mainatianed so the analogy of classical epidemic model to quantum tight-binding model is deep.
\section{Case of constant occupacy of 2 eigenergy levels in classical epidemic model}
We consider the case of $p_{I}(t)=constant_1$ and $p_{II}(t)=constant_{II}$. We have time-dependent parameters $s_{11}$, $s_{22}$, $s_{12}$ and $s_{21}$ and we obtain the following equations of motion
\begin{eqnarray}
(E_1(t)\ket{\psi_{E_1}}_t\bra{\psi_{E_1}}_t+E_2(t)\ket{\psi_{E_2}}_t\bra{\psi_{E_2}}_t)(p_{I}\ket{\psi_{E_1}}_t+p_{II}\ket{\psi_{E_2}}_t).
=\frac{d}{dt}(p_{I}\ket{\psi_{E_1}}_t+p_{II}\ket{\psi_{E_2}}_t)=(p_{I}\frac{d}{dt})\ket{\psi_{E_1}}_t+(p_{II}\frac{d}{dt})\ket{\psi_{E_2}}_t).
\end{eqnarray}
We obtain the set of 2 equations
\begin{eqnarray}
E_1(t) p_I=(p_{I}\bra{\psi_{E_1}}\frac{d}{dt})\ket{\psi_{E_1}}+(p_{II}\bra{\psi_{E_1}}\frac{d}{dt})\ket{\psi_{E_2}}),  \\
E_2(t) p_{II}=(p_{I}\bra{\psi_{E_2}}\frac{d}{dt})\ket{\psi_{E_1}}+(p_{II}\bra{\psi_{E_2}}\frac{d}{dt})\ket{\psi_{E_2}}).
\end{eqnarray}
Consequently we obtain
\begin{eqnarray}
\frac{p_I}{p_{II}}=\frac{\bra{\psi_{E_1}}\frac{d}{dt}\ket{\psi_{E_2}}}{E_1(t)-\bra{\psi_{E_1}}\frac{d}{dt}\ket{\psi_{E_1}}}, \nonumber \\ 
\frac{p_I}{p_{II}}=\frac{(E_2(t)-\bra{\psi_{E_2}}\frac{d}{dt}\ket{\psi_{E_2}})}{\bra{\psi_{E_2}}\frac{d}{dt}\ket{\psi_{E_1}}} 
\end{eqnarray}
and it implies
\begin{eqnarray}
\frac{\bra{\psi_{E_1}}\frac{d}{dt}\ket{\psi_{E_2}}}{(E_1(t)-\bra{\psi_{E_1}}\frac{d}{dt}\ket{\psi_{E_1}})}=\frac{(E_2(t)-\bra{\psi_{E_2}}\frac{d}{dt}\ket{\psi_{E_2}})}{\bra{\psi_{E_2}}\frac{d}{dt}\ket{\psi_{E_1}}} 
\end{eqnarray}
\section{Equations of motion for classical epidemic model in projector representation (Dirac notation) and Rabi oscillations in classical epidemic model}
Let us consider the equations of motion in the following way:
\begin{eqnarray}
(E_1(t)\ket{\psi_{E1}}\frac{1}{n_{E_1}}\bra{\psi_{E1}}+E_2(t)\ket{\psi_{E2}}\bra{\psi_{E2}}+e_{12}(t)\ket{\psi_{E2}}\bra{\psi_{E1}}+e_{21}(t)\ket{\psi_{E1}}\bra{\psi_{E2}})(p_{I}\ket{\psi_{E1}}+p_{II}\ket{\psi_{E2}})=\nonumber \\
=\frac{d}{dt}(p_{I}\ket{\psi_{E1}}+p_{II}\ket{\psi_{E2}}).
\end{eqnarray}
This equation is equivalent to the set of 2 coupled ordinary differential equations given as
\begin{eqnarray}
E_1(t)p_{I}(t)+e_{21}(t)p_{II}(t)=\bra{\psi_{E1}(t)}\frac{d}{dt}(p_{I}(t)\ket{\psi_{E1}(t)})+\bra{\psi_{E1(t)}}\frac{d}{dt}(p_{II}(t)\ket{\psi_{E2}(t)}), \nonumber \\ 
E_2(t)p_{II}(t)+e_{12}(t)p_{I}(t)=\bra{\psi_{E2}(t)}\frac{d}{dt}(p_{I}(t)\ket{\psi_{E1}(t)})+\bra{\psi_{E2}(t)}\frac{d}{dt}(p_{II}(t)\ket{\psi_{E2}(t)}).
\end{eqnarray}
and can be rewritten to be as
\begin{eqnarray}
E_1(t)p_{I}(t)+e_{21}(t)p_{II}(t)=\frac{d}{dt}p_{I}(t)+p_{I}(t)(\bra{\psi_{E1}(t)}\frac{d}{dt}\ket{\psi_{E1}(t)}) 
+p_{II}(t)\bra{\psi_{E1(t)}}\frac{d}{dt}(\ket{\psi_{E2}(t)}), \nonumber \\ 
+e_{12}(t)p_{I}(t)+E_2(t)p_{II}(t)=\frac{d}{dt}p_{II}(t)+p_{II}(t)(\bra{\psi_{E2}(t)}\frac{d}{dt}\ket{\psi_{E2}(t)}) 
+p_{I}(t)\bra{\psi_{E2(t)}}\frac{d}{dt}(\ket{\psi_{E1}(t)}).
\end{eqnarray}
It will lead to further simplification that can be written as
\begin{eqnarray}
+[E_1(t)-(\bra{\psi_{E1}(t)}\frac{d}{dt}\ket{\psi_{E1}(t)})]p_{I}(t)+[e_{21}(t)-p_{II}(t)\bra{\psi_{E1(t)}}\frac{d}{dt}(\ket{\psi_{E2}(t)})]p_{II}(t)=\frac{d}{dt}p_{I}(t) 
\nonumber \\ 
+[e_{12}(t)-\bra{\psi_{E2}(t)}\frac{d}{dt}(\ket{\psi_{E1}(t)})]p_{I}(t)+[E_2(t)-(\bra{\psi_{E2}(t)}\frac{d}{dt}\ket{\psi_{E2}(t)})]p_{II}(t)=\frac{d}{dt}p_{II}(t). 
\end{eqnarray}
We can write it in the compact form as
\begin{eqnarray}
\begin{pmatrix}
[E_1(t)-(\bra{\psi_{E1}(t)}\frac{d}{dt}\ket{\psi_{E1}(t)})] & [e_{21}(t)-\bra{\psi_{E1(t)}}\frac{d}{dt}(\ket{\psi_{E2}(t)})] \\
[e_{12}(t)-\bra{\psi_{E2}(t)}\frac{d}{dt}(\ket{\psi_{E1}(t)})] & [E_2(t)-(\bra{\psi_{E2}(t)}\frac{d}{dt}\ket{\psi_{E2}(t)})] \\
\end{pmatrix}
\begin{pmatrix}
p_{I}(t) \\
p_{II}(t) \\
\end{pmatrix}=
\frac{d}{dt}
\begin{pmatrix}
p_{I}(t) \\
p_{II}(t) \\
\end{pmatrix}.
\end{eqnarray}
The solution is analytical given as
\begin{eqnarray}
exp
\begin{pmatrix}
\int_{t_0}^{t}dt'[E_1(t')-(\bra{\psi_{E1}(t')}\frac{d}{dt'}\ket{\psi_{E1}(t')})] & \int_{t_0}^{t}dt'[e_{21}(t')-\bra{\psi_{E1(t')}}\frac{d}{dt'}(\ket{\psi_{E2}(t')})] \\
\int_{t_0}^{t}dt'[e_{12}(t')-\bra{\psi_{E2}(t')}\frac{d}{dt'}(\ket{\psi_{E1}(t')})] & \int_{t_0}^{t}dt'[E_2(t')-(\bra{\psi_{E2}(t')}\frac{d}{dt'}\ket{\psi_{E2}(t')})] \\
\end{pmatrix}
\begin{pmatrix}
p_{I}(t_0) \\
p_{II}(t_0) \\
\end{pmatrix}=
\begin{pmatrix}
p_{I}(t) \\
p_{II}(t) \\
\end{pmatrix}
\end{eqnarray}
and can be written as
\begin{eqnarray}
\hat{G}(t,t_0)
\begin{pmatrix}
p_{I}(t_0) \\
p_{II}(t_0) \\
\end{pmatrix}=exp
\begin{pmatrix}
g_{1,1}(t,t_0)  & g_{1,2}(t,t_0) \\
g_{2,1})(t,t_0) & g_{2,1}(t,t_0) \\
\end{pmatrix}
\begin{pmatrix}
p_{I}(t_0) \\
p_{II}(t_0) \\
\end{pmatrix}=
\begin{pmatrix}
p_{I}(t) \\
p_{II}(t) \\
\end{pmatrix}
=
\begin{pmatrix}
G_{1,1}(t,t_0) & G_{1,2}(t,t_0) \\
G_{2,1})(t,t_0) & G_{2,1}(t,t_0) \\
\end{pmatrix}
\begin{pmatrix}
p_{I}(t_0) \\
p_{II}(t_0) \\
\end{pmatrix}=
\begin{pmatrix}
p_{I}(t) \\
p_{II}(t) \\
\end{pmatrix},
\end{eqnarray}
where
\begin{eqnarray}
g_{1,1}(t,t_0)=\int_{t_0}^{t}dt'[E_1(t')-(\bra{\psi_{E1}(t')}\frac{d}{dt'}\ket{\psi_{E1}(t')})] ,  \\
g_{1,2}(t,t_0)=\int_{t_0}^{t}dt'[e_{21}(t')-\bra{\psi_{E1}(t')}\frac{d}{dt'}(\ket{\psi_{E2}(t')})] ,  \\
g_{2,1}(t,t_0)=\int_{t_0}^{t}dt'[e_{12}(t')-\bra{\psi_{E2}(t')}\frac{d}{dt'}(\ket{\psi_{E1}(t')})] ,  \\
g_{2,2}(t,t_0)=\int_{t_0}^{t}dt'[E_2(t')-(\bra{\psi_{E2}(t')}\frac{d}{dt'}\ket{\psi_{E2}(t')})].
\end{eqnarray}
 and we have the corresponding classical eigenstates
\begin{eqnarray}
\bra{\psi_{E_1}(t')}(\frac{d}{dt'}\ket{\psi_{E_1}(t')})=\nonumber \\
\Bigg[\frac{1}{2 s_{21}+(-\sqrt{(s_{11}-s_{22})^2+4s_{12}s_{21}}+s_{11}-s_{22})}
\begin{pmatrix}
-\sqrt{(s_{11}-s_{22})^2+4s_{12}s_{21}}+s_{11}-s_{22} & 2 s_{21} 
\end{pmatrix}\Bigg] \times \nonumber \\ \times
\frac{d}{dt}\Bigg[\frac{1}{2 s_{21}+(-\sqrt{(s_{11}-s_{22})^2+4s_{12}s_{21}}+s_{11}-s_{22})}
\begin{pmatrix}
-\sqrt{(s_{11}-s_{22})^2+4s_{12}s_{21}}+s_{11}-s_{22} \\ 2 s_{21} 
\end{pmatrix}\Bigg], \nonumber \\
\bra{\psi_{E_1}(t')}(\frac{d}{dt'}\ket{\psi_{E_2}(t')})=\nonumber \\
\Bigg[\frac{1}{2 s_{21}+(-\sqrt{(s_{11}-s_{22})^2+4s_{12}s_{21}}+s_{11}-s_{22})}
\begin{pmatrix}
-\sqrt{(s_{11}-s_{22})^2+4s_{12}s_{21}}+s_{11}-s_{22} & 2 s_{21} 
\end{pmatrix}\Bigg] \times \nonumber \\ \times
\frac{d}{dt'}\Bigg[\frac{1}{2 s_{21}+(+\sqrt{(s_{11}-s_{22})^2+4s_{12}s_{21}}+s_{11}-s_{22})}
\begin{pmatrix}
+\sqrt{(s_{11}-s_{22})^2+4s_{12}s_{21}}+s_{11}-s_{22} \\ 2 s_{21} 
\end{pmatrix}\Bigg],
\end{eqnarray}
\begin{eqnarray}
\bra{\psi_{E_1}(t')}(\frac{d}{dt'}\ket{\psi_{E_2}(t')})=\nonumber \\
\Bigg[\frac{1}{2 s_{21}+(-\sqrt{(s_{11}-s_{22})^2+4s_{12}s_{21}}+s_{11}-s_{22})}
\begin{pmatrix}
-\sqrt{(s_{11}-s_{22})^2+4s_{12}s_{21}}+s_{11}-s_{22} & 2 s_{21} 
\end{pmatrix}\Bigg] \times \nonumber \\ \times
\frac{d}{dt'}\Bigg[\frac{1}{2 s_{21}+(+\sqrt{(s_{11}-s_{22})^2+4s_{12}s_{21}}+s_{11}-s_{22})}
\begin{pmatrix}
+\sqrt{(s_{11}-s_{22})^2+4s_{12}s_{21}}+s_{11}-s_{22} \\ 2 s_{21} 
\end{pmatrix}\Bigg], \nonumber \\
\bra{\psi_{E_2}(t')}(\frac{d}{dt'}\ket{\psi_{E_2}(t')})=\nonumber \\
\Bigg[\frac{1}{2 s_{21}+(+\sqrt{(s_{11}-s_{22})^2+4s_{12}s_{21}}+s_{11}-s_{22})}
\begin{pmatrix}
+\sqrt{(s_{11}-s_{22})^2+4s_{12}s_{21}}+s_{11}-s_{22} & 2 s_{21} 
\end{pmatrix}\Bigg] \times \nonumber \\ \times
\frac{d}{dt'}\Bigg[\frac{1}{2 s_{21}+(+\sqrt{(s_{11}-s_{22})^2+4s_{12}s_{21}}+s_{11}-s_{22})}
\begin{pmatrix}
+\sqrt{(s_{11}-s_{22})^2+4s_{12}s_{21}}+s_{11}-s_{22} \\ 2 s_{21} 
\end{pmatrix}\Bigg],
\end{eqnarray}

and
\begin{eqnarray}
  \int_{t_0}^{t}dt'E_1(t') &=& \int_{t_0}^{t}dt'\frac{1}{2} \left(-\sqrt{(s_{11}(t')-s_{22}(t'))^2+4s_{12}(t')s_{21}(t')}+s_{11}(t')+s_{22}(t')\right) \\
  \int_{t_0}^{t}dt'E_2(t') &=& \int_{t_0}^{t}dt'\frac{1}{2} \left(+\sqrt{(s_{11}(t')-s_{22}(t'))^2+4s_{12}(t')s_{21}(t')}+s_{11}(t')+s_{22}(t')\right).
\end{eqnarray}
and
\begin{eqnarray}
G_{1,1}(t,t_0)=e^{\frac{g_{11}(t,t_0)+g_{22}(t,t_0)}{2}}\Bigg[+\frac{(g_{11}(t,t_0)-g_{22}(t,t_0)) \sinh \left(\frac{1}{2} \sqrt{(g_{11}(t,t_0)-g_{22}(t,t_0))^2+4 g_{12}(t,t_0)
   g_{21}(t,t_0)}\right)}{\sqrt{(g_{11}(t,t_0)-g_{22}(t,t_0))^2+4 g_{12}(t,t_0) g_{21}(t,t_0)}}+\nonumber \\ +\cosh \left(\frac{1}{2} \sqrt{(g_{11}(t,t_0)-g_{22}(t,t_0))^2+4
   g_{12}(t,t_0)g_{21}(t,t_0)}\right)\Bigg]
\end{eqnarray}
\begin{eqnarray}
G_{2,2}(t,t_0)=e^{\frac{g_{11}(t,t_0)+g_{22}(t,t_0)}{2}} \Bigg[-\frac{(g_{11}(t,t_0)-g_{22}(t,t_0)) \sinh \left(\frac{1}{2} \sqrt{(g_{11}(t,t_0)-g_{22}(t,t_0))^2+4 g_{12}(t,t_0)
   g_{21}(t,t_0)}\right)}{\sqrt{(g_{11}(t,t_0)-g_{22}(t,t_0))^2+4 g_{12}(t,t_0)g_{21}(t,t_0)}}+ \nonumber \\ + \cosh \left(\frac{1}{2} \sqrt{(g_{11}(t,t_0)-g_{22}(t,t_0))^2+4
   g_{12}(t,t_0) g_{21}(t,t_0)}\right)\Bigg]
\end{eqnarray}
\begin{eqnarray}
G_{1,2}(t,t_0)=\frac{2 g_{12}(t,t_0) e^{\frac{g_{11}(t,t_0)+g_{22}(t,t_0)}{2}} \sinh \left(\frac{1}{2} \sqrt{(g_{11}(t,t_0)-g_{22}(t,t_0))^2+4 g_{12}(t,t_0)
   g_{21}(t,t_0)}\right)}{\sqrt{(g_{11}(t,t_0)-g_{22}(t,t_0))^2+4 g_{12}(t,t_0) g_{21}(t,t_0)}},
\end{eqnarray}
\begin{eqnarray}
G_{2,1}(t,t_0)=\frac{2 g_{21}(t,t_0) e^{\frac{g_{11}(t,t_0)+g_{22}(t,t_0)}{2}} \sinh \left(\frac{1}{2} \sqrt{(g_{11}(t,t_0)-g_{22}(t,t_0))^2+4g_{12}(t,t_0)
   g_{21}(t,t_0)}\right)}{\sqrt{(g_{11}(t,t_0)-g_{22}(t,t_0))^2+4g_{12}(t,t_0)g_{21}(t,t_0)}}
\end{eqnarray}

\section{Analogy of quantum entanglement in classical epidemic model}
 It is now easy to generalize our considerations for 2 coupled statistical ensembles as corresponding to cities with flight traffic.
 We have
 \begin{eqnarray}
\begin{pmatrix}
s_{11}(t)_A & s_{12}(t)_A & 0 & s_{1A2B}(t) \\
s_{21}(t)_A & s_{22}(t)_A & s_{2A2B}(t) & 0 \\
0 & s_{2A1B}(t) & s_{11}(t)_B & s_{12}(t)_B \\
s_{2A1B}(t) & 0 & s_{21}(t)_B & s_{22}(t)_B \\
\end{pmatrix}
\begin{pmatrix}
p_{1}(t)_A \\
p_{2}(t)_A \\
p_{1}(t)_B \\
p_{2}(t)_B \\
\end{pmatrix}=
\hat{S}|\psi_{classical}>
=\frac{d}{dt}
\begin{pmatrix}
p_{1}(t)_A \\
p_{2}(t)_A \\
p_{1}(t)_B \\
p_{2}(t)_B \\
\end{pmatrix}=\nonumber \\
=\hat{S}_t(p_{1A}(t)\ket{1}_A|\ket{1}_B+p_{2A}(t)\ket{1}_A\ket{2}_B+p_{1B}(t)\ket{2}_A|\ket{1}_B+p_{2B}(t)\ket{2}_A\ket{2}_B)=\frac{d}{dt}|\psi_{classical}>
\end{eqnarray}
We make the analytic simplifications by assumption of two symmetric systems A and B interacting interacting in asymmetric way as
 \begin{eqnarray}
\begin{pmatrix}
s_{11}(t) & s_{12}(t) & 0 & s(t) \\
s_{21}(t) & s_{22}(t) & s(t) & 0 \\
0 & s(t) & s_{11}(t) & s_{12}(t) \\
s(t) & 0 & s_{21}(t) & s_{22}(t) \\
\end{pmatrix}
\begin{pmatrix}
p_{1}(t)_A \\
p_{2}(t)_A \\
p_{1}(t)_B \\
p_{2}(t)_B \\
\end{pmatrix}=\frac{d}{dt}
\begin{pmatrix}
p_{1}(t)_A \\
p_{2}(t)_A \\
p_{1}(t)_B \\
p_{2}(t)_B \\
\end{pmatrix}
.
\end{eqnarray}
We have 4 eigenstates
\begin{eqnarray}
\ket{V_1(t)}=
\begin{pmatrix}
-\frac{\sqrt{4 (S-\text{S12}) (S-\text{S21})+(\text{S11}-\text{S22})^2}-\text{S11}+\text{S22}}{2 (S-\text{S21})}, \\
-1, \\
\frac{\sqrt{4(S-\text{S12}) (S-\text{S21})+(\text{S11}-\text{S22})^2}-\text{S11}+\text{S22}}{2 (S-\text{S21})}, \\
   1
\end{pmatrix}
\end{eqnarray}

\begin{eqnarray}
\ket{V_2(t)}=
\begin{pmatrix}
\frac{\sqrt{4 (S-\text{S12}) (S-\text{S21})+(\text{S11}-\text{S22})^2}+\text{S11}-\text{S22}}{2 (S-\text{S21})}, \\
-1, \\
-\frac{\sqrt{4(S-\text{S12}) (S-\text{S21})+(\text{S11}-\text{S22})^2}+\text{S11}-\text{S22}}{2 (S-\text{S21})}, \\
   1
\end{pmatrix}
\end{eqnarray}
\begin{eqnarray}
\ket{V_3(t)}=
\begin{pmatrix}
\frac{-\sqrt{4 (S+\text{S12}) (S+\text{S21})+(\text{S11}-\text{S22})^2}+\text{S11}-\text{S22}}{2 (S+\text{S21})}, \\
1, \\
\frac{-\sqrt{4
   (S+\text{S12}) (S+\text{S21})+(\text{S11}-\text{S22})^2}+\text{S11}-\text{S22}}{2 (S+\text{S21})}, \\
   1
\end{pmatrix}
\end{eqnarray}

\begin{eqnarray}
\ket{V_4(t)}=
\begin{pmatrix}
\frac{\sqrt{4 (S+\text{S12}) (S+\text{S21})+(\text{S11}-\text{S22})^2}+\text{S11}-\text{S22}}{2 (S+\text{S21})}, \\
1, \\
\frac{\sqrt{4
   (S+\text{S12}) (S+\text{S21})+(\text{S11}-\text{S22})^2}+\text{S11}-\text{S22}}{2 (S+\text{S21})},\\
1
\end{pmatrix}
\end{eqnarray}
and $\ket{V_3(t)}$ and $\ket{V_4(t)}$ are physically justifiable in framework of epidemic model that can be used for classical entanglement.
We have 4 projectors corresponding to the measurement of $p_{1A}$, $p_{2A}$, $p_{1B}$ and $p_{2B}$ represented as matrices
\begin{eqnarray}
\hat{P}_{p1A}=
\begin{pmatrix}
1 & 0 & 0 & 0 \\
0 & 0 & 0 & 0 \\
0 & 0 & 1 & 0 \\
0 & 0 & 0 & 1 \\
\end{pmatrix},
\hat{P}_{p2A}=
\begin{pmatrix}
0 & 0 & 0 & 0 \\
0 & 1 & 0 & 0 \\
0 & 0 & 1 & 0 \\
0 & 0 & 0 & 1 \\
\end{pmatrix},
\hat{P}_{p1B}=
\begin{pmatrix}
1 & 0 & 0 & 0 \\
0 & 1 & 0 & 0 \\
0 & 0 & 1 & 0 \\
0 & 0 & 0 & 0 \\
\end{pmatrix},
\hat{P}_{p2B}=
\begin{pmatrix}
1 & 0 & 0 & 0 \\
0 & 1 & 0 & 0 \\
0 & 0 & 0 & 0 \\
0 & 0 & 0 & 1 \\
\end{pmatrix}
\end{eqnarray}
Measurement conducted on system A also brings the change of system B state what is due to presence of non-diagonal matrix elements in system evolution equations (generalized
epidemic model). It is therefore analogical to measurement of quantum entangled state. After the measurements we obtain the following classical states

\begin{eqnarray}
\hat{P}_{p1A}\ket{\psi_{classical}}=
\begin{pmatrix}
1 & 0 & 0 & 0 \\
0 & 0 & 0 & 0 \\
0 & 0 & 1 & 0 \\
0 & 0 & 0 & 1 \\
\end{pmatrix}
\begin{pmatrix}
p_{A1}(t) \\
p_{A2}(t) \\
p_{B1}(t) \\
p_{B2}(t) \\
\end{pmatrix}
=
\begin{pmatrix}
1 \\
0 \\
p_{B1}(t) \\
p_{B2}(t) \\
\end{pmatrix}
, \nonumber \\
\hat{P}_{p2A}\ket{\psi_{classical}}=
\begin{pmatrix}
0 & 0 & 0 & 0 \\
0 & 1 & 0 & 0 \\
0 & 0 & 1 & 0 \\
0 & 0 & 0 & 1 \\
\end{pmatrix}
\begin{pmatrix}
p_{A1}(t) \\
p_{A2}(t) \\
p_{B1}(t) \\
p_{B2}(t) \\
\end{pmatrix}
=
\begin{pmatrix}
0 \\
1 \\
p_{B1}(t) \\
p_{B2}(t) \\
\end{pmatrix}
\nonumber \\
,
\hat{P}_{p1B}\ket{\psi_{classical}}=
\begin{pmatrix}
1 & 0 & 0 & 0 \\
0 & 1 & 0 & 0 \\
0 & 0 & 1 & 0 \\
0 & 0 & 0 & 0 \\
\end{pmatrix}
\begin{pmatrix}
p_{A1}(t) \\
p_{A2}(t) \\
p_{B1}(t) \\
p_{B2}(t) \\
\end{pmatrix}
=
\begin{pmatrix}
p_{A1}(t) \\
p_{A2}(t) \\
1 \\
0 \\
\end{pmatrix}
, \nonumber \\
\hat{P}_{p2B}\ket{\psi_{classical}}=
\begin{pmatrix}
1 & 0 & 0 & 0 \\
0 & 1 & 0 & 0 \\
0 & 0 & 0 & 0 \\
0 & 0 & 0 & 1 \\
\end{pmatrix}
\begin{pmatrix}
p_{A1}(t) \\
p_{A2}(t) \\
p_{B1}(t) \\
p_{B2}(t) \\
\end{pmatrix}
=
\begin{pmatrix}
p_{A1}(t) \\
p_{A2}(t) \\
0 \\
1 \\
\end{pmatrix}
\end{eqnarray}

\section{2 classical statistical systems interacting in quantum mechanical way}
We are going to reformulate the description of 2 classical noninteracting systems A and B.
We have
\begin{eqnarray}
\hat{H}_0(t)=
\begin{pmatrix}
s_{11A}(t) & s_{12A}(t) \\
s_{21A}(t) & s_{22A}(t) \\
\end{pmatrix} \times
\begin{pmatrix}
1 & 0 \\
0 & 1 \\
\end{pmatrix} +
\begin{pmatrix}
1 & 0 \\
0 & 1 \\
\end{pmatrix} \times
\begin{pmatrix}
s_{11B}(t) & s_{12B}(t) \\
s_{21B}(t) & s_{22B}(t) \\
\end{pmatrix}= \nonumber \\
=
\begin{pmatrix}
s_{11A}(t)              & 0                   & s_{12A}(t) & 0          \\
0                       & s_{11A}(t)          & 0          & s_{12A}(t) \\
s_{21A}(t)              & 0                   & s_{22A}(t) & 0          \\
0                       & s_{21A}(t)          & 0          & s_{22A}(t) \\
\end{pmatrix} +
\begin{pmatrix}
s_{11B}(t)              & s_{12B}(t)          & 0          & 0          \\
s_{21B}(t)              & s_{22B}(t)          & 0          & 0          \\
0                       & 0                   & s_{11B}(t) & s_{12B}(t) \\
0                       & 0                   & s_{21B}(t) & s_{22B}(t) \\
\end{pmatrix}= \nonumber \\
=
\begin{pmatrix}
s_{11A}(t) +s_{11B}(t)  & s_{12B}(t)                     & s_{12A}(t)              & 0                     \\
s_{21B}(t)              & s_{11A}(t)+s_{22B}(t)          & 0                       & s_{12A}(t)            \\
s_{21A}(t)              & 0                              &  s_{22A}(t)+ s_{11B}(t) & s_{12B}(t)            \\
0                       & s_{21A}(t)                     & s_{21B}(t)              & s_{22B}(t)+s_{22A}(t) \\
\end{pmatrix}= \nonumber \\
=(E_{1A}(t)\ket{\psi_{E_{1A}}(t)}\bra{\psi_{E_{1A}}(t)}+E_{2A}(t)\ket{\psi_{E_{2A}}(t)}\bra{\psi_{E_{2A}}(t)})(\ket{\psi_{E_{1B}}(t)}\bra{\psi_{E_{1B}}(t)}+\ket{\psi_{E_{2B}}(t)}\bra{\psi_{E_{2B}}(t)})+\nonumber \\
+(\ket{\psi_{E_{1A}}(t)}\bra{\psi_{E_{1A}}(t)}+\ket{\psi_{E_{2A}}(t)}\bra{\psi_{E_{2A}}(t)})(E_{1B}(t)\ket{\psi_{E_{1B}}(t)}\bra{\psi_{E_{1B}}(t)}+E_{2B}(t)\ket{\psi_{E_{2B}}(t)}\bra{\psi_{E_{2B}}(t)}).
\end{eqnarray}
and we have the
\begin{eqnarray}
\ket{\psi(t)}=\gamma_1(t)\ket{\psi_{E_{1A}}(t)}\ket{\psi_{E_{1B}}(t)}+ \gamma_2(t)\ket{\psi_{E_{1A}}(t)}\ket{\psi_{E_{2B}}(t)}+\gamma_3(t)\ket{\psi_{E_{2A}}(t)}\ket{\psi_{E_{1B}}(t)}+\gamma_4(t)\ket{\psi_{E_{2A}}(t)}\ket{\psi_{E_{2B}}(t)}= \nonumber \\
=(p_{1A}(t)\ket{x_{1A}}+p_{2A}(t)\ket{x_{2A}})(p_{1B}(t)\ket{x_{1B}}+p_{2B}(t)\ket{x_{2B}})= \nonumber \\
=p_{1A}(t)p_{1B}(t)\ket{x_{1A}}\ket{x_{1B}}+p_{1A}(t)p_{2B}(t)\ket{x_{1A}}\ket{x_{2B}}+p_{2A}(t)p_{1B}(t)\ket{x_{2A}}\ket{x_{1B}}+p_{2A}(t)p_{2B}(t)\ket{x_{2A}}\ket{x_{2B}},
\end{eqnarray}
where $\ket{x_{kA}}\ket{x_{lB}}$ are time-independent and k and l are 1 or 2 and $\gamma_1(t)+\gamma_2(t)+\gamma_3(t)+\gamma_4(t)=1$. We observe that $\hat{H}_0(t)\ket{\psi}_t=\frac{d}{dt}\ket{\psi}_t$ explicitly given as
\begin{eqnarray}
\frac{d}{dt}
\begin{pmatrix}
p_{1A}(t)p_{1B}(t) \\
p_{1A}(t)p_{2B}(t) \\
p_{2A}(t)p_{1B}(t) \\
p_{2A}(t)p_{2B}(t) \\
\end{pmatrix}=
\begin{pmatrix}
s_{11A}(t) +s_{11B}(t)  & s_{12B}(t)                     & s_{12A}(t)              & 0                     \\
s_{21B}(t)              & s_{11A}(t)+s_{22B}(t)          & 0                       & s_{12A}(t)            \\
s_{21A}(t)              & 0                              &  s_{22A}(t)+ s_{11B}(t) & s_{12B}(t)            \\
0                       & s_{21A}(t)                     & s_{21B}(t)              & s_{22B}(t)+s_{22A}(t) \\
\end{pmatrix} 
\begin{pmatrix}
p_{1A}(t)p_{1B}(t) \\
p_{1A}(t)p_{2B}(t) \\
p_{2A}(t)p_{1B}(t) \\
p_{2A}(t)p_{2B}(t) \\
\end{pmatrix}= \nonumber \\
=\begin{pmatrix}
s_{11A}(t) +s_{11B}(t)  & s_{12B}(t)                     & s_{12A}(t)              & 0                     \\
s_{21B}(t)              & s_{11A}(t)+s_{22B}(t)          & 0                       & s_{12A}(t)            \\
s_{21A}(t)              & 0                              &  s_{22A}(t)+ s_{11B}(t) & s_{12B}(t)            \\
0                       & s_{21A}(t)                     & s_{21B}(t)              & s_{22B}(t)+s_{22A}(t) \\
\end{pmatrix} 
\begin{pmatrix}
p_{IQ}(t)   \\
p_{IIQ}(t)  \\
p_{IIIQ}(t) \\
p_{IVQ}(t)  \\
\end{pmatrix}
=\frac{d}{dt}
\begin{pmatrix}
p_{IQ}(t)   \\
p_{IIQ}(t)  \\
p_{IIIQ}(t) \\
p_{IVQ}(t)  \\
\end{pmatrix}
,
\end{eqnarray}
where $p_{IQ}(t)$, $p_{IIQ}(t)$, $p_{IIIQ}(t)$ and  $p_{IVQ}(t)$ (with $p_{IQ}(t)=p_{1A}(t)p_{1B}(t)$, $p_{IIQ}(t)=p_{1A}(t)p_{2B}(t)$, $p_{IIIQ}(t)=p_{2A}(t)p_{1B}(t)$, $p_{IVQ}(t)=p_{2A}(t)p_{2B}(t)$) are describing probabilities for 4 different states of statistic finite machine.
Similar situation occurs in the case of Schroedinger equation written for 2 non-interacting systems, but instead of probabilities we have square root of probabilities times phase factor.
In general case of interaction between A and B systems in classical epidemic model we have
\begin{eqnarray}
\hat{H}_0(t)+\hat{H}_{A-B}(t)
=E_{I}(t)\ket{\psi_{E_{1A}}(t)}\ket{\psi_{E_{1B}}(t)}\bra{\psi_{E_{1A}}(t)}\bra{\psi_{E_{1B}}(t)} 
+E_{II}(t)\ket{\psi_{E_{1A}}(t)}\ket{\psi_{E_{2B}}(t)}\bra{\psi_{E_{1A}}(t)}\bra{\psi_{E_{2B}}(t)}+ \nonumber \\
+E_{III}(t)\ket{\psi_{E_{2A}}(t)}\ket{\psi_{E_{1B}}(t)}\bra{\psi_{E_{2A}}(t)}\bra{\psi_{E_{1B}}(t)} 
+E_{IV}(t)\ket{\psi_{E_{2A}}(t)}\ket{\psi_{E_{2B}}(t)}\bra{\psi_{E_{2A}}(t)}\bra{\psi_{E_{2B}}(t)}+ \nonumber \\
+e_{(1A,1B) \rightarrow (1A,2B)}(t)\ket{\psi_{E_{1A}}(t)}\ket{\psi_{E_{1B}}(t)}\bra{\psi_{E_{1A}}(t)}\bra{\psi_{E_{2B}}(t)}
+e_{(1A,2B) \rightarrow (1A,1B)}(t)\ket{\psi_{E_{1A}}(t)}\ket{\psi_{E_{2B}}(t)}\bra{\psi_{E_{1A}}(t)}\bra{\psi_{E_{1B}}(t)}+ \nonumber \\
+e_{(2A,1B) \rightarrow (1A,1B)}(t)\ket{\psi_{E_{2A}}(t)}\ket{\psi_{E_{1B}}(t)}\bra{\psi_{E_{1A}}(t)}\bra{\psi_{E_{1B}}(t)}
+e_{(1A,1B) \rightarrow (2A,1B)}(t)\ket{\psi_{E_{1A}}(t)}\ket{\psi_{E_{1B}}(t)}\bra{\psi_{E_{2A}}(t)}\bra{\psi_{E_{1B}}(t)}+
\nonumber \\
+e_{(1A,2B) \rightarrow (2A,2B)}(t)\ket{\psi_{E_{1A}}(t)}\ket{\psi_{E_{2B}}(t)}\bra{\psi_{E_{2A}}(t)}\bra{\psi_{E_{2B}}(t)}+
e_{(2A,2B) \rightarrow (1A,2B)}(t)\ket{\psi_{E_{2A}}(t)}\ket{\psi_{E_{2B}}(t)}\bra{\psi_{E_{1A}}(t)}\bra{\psi_{E_{2B}}(t)}+ \nonumber \\
+e_{(2A,1B) \rightarrow (2A,2B)}(t)\ket{\psi_{E_{2A}}(t)}\ket{\psi_{E_{1B}}(t)}\bra{\psi_{E_{2A}}(t)}\bra{\psi_{E_{2B}}(t)}+
e_{(2A,2B) \rightarrow (2A,1B)}(t)\ket{\psi_{E_{2A}}(t)}\ket{\psi_{E_{2B}}(t)}\bra{\psi_{E_{2A}}(t)}\bra{\psi_{E_{1B}}(t)}+
\nonumber \\
+e_{(1A,1B) \rightarrow (2A,2B)}(t)\ket{\psi_{E_{1A}}(t)}\ket{\psi_{E_{1B}}(t)}\bra{\psi_{E_{2A}}(t)}\bra{\psi_{E_{2B}}(t)}+
e_{(2A,2B) \rightarrow (1A,1B)}(t)\ket{\psi_{E_{2A}}(t)}\ket{\psi_{E_{2B}}(t)}\bra{\psi_{E_{1A}}(t)}\bra{\psi_{E_{1B}}(t)}.
\end{eqnarray}
The presented analytical approach can be applied for system A with 4 distinct states as well as for system B with 4 distinct states since isolated system A(B) with 4 states can be described by evolution matrix 4 by 4 that has 4 analytical eigenvalues and eigenstates and becomes non-analytical for 5 and more distinct states due to fact that roots of polynomial of higher order than 4 becomes non-analytical and becomes numerical with some limited exceptions.
We can write the matrix
\begin{eqnarray}
\hat{H}_{E_{IQ},..,E_{IVQ}}=
\begin{pmatrix}
E_{IQ}(t) & e_{(1A,2B) \rightarrow (1A,1B)} & e_{(2A,1B) \rightarrow (1A,1B)} & e_{(2A,2B) \rightarrow (1A,1B)} \\
e_{(1A,1B) \rightarrow (1A,2B)} & E_{IIQ}(t) & e_{(2A,1B) \rightarrow (1A,2B)} & e_{(2A,2B) \rightarrow (1A,2B)} \\
e_{(1A,1B) \rightarrow (2A,1B)} & e_{(1A,1B) \rightarrow (2A,1B)} & E_{IIIQ}(t) & e_{(2A,2B) \rightarrow (1A,2B)} \\
e_{(1A,1B) \rightarrow (2A,2B)} & e_{(1A,1B) \rightarrow 2A,2B)} & e_{(2A,1B) \rightarrow (2A,2B)} & E_{IVQ}(t) \\
\end{pmatrix}
\end{eqnarray}
\section{From epidemic model to tight-binding equations}
\subsection{Case of 2 level classical stochastic finite state machine}
Let us be motivated by work on single electron devices by \cite{Fujisawa}, \cite{Likharev}, \cite{Petta}, \cite{Dirk} and \cite{Bashir19}.
Instead of probabilities it will be useful to operate with square root of probabilities as they are present in quantum mechanics and in Schroedinger or Dirac equation. Since
\begin{eqnarray}
\frac{d}{dt}(\sqrt{p_1}\sqrt{p_1})=2 \sqrt{p_1(t)}\frac{d}{dt}\sqrt{p_1(t)}, \nonumber \\
\frac{d}{dt}(\sqrt{p_2}\sqrt{p_2})=2 \sqrt{p_2(t)}\frac{d}{dt}\sqrt{p_2(t)}.
\end{eqnarray}
we can rewrite the epidemic equation as
\begin{eqnarray}
\begin{pmatrix}
s_{11}(t) & s_{12}(t) \\
s_{21}(t) & s_{22}(t) \\
\end{pmatrix}
\begin{pmatrix}
p_{1}(t) \\
p_{2}(t) \\
\end{pmatrix}=
\begin{pmatrix}
2 \sqrt{p_{1}(t)}\frac{d}{dt}\sqrt{p_1(t)} \\
2 \sqrt{p_{2}(t)}\frac{d}{dt}\sqrt{p_2(t)} \\
\end{pmatrix}
\end{eqnarray}
This equation is equivalent to
\begin{eqnarray}
\begin{pmatrix}
\frac{1}{2}s_{11}(t) & \frac{1}{2}\sqrt{\frac{p2(t)}{p1(t)}}s_{12}(t)\\ 
\frac{1}{2}\sqrt{\frac{p1(t)}{p2(t)}}s_{21}(t) & \frac{1}{2}s_{22}(t) \\ 
\end{pmatrix}
\begin{pmatrix}
\sqrt{p_1(t)} \\
\sqrt{p_2(t)} \\
\end{pmatrix}
=
\begin{pmatrix}
\frac{d}{dt}\sqrt{p_1(t)} \\
\frac{d}{dt}\sqrt{p_2(t)} \\
\end{pmatrix}=
\frac{i\hbar}{i\hbar}
\frac{d}{dt}
\begin{pmatrix}
\sqrt{p_1(t)} \\
\sqrt{p_2(t)} \\
\end{pmatrix}
\end{eqnarray}
and we set $t_1=i \hbar t$ so one has $dt_1=i \hbar dt$ and $t=\frac{t_1}{i\hbar}$ so we have \small
\begin{eqnarray}
\begin{pmatrix}
\frac{1}{2}s_{11}(\frac{t_1}{i\hbar}) & \frac{1}{2}\sqrt{\frac{p_1(t)}{p_2(t)}}s_{12}(\frac{t_1}{i\hbar})\\ 
\frac{1}{2}\sqrt{\frac{p_2(t)}{p_1(t)}}s_{21}(\frac{t_1}{i\hbar}) & \frac{1}{2}s_{22}(\frac{t_1}{i\hbar}) \\ 
\end{pmatrix} 
\begin{pmatrix}
\sqrt{p_1(\frac{t_1}{i\hbar})} \\
\sqrt{p_2(\frac{t_1}{i\hbar})} \\
\end{pmatrix}
=
i\hbar
\frac{d}{dt_1}
\begin{pmatrix}
\sqrt{p_1(\frac{t_1}{i\hbar})} \\
\sqrt{p_2(\frac{t_1}{i\hbar})} \\
\end{pmatrix}
\end{eqnarray}
\normalsize
The following notation is introduced:
\begin{eqnarray}
S_{12}[\frac{t_0}{i\hbar},t_1]=\frac{1}{2}\sqrt{\frac{p_1(t)}{p_2(t)}}s_{12}(\frac{t_1}{i\hbar})=S_{12R}[\frac{t_0}{i\hbar},t_1]+iS_{12I}[\frac{t_0}{i\hbar},t_1] \end{eqnarray}, 
\begin{eqnarray}
S_{21}[\frac{t_0}{i\hbar},t_1]=\frac{1}{2}\sqrt{\frac{p_2(t)}{p_1(t)}}s_{21}(\frac{t_1}{i\hbar})=S_{21R}[\frac{t_0}{i\hbar},t_1]+iS_{21I}[\frac{t_0}{i\hbar},t_1]
\end{eqnarray}
 and one obtains
\begin{eqnarray*}
\begin{pmatrix}
\frac{1}{2}s_{11}(\frac{t_1}{i\hbar})-\hbar \frac{d}{dt_1}\Theta_1(t) & S_{12R}[\frac{t_0}{i\hbar},t_1]+iS_{12R}[\frac{t_0}{i\hbar},t_1]e^{i(\Theta_1(t)-\Theta_2(t))}\\ 
S_{21R}[\frac{t_0}{i\hbar},t_1]+iS_{21I}[\frac{t_0}{i\hbar},t_1]e^{i((\Theta_2(t)-\Theta_1(t)))} & \frac{1}{2}s_{22}(\frac{t_1}{i\hbar})-\hbar \frac{d}{dt_1}\Theta_2(t) \\ 
\end{pmatrix}
\begin{pmatrix}
\sqrt{p_{1}(\frac{t_1}{i\hbar})}e^{i\Theta_1(t)} \\
\sqrt{p_{2}(\frac{t_1}{i\hbar})}e^{i\Theta_2(t)} \\
\end{pmatrix}
= \nonumber \\ =i \hbar \frac{d}{dt}
\begin{pmatrix}
\sqrt{p_{1}(\frac{t_1}{i\hbar})}e^{i\Theta_1(t)}\\
\sqrt{p_{2}(\frac{t_1}{i\hbar})}e^{i\Theta_2(t)} \\
\end{pmatrix}.
\end{eqnarray*}
Let us start from quantum mechanical perspective
\begin{eqnarray}
\begin{pmatrix}
E_{p1} & t_{sR}+it_{sI}\\
t_{sR}-it_{sI} & E_{p2}
\end{pmatrix}
\begin{pmatrix}
\sqrt{p_{1}}cos(\Theta_1)+i\sqrt{p_{1}}sin(\Theta_1)    \\
\sqrt{p_{2}}cos(\Theta_2)+i\sqrt{p_{1}}sin(\Theta_2)  \\
\end{pmatrix}=i\hbar \frac{d}{dt_1}
\begin{pmatrix}
\sqrt{p_{1}}cos(\Theta_1)+i\sqrt{p_{1}}sin(\Theta_1)  \\
\sqrt{p_{2}}cos(\Theta_2)+i\sqrt{p_{2}}sin(\Theta_2)  \\
\end{pmatrix}
\end{eqnarray}
and we obtain the set of 4 equations
\begin{eqnarray}
\begin{pmatrix}
E_{p1} & t_{sR}+it_{sI}\\
t_{sR}-it_{sI} & E_{p2}
\end{pmatrix}
\begin{pmatrix}
\sqrt{p_{1}}cos(\Theta_1)+i\sqrt{p_{1}}sin(\Theta_1)    \\
\sqrt{p_{2}}cos(\Theta_2)+i\sqrt{p_{2}}sin(\Theta_2)  \\
\end{pmatrix}=i\hbar \frac{d}{dt_1}
\begin{pmatrix}
\sqrt{p_{1}}cos(\Theta_1)+i\sqrt{p_{1}}sin(\Theta_1)  \\
\sqrt{p_{2}}cos(\Theta_2)+i\sqrt{p_{2}}sin(\Theta_2)  \\
\end{pmatrix}
\end{eqnarray}
that can be rewritten
\begin{eqnarray}
E_{p1}(\sqrt{p_{1}}cos(\Theta_1)+i\sqrt{p_{1}}sin(\Theta_1))+(t_{sR}+it_{sI})(\sqrt{p_{2}}cos(\Theta_1)+i\sqrt{p_{2}}sin\Theta_1))=\nonumber \\ =i\hbar \frac{d}{dt_1}(\sqrt{p_{1}}cos(\Theta_1)+i\sqrt{p_{1}}sin(\Theta_1)) \\
(t_{sR}-it_{sI})(\sqrt{p_{1}}cos(\Theta_1)+i\sqrt{p_{1}}sin\Theta_1)+E_{p2}(\sqrt{p_{2}}cos(\Theta_2)+i\sqrt{p_{2}}sin(\Theta_2))=\nonumber \\ =i\hbar \frac{d}{dt_1}(\sqrt{p_{2}}cos(\Theta_2)+i\sqrt{p_{2}}sin(\Theta_2)) \\
\end{eqnarray}
and that can be translated into
\begin{eqnarray}
E_{p1}\sqrt{p_{1}}cos(\Theta_1)+t_{sR}\sqrt{p_{2}}cos(\Theta_2)-t_{sI}\sqrt{p_{2}}sin(\Theta_2)=-\hbar \frac{d}{dt_1}\sqrt{p_{1}}sin(\Theta_1) \\
E_{p1}\sqrt{p_{1}}sin(\Theta_1)+t_{sR}\sqrt{p_{2}}sin(\Theta_2)+t_{sI}\sqrt{p_{2}}cos(\Theta_2)=+\hbar \frac{d}{dt_1}\sqrt{p_{1}}cos(\Theta_1) \\
E_{p2}\sqrt{p_{2}}cos(\Theta_2)+t_{sR}\sqrt{p_{1}}cos(\Theta_1)+t_{sI}\sqrt{p_{1}}sin(\Theta_1)=-\hbar \frac{d}{dt_1}\sqrt{p_{2}}sin(\Theta_2) \\
E_{p2}\sqrt{p_{2}}sin(\Theta_2)+t_{sR}\sqrt{p_{1}}sin(\Theta_1)-t_{sI}\sqrt{p_{1}}cos(\Theta_1)=+\hbar \frac{d}{dt_1}\sqrt{p_{2}}cos(\Theta_2) \\
\end{eqnarray}
so one can write

\begin{eqnarray}
\frac{1}{\hbar}
\begin{pmatrix}
0 & E_{p1} & +t_{sI} & t_{sR} \\
-E_{p1} & 0 & -t_{sR} & t_{sI}\\
-t_{sI} & t_{sR} & 0 & E_{p2}\\
-t_{sR} &
-it_{sI} &  -E_{p2} & 0\\
\end{pmatrix}
\begin{pmatrix}
\sqrt{p_{1}}cos(\Theta_1) \\
\sqrt{p_{1}}sin(\Theta_1)    \\
\sqrt{p_{2}}cos(\Theta_1)\\
\sqrt{p_{1}}sin(\Theta_1)  \\
\end{pmatrix}= \frac{d}{dt_1}
\begin{pmatrix}
\sqrt{p_{1}}cos(\Theta_1)\\
\sqrt{p_{1}}sin(\Theta_1)  \\
\sqrt{p_{2}}cos(\Theta_2) \\
\sqrt{p_{2}}sin(\Theta_2)  \\
\end{pmatrix}
\end{eqnarray}
what can be written as
\begin{eqnarray}
\hat{A}(t)
\begin{pmatrix}
\sqrt{p_{1}(t)}cos(\Theta_1(t)) \\
\sqrt{p_{1}(t)}sin(\Theta_1(t))    \\
\sqrt{p_{2}(t)}cos(\Theta_1(t))\\
\sqrt{p_{1}(t)}sin(\Theta_1(t))  \\
\end{pmatrix}= \frac{d}{dt}
\begin{pmatrix}
\sqrt{p_{1}(t)}cos(\Theta_1(t))\\
\sqrt{p_{1}(t)}sin(\Theta_1(t))  \\
\sqrt{p_{2}(t)}cos(\Theta_2(t)) \\
\sqrt{p_{2}(t)}sin(\Theta_2(t))  \\
\end{pmatrix},
\end{eqnarray}
what implies
\begin{eqnarray}
e^{\int_{t0}^{t}\hat{A}(t')dt'}
\begin{pmatrix}
\sqrt{p_{1}(t_0)}cos(\Theta_1(t_0))  \\
\sqrt{p_{1}(t_0)}sin(\Theta_1(t_0))  \\
\sqrt{p_{2}(t_0)}cos(\Theta_2(t_0))  \\
\sqrt{p_{2}(t_0)}sin(\Theta_2(t_0))  \\
\end{pmatrix}=
\begin{pmatrix}
\sqrt{p_{1}(t)}cos(\Theta_1(t))\\
\sqrt{p_{1}(t)}sin(\Theta_1(t))  \\
\sqrt{p_{2}(t)}cos(\Theta_2(t)) \\
\sqrt{p_{2}(t)}sin(\Theta_2(t))  \\
\end{pmatrix} 
\end{eqnarray}
Furthermore we can obtain
\small
\begin{eqnarray*}
2
\begin{pmatrix}
\sqrt{p_{1}(t)cos(\Theta_1(t))} & 0 & 0 & 0 \\
0 & \sqrt{p_{1}(t)sin(\Theta_1(t))}& 0 & 0   \\
0 & 0 & \sqrt{p_{2}(t)cos(\Theta_2(t))} & 0 \\
0 & 0 & 0 & \sqrt{p_{2}(t)sin(\Theta_2(t))}  \\
\end{pmatrix}
e^{\int_{t0}^{t}\hat{A}(t')dt'} \times \\ \times
\begin{pmatrix}
\frac{1}{\sqrt{p_{1}(t)cos(\Theta_1(t))^2}} & 0 & 0 & 0 \\
0 & \frac{1}{\sqrt{p_{1}(t)sin(\Theta_1(t))^2}} & 0 & 0   \\
0 & 0 & \frac{1}{\sqrt{p_{2}(t)cos(\Theta_2(t))^2}} & 0 \\
0 & 0 & 0 & \frac{1}{\sqrt{p_{2}(t)sin(\Theta_2(t))^2}}  \\
\end{pmatrix} \times \nonumber \\
\begin{pmatrix}
\sqrt{p_{1}(t)cos(\Theta_1(t))^2} & 0 & 0 & 0 \\
0 & \sqrt{p_{1}(t)sin(\Theta_1(t))^2}& 0 & 0   \\
0 & 0 & \sqrt{p_{2}(t)cos(\Theta_2(t))^2} & 0 \\
0 & 0 & 0 & \sqrt{p_{2}(t)sin(\Theta_2(t))^2}  \\
\end{pmatrix}
\begin{pmatrix}
\sqrt{p_{1}(t_0)cos(\Theta_1(t_0))}  \\
\sqrt{p_{1}(t_0)sin(\Theta_1(t_0))}  \\
\sqrt{p_{2}(t_0)cos(\Theta_2(t_0))}  \\
\sqrt{p_{2}(t_0)sin(\Theta_2(t_0))}  \\
\end{pmatrix}=\nonumber \\
2
\begin{pmatrix}
\sqrt{p_{1}(t)cos(\Theta_1(t))^2} & 0 & 0 & 0 \\
0 & \sqrt{p_{1}(t)sin(\Theta_1(t))^2}& 0 & 0   \\
0 & 0 & \sqrt{p_{2}(t)cos(\Theta_2(t)^2)} & 0 \\
0 & 0 & 0 & \sqrt{p_{2}(t)sin(\Theta_2(t)^2)}  \\
\end{pmatrix}
\begin{pmatrix}
\frac{d}{dt}\sqrt{p_{1}(t)cos(\Theta_1(t))^2}\\
\frac{d}{dt}\sqrt{p_{1}(t)sin(\Theta_1(t))^2}  \\
\frac{d}{dt}\sqrt{p_{2}(t)cos(\Theta_2(t))^2} \\
\frac{d}{dt}\sqrt{p_{2}(t)sin(\Theta_2(t))^2}  \\
\end{pmatrix}.
\end{eqnarray*}
\normalsize
The last set of equations is equivalent to
\small
\begin{eqnarray*}
\Bigg[
2
\begin{pmatrix}
\sqrt{p_{1}(t)(cos(\Theta_1(t)))^2} & 0 & 0 & 0 \\
0 & \sqrt{p_{1}(t)(sin(\Theta_1(t)))^2}& 0 & 0   \\
0 & 0 & \sqrt{p_{2}(t)(cos(\Theta_2(t)))^2} & 0 \\
0 & 0 & 0 & \sqrt{p_{2}(t)(sin(\Theta_2(t)))^2}  \\
\end{pmatrix}
\Bigg[
e^{\int_{t0}^{t}\hat{A}(t')dt'} \Bigg] \times \\ \times
\begin{pmatrix}
\frac{1}{\sqrt{p_{1}(t)(cos(\Theta_1(t)))^2}} & 0 & 0 & 0 \\
0 & \frac{1}{\sqrt{p_{1}(t)(sin(\Theta_1(t)))^2}} & 0 & 0   \\
0 & 0 & \frac{1}{\sqrt{p_{2}(t)(cos(\Theta_2(t)))^2}} & 0 \\
0 & 0 & 0 & \frac{1}{\sqrt{p_{2}(t)(sin(\Theta_2(t)))^2}}  \\
\end{pmatrix}
\Bigg]
\begin{pmatrix}
p_{1}(t_0)(cos(\Theta_1(t_0)))^2=p_{1R}(t)  \\
p_{1}(t_0)(sin(\Theta_1(t_0)))^2=p_{1I}(t)  \\
p_{2}(t_0)(cos(\Theta_2(t_0)))^2=p_{2R}(t)  \\
p_{2}(t_0)(sin(\Theta_2(t_0)))^2=p_{2I}(t)  \\
\end{pmatrix}=\nonumber \\
=\frac{d}{dt}
\begin{pmatrix}
p_{1}(t)(cos(\Theta_1(t))^2)\\
p_{1}(t)(sin(\Theta_1(t))^2)  \\
p_{2}(t)(cos(\Theta_2(t))^2) \\
p_{2}(t)(sin(\Theta_2(t))^2)  \\
\end{pmatrix} =
\frac{d}{dt}
\begin{pmatrix}
p_{1R}(t)  \\
p_{1I}(t)  \\
p_{2R}(t)  \\
p_{2I}(t)  \\
\end{pmatrix}
\end{eqnarray*}
\normalsize




\subsection{Case of 2 coupled 2 level classical stochastic finite state machines}
Instead of probabilities it will be useful to operate with square root of probabilities as they are present in quantum mechanics and in Schroedinger or Dirac equation. Since
\begin{eqnarray}
\frac{d}{dt}(\sqrt{p_1}\sqrt{p_1})=2 \sqrt{p_1(t)}\frac{d}{dt}\sqrt{p_1(t)}, \nonumber \\
\frac{d}{dt}(\sqrt{p_2}\sqrt{p_2})=2 \sqrt{p_2(t)}\frac{d}{dt}\sqrt{p_2(t)}, \nonumber \\
\frac{d}{dt}(\sqrt{p_3}\sqrt{p_3})=2 \sqrt{p_3(t)}\frac{d}{dt}\sqrt{p_3(t)}, \nonumber \\
\frac{d}{dt}(\sqrt{p_4}\sqrt{p_4})=2 \sqrt{p_4(t)}\frac{d}{dt}\sqrt{p_4(t)}, \nonumber \\
\end{eqnarray}
we can rewrite the epidemic equation as
\begin{eqnarray}
\begin{pmatrix}
s_{11}(t) & s_{12}(t) & s_{13}(t) & s_{14}(t)\\
s_{21}(t) & s_{22}(t) & s_{23}(t) & s_{24}(t)\\
s_{31}(t) & s_{32}(t) & s_{33}(t) & s_{34}(t)\\
s_{41}(t) & s_{42}(t) & s_{43}(t) & s_{44}(t)\\
\end{pmatrix}
\begin{pmatrix}
p_{1}(t) \\
p_{2}(t) \\
p_{3}(t) \\
p_{4}(t) \\
\end{pmatrix}=
\begin{pmatrix}
2 \sqrt{p_{1}(t)}\frac{d}{dt}\sqrt{p_1(t)} \\
2 \sqrt{p_{2}(t)}\frac{d}{dt}\sqrt{p_2(t)} \\
2 \sqrt{p_{3}(t)}\frac{d}{dt}\sqrt{p_3(t)} \\
2 \sqrt{p_{4}(t)}\frac{d}{dt}\sqrt{p_4(t)} \\
\end{pmatrix}
\end{eqnarray}
Equivalently we have
\begin{eqnarray}
\begin{pmatrix}
s_{11}(t) & s_{12}(t) & s_{13}(t) & s_{14}(t)\\
s_{21}(t) & s_{22}(t) & s_{23}(t) & s_{24}(t)\\
s_{31}(t) & s_{32}(t) & s_{33}(t) & s_{34}(t)\\
s_{41}(t) & s_{42}(t) & s_{43}(t) & s_{44}(t)\\
\end{pmatrix}
\begin{pmatrix}
\sqrt{p_1} & 0 & 0 & 0\\
0 & \sqrt{p_{2}(t)} & 0 & 0\\
0 & 0 & \sqrt{p_{3}(t)} & 0\\
0 & 0 & 0 & \sqrt{p_{4}(t)}\\
\end{pmatrix}
\begin{pmatrix}
\sqrt{p_{1}(t)} \\
\sqrt{p_{2}(t)} \\
\sqrt{p_{3}(t)} \\
\sqrt{p_{4}(t)} \\
\end{pmatrix}=
\begin{pmatrix}
2 \sqrt{p_{1}(t)}\frac{d}{dt}\sqrt{p_1(t)} \\
2 \sqrt{p_{2}(t)}\frac{d}{dt}\sqrt{p_2(t)} \\
2 \sqrt{p_{3}(t)}\frac{d}{dt}\sqrt{p_3(t)} \\
2 \sqrt{p_{4}(t)}\frac{d}{dt}\sqrt{p_4(t)} \\
\end{pmatrix}
\end{eqnarray}
and we have
\begin{eqnarray}
\frac{1}{2}
\begin{pmatrix}
\frac{1}{\sqrt{p_1}} & 0 & 0 & 0\\
0 & \frac{1}{\sqrt{p_{2}(t)}} & 0 & 0\\
0 & 0 & \frac{1}{\sqrt{p_{3}(t)}} & 0\\
0 & 0 & 0 & \frac{1}{\sqrt{p_{4}(t)}}\\
\end{pmatrix}
\begin{pmatrix}
s_{11}(t) & s_{12}(t) & s_{13}(t) & s_{14}(t)\\
s_{21}(t) & s_{22}(t) & s_{23}(t) & s_{24}(t)\\
s_{31}(t) & s_{32}(t) & s_{33}(t) & s_{34}(t)\\
s_{41}(t) & s_{42}(t) & s_{43}(t) & s_{44}(t)\\
\end{pmatrix}
\begin{pmatrix}
\sqrt{p_1} & 0 & 0 & 0\\
0 & \sqrt{p_{2}(t)} & 0 & 0\\
0 & 0 & \sqrt{p_{3}(t)} & 0\\
0 & 0 & 0 & \sqrt{p_{4}(t)}\\
\end{pmatrix}
\begin{pmatrix}
\sqrt{p_{1}(t)} \\
\sqrt{p_{2}(t)} \\
\sqrt{p_{3}(t)} \\
\sqrt{p_{4}(t)} \\
\end{pmatrix}= \nonumber \\ = \frac{1}{2}
\begin{pmatrix}
\frac{1}{\sqrt{p_1}} & 0 & 0 & 0\\
0 & \frac{1}{\sqrt{p_{2}(t)}} & 0 & 0\\
0 & 0 & \frac{1}{\sqrt{p_{3}(t)}} & 0\\
0 & 0 & 0 & \frac{1}{\sqrt{p_{4}(t)}}\\
\end{pmatrix}
\begin{pmatrix}
2 \sqrt{p_{1}(t)}\frac{d}{dt}\sqrt{p_1(t)} \\
2 \sqrt{p_{2}(t)}\frac{d}{dt}\sqrt{p_2(t)} \\
2 \sqrt{p_{3}(t)}\frac{d}{dt}\sqrt{p_3(t)} \\
2 \sqrt{p_{4}(t)}\frac{d}{dt}\sqrt{p_4(t)} \\
\end{pmatrix}
\end{eqnarray}

In very real way we obtain the following classical physics Hamiltonian and we obtain

\begin{eqnarray}
\hat{H}_{classical}(t)=
\frac{1}{2}
\begin{pmatrix}
\frac{1}{\sqrt{p_1(t)}} & 0 & 0 & 0\\
0 & \frac{1}{\sqrt{p_{2}(t)}} & 0 & 0\\
0 & 0 & \frac{1}{\sqrt{p_{3}(t)}} & 0\\
0 & 0 & 0 & \frac{1}{\sqrt{p_{4}(t)}}\\
\end{pmatrix}
\begin{pmatrix}
s_{11}(t) & s_{12}(t) & s_{13}(t) & s_{14}(t)\\
s_{21}(t) & s_{22}(t) & s_{23}(t) & s_{24}(t)\\
s_{31}(t) & s_{32}(t) & s_{33}(t) & s_{34}(t)\\
s_{41}(t) & s_{42}(t) & s_{43}(t) & s_{44}(t)\\
\end{pmatrix}
\begin{pmatrix}
\sqrt{p_1(t)} & 0 & 0 & 0\\
0 & \sqrt{p_{2}(t)} & 0 & 0\\
0 & 0 & \sqrt{p_{3}(t)} & 0\\
0 & 0 & 0 & \sqrt{p_{4}(t)}\\
\end{pmatrix}= \nonumber \\
=
\frac{1}{2}
\begin{pmatrix}
\frac{1}{\sqrt{p_1}(t)} & 0 & 0 & 0\\
0 & \frac{1}{\sqrt{p_{2}(t)}} & 0 & 0\\
0 & 0 & \frac{1}{\sqrt{p_{3}(t)}} & 0\\
0 & 0 & 0 & \frac{1}{\sqrt{p_{4}(t)}}\\
\end{pmatrix}
\begin{pmatrix}
\sqrt{p_1(t)}s_{11}(t) & \sqrt{p_2}(t)s_{12}(t) & \sqrt{p_3(t)}s_{13}(t) & \sqrt{p_4(t)}s_{14}(t)\\
\sqrt{p_1(t)}s_{21}(t) & \sqrt{p_2(t)}s_{22}(t) & \sqrt{p_3(t)}s_{23}(t) & \sqrt{p_4}(t)s_{24}(t)\\
\sqrt{p_1(t)}s_{31}(t) & \sqrt{p_2(t)}s_{32}(t) & \sqrt{p_3(t)}s_{33}(t) & \sqrt{p_4}(t)s_{34}(t)\\
\sqrt{p_1(t)}s_{41}(t) & \sqrt{p_2(t)}s_{42}(t) & \sqrt{p_3(t)}s_{43}(t) & \sqrt{p_4}(t)s_{44}(t)\\
\end{pmatrix}= \nonumber \\
=\frac{1}{2}
\begin{pmatrix}
s_{11}(t) & \frac{\sqrt{p_2}(t)}{\sqrt{p_1}(t)}s_{12}(t) & \frac{\sqrt{p_3}(t)}{\sqrt{p_1}(t)}s_{13}(t) & \frac{\sqrt{p_4}(t)}{\sqrt{p_1}(t)}s_{14}(t)\\
\frac{\sqrt{p_1(t)}s_{21}(t)}{\sqrt{p_2}} & s_{22}(t) & \frac{\sqrt{p_3}(t)}{\sqrt{p_2(t)}}s_{23}(t) & \frac{\sqrt{p_4(t)}}{\sqrt{p_2(t)}}s_{24}(t)\\
\frac{\sqrt{p_1(t)}s_{31}(t)}{\sqrt{p_3(t)}} & \frac{\sqrt{p_2(t)}s_{32}(t)}{\sqrt{p_3(t)}} & s_{33}(t) & \frac{\sqrt{p_4(t)}s_{34}(t)}{\sqrt{p_3(t)}}\\
\frac{\sqrt{p_1(t)}s_{41}(t)}{\sqrt{p_4(t)}} & \frac{\sqrt{p_2(t)}s_{42}(t)}{\sqrt{p_4(t)}} & \frac{\sqrt{p_3(t)}s_{43}(t)}{\sqrt{p_4(t)}} & s_{44}(t)\\
\end{pmatrix}
\end{eqnarray}

This equation is equivalent to
\begin{eqnarray}
\begin{pmatrix}
\frac{1}{2}s_{11}(t) & \frac{1}{2}\sqrt{\frac{p_2(t)}{p_1(t)}}s_{12}(t) & \frac{1}{2}\sqrt{\frac{p_3(t)}{p_1(t)}}s_{13}(t) & \frac{1}{2}\sqrt{\frac{p_4(t)}{p_1(t)}}s_{14}(t) \\ 
\frac{1}{2}\sqrt{\frac{p_1(t)}{p_2(t)}}s_{21}(t) & \frac{1}{2}s_{22}(t) & \frac{1}{2}\sqrt{\frac{p_3(t)}{p_2(t)}}s_{23}(t) & \frac{1}{2}\sqrt{\frac{p_4(t)}{p_2(t)}}s_{24}(t) \\ 
\frac{1}{2}\sqrt{\frac{p_1(t)}{p_3(t)}}s_{31}(t) & \frac{1}{2}\sqrt{\frac{p_2(t)}{p_3(t)}}s_{32}(t) & \frac{1}{2}s_{33}(t) & \frac{1}{2}\sqrt{\frac{p_4(t)}{p_3(t)}}s_{34}(t) \\ 
\frac{1}{2}\sqrt{\frac{p_1(t)}{p_4(t)}}s_{41}(t) & \frac{1}{2}\sqrt{\frac{p_2(t)}{p_4(t)}}s_{42}(t) & \frac{1}{2}\sqrt{\frac{p_3(t)}{p_4(t)}}s_{43}(t) & s_{44}(t) \\ 
\end{pmatrix}
\begin{pmatrix}
\sqrt{p_1(t)} \\
\sqrt{p_2(t)} \\
\sqrt{p_3(t)} \\
\sqrt{p_4(t)} \\
\end{pmatrix}
=\frac{d}{dt}
\begin{pmatrix}
\sqrt{p_1(t)} \\
\sqrt{p_2(t)} \\
\sqrt{p_3(t)} \\
\sqrt{p_4(t)} \\
\end{pmatrix}=
\frac{i\hbar}{i\hbar}
\frac{d}{dt}
\begin{pmatrix}
\sqrt{p_1(t)} \\
\sqrt{p_2(t)} \\
\sqrt{p_3(t)} \\
\sqrt{p_4(t)} \\
\end{pmatrix} = \nonumber \\
=\frac{d}{dt}\ket{\psi}_{classical}(t)=\hat{H}_{classical}(t)\ket{\psi}_{classical}(t).
\end{eqnarray}
At this stage we can identify analytical solutions given as

\begin{eqnarray}
exp \Bigg[
\begin{pmatrix}
\int_{t_0}^{t}dt'\frac{1}{2}s_{11}(t') & \int_{t_0}^{t}dt'\frac{1}{2}\sqrt{\frac{p_2(t')}{p_1(t')}}s_{12}(t') & \int_{t_0}^{t}dt'\frac{1}{2}\sqrt{\frac{p_3(t')}{p_1(t')}}s_{13}(t') & \int_{t_0}^{t}dt'\frac{1}{2}\sqrt{\frac{p_4(t')}{p_1(t')}}s_{14}(t') \\ 
\int_{t_0}^{t}dt'\frac{1}{2}\sqrt{\frac{p_1(t')}{p_2(t')}}s_{21}(t') & \int_{t_0}^{t}dt'\frac{1}{2}s_{22}(t') &\int_{t_0}^{t}dt' \frac{1}{2}\sqrt{\frac{p_3(t')}{p_2(t')}}s_{23}(t') & \int_{t_0}^{t}dt'\frac{1}{2}\sqrt{\frac{p_4(t')}{p_2(t')}}s_{24}(t') \\ 
\int_{t_0}^{t}dt'\frac{1}{2}\sqrt{\frac{p_1(t')}{p_3(t')}}s_{31}(t') & \int_{t_0}^{t}dt'\frac{1}{2}\sqrt{\frac{p_2(t')}{p_3(t')}}s_{32}(t') & \int_{t_0}^{t}dt'\frac{1}{2}s_{33}(t') & \int_{t_0}^{t}dt'\frac{1}{2}\sqrt{\frac{p_4(t')}{p_3(t')}}s_{34}(t') \\ 
\int_{t_0}^{t}dt'\frac{1}{2}\sqrt{\frac{p_1(t')}{p_4(t')}}s_{41}(t') & \int_{t_0}^{t}dt'\frac{1}{2}\sqrt{\frac{p_2(t')}{p_4(t')}}s_{42}(t') & \int_{t_0}^{t}dt'\frac{1}{2}\sqrt{\frac{p_3(t')}{p_4(t')}}s_{43}(t') & \int_{t_0}^{t}dt's_{44}(t') \\ 
\end{pmatrix} \Bigg]
\begin{pmatrix}
\sqrt{p_1(t_0)} \\
\sqrt{p_2(t_0)} \\
\sqrt{p_3(t_0)} \\
\sqrt{p_4(t_0)} \\
\end{pmatrix}  
=
\begin{pmatrix}
\sqrt{p_1(t)} \\
\sqrt{p_2(t)} \\
\sqrt{p_3(t)} \\
\sqrt{p_4(t)} \\
\end{pmatrix}
\end{eqnarray}

We have
\begin{eqnarray}
\frac{d}{dt}\bra{\psi(t)}_{classical}=\bra{\psi(t)}_{classical}[\hat{H(t)}_{classical}]^{T}.
\end{eqnarray}
and it implies that
\begin{eqnarray}
\frac{d}{dt}(\ket{\psi(t)}(t)\bra{\psi}_{classical})=(\frac{d}{dt}(\ket{\psi(t)})\bra{\psi}_{classical}+(\ket{\psi(t)})\frac{d}{dt}(\bra{\psi}_{classical})) = \nonumber \\=\hat{H(t)}_{classical}\ket{\psi(t)}\bra{\psi(t)}_{classical}+\ket{\psi(t)}\bra{\psi(t)}_{classical}[\hat{H(t)}_{classical}]^{T}=\Bigg[\hat{H(t)}_{classical}, \ket{\psi(t)}\bra{\psi(t)}_{classical} \Bigg]_{+},
\end{eqnarray}
where $\Bigg[\hat{A},\hat{B}\Bigg]_{+}=\hat{A}\hat{B}+\hat{B}\hat{A}$ is anticommutator. We notice that in such situation we have commutator in Quantum Mechanics defined as
$\Bigg[\hat{A},\hat{B}\Bigg]_{-}=\hat{A}\hat{B}-\hat{B}\hat{A}$ and thus
\begin{eqnarray}
i \hbar \frac{d}{dt}(\ket{\psi(t)}(t)\bra{\psi}_{quantum})=i \hbar(\frac{d}{dt}(\ket{\psi(t)})\bra{\psi}_{quantum}+i \hbar(\ket{\psi(t)})\frac{d}{dt}(\bra{\psi}_{classical})) = \nonumber \\=\hat{H}(t)_{quantum}\ket{\psi(t)}\bra{\psi(t)}_{quantum}-\ket{\psi(t)}\bra{\psi(t)}_{quantum}[\hat{H}(t)_{quantum}]^{\dag}=\Bigg[\hat{H}(t)_{quantum}, \ket{\psi(t)}\bra{\psi(t)}_{quantum} \Bigg]_{-}= \nonumber \\
=\Bigg[\hat{H}(t)_{quantum}, \rho_{quantum} \Bigg]_{-}.
\end{eqnarray}

.
We recognize analogies with Schroedinger equation that are occurring only in case of real valued wave functions.
We can introduce the object similar to quantum density matrix that is classical density matrix. We have
\begin{eqnarray}
\hat{\rho}_{classical}(t)=
\begin{pmatrix}
\sqrt{p_1(t)} \\
\sqrt{p_2(t)} \\
\sqrt{p_3(t)} \\
\sqrt{p_4(t)} \\
\end{pmatrix}
\begin{pmatrix}
\sqrt{p_1(t)} & \sqrt{p_2(t)} & \sqrt{p_3(t)} & \sqrt{p_4(t)} \\
\end{pmatrix}=\nonumber \\ =
\begin{pmatrix}
p_1(t) & \sqrt{p_1(t)}\sqrt{p_2(t)} & \sqrt{p_1(t)}\sqrt{p_3(t)} & \sqrt{p_1(t)}\sqrt{p_4(t)}\\
\sqrt{p_2(t)}\sqrt{p_1(t)} & p_2(t)& \sqrt{p_2(t)}\sqrt{p_3(t)} & \sqrt{p_2(t)}\sqrt{p_4(t)} \\
\sqrt{p_3(t)}\sqrt{p_1(t)} & \sqrt{p_3(t)}\sqrt{p_2(t)}& p_3(t) & \sqrt{p_3(t)}\sqrt{p_4(t)} \\
\sqrt{p_4(t)}\sqrt{p_1(t)}& \sqrt{p_4(t)}\sqrt{p_2(t)} & \sqrt{p_4(t)}\sqrt{p_3(t)} & p_4(t)\\
\end{pmatrix}=\hat{\rho}_{classical}(t)^{T}
\end{eqnarray}
We can obtain the analytical solution for such system in quite straighforward way
\begin{eqnarray}
\hat{\rho}_{classical}(t)=
exp \Bigg[
\begin{pmatrix}
\int_{t_0}^{t}dt'\frac{1}{2}s_{11}(t') & \int_{t_0}^{t}dt'\frac{1}{2}\sqrt{\frac{p_2(t')}{p_1(t')}}s_{12}(t') & \int_{t_0}^{t}dt'\frac{1}{2}\sqrt{\frac{p_3(t')}{p_1(t')}}s_{13}(t') & \int_{t_0}^{t}dt'\frac{1}{2}\sqrt{\frac{p_4(t')}{p_1(t')}}s_{14}(t') \\ 
\int_{t_0}^{t}dt'\frac{1}{2}\sqrt{\frac{p_1(t')}{p_2(t')}}s_{21}(t') & \int_{t_0}^{t}dt'\frac{1}{2}s_{22}(t') &\int_{t_0}^{t}dt' \frac{1}{2}\sqrt{\frac{p_3(t')}{p_2(t')}}s_{23}(t') & \int_{t_0}^{t}dt'\frac{1}{2}\sqrt{\frac{p_4(t')}{p_2(t')}}s_{24}(t') \\ 
\int_{t_0}^{t}dt'\frac{1}{2}\sqrt{\frac{p_1(t')}{p_3(t')}}s_{31}(t') & \int_{t_0}^{t}dt'\frac{1}{2}\sqrt{\frac{p_2(t')}{p_3(t')}}s_{32}(t') & \int_{t_0}^{t}dt'\frac{1}{2}s_{33}(t') & \int_{t_0}^{t}dt'\frac{1}{2}\sqrt{\frac{p_4(t')}{p_3(t')}}s_{34}(t') \\ 
\int_{t_0}^{t}dt'\frac{1}{2}\sqrt{\frac{p_1(t')}{p_4(t')}}s_{41}(t') & \int_{t_0}^{t}dt'\frac{1}{2}\sqrt{\frac{p_2(t')}{p_4(t')}}s_{42}(t') & \int_{t_0}^{t}dt'\frac{1}{2}\sqrt{\frac{p_3(t')}{p_4(t')}}s_{43}(t') & \int_{t_0}^{t}dt's_{44}(t') \\ 
\end{pmatrix} \Bigg] \times
\hat{\rho}_{classical}(t_0) \times \nonumber \\
\times \exp \Bigg[ \Bigg[
\begin{pmatrix}
\int_{t_0}^{t}dt'\frac{1}{2}s_{11}(t') & \int_{t_0}^{t}dt'\frac{1}{2}\sqrt{\frac{p_2(t')}{p_1(t')}}s_{12}(t') & \int_{t_0}^{t}dt'\frac{1}{2}\sqrt{\frac{p_3(t')}{p_1(t')}}s_{13}(t') & \int_{t_0}^{t}dt'\frac{1}{2}\sqrt{\frac{p_4(t')}{p_1(t')}}s_{14}(t') \\ 
\int_{t_0}^{t}dt'\frac{1}{2}\sqrt{\frac{p_1(t')}{p_2(t')}}s_{21}(t') & \int_{t_0}^{t}dt'\frac{1}{2}s_{22}(t') &\int_{t_0}^{t}dt' \frac{1}{2}\sqrt{\frac{p_3(t')}{p_2(t')}}s_{23}(t') & \int_{t_0}^{t}dt'\frac{1}{2}\sqrt{\frac{p_4(t')}{p_2(t')}}s_{24}(t') \\ 
\int_{t_0}^{t}dt'\frac{1}{2}\sqrt{\frac{p_1(t')}{p_3(t')}}s_{31}(t') & \int_{t_0}^{t}dt'\frac{1}{2}\sqrt{\frac{p_2(t')}{p_3(t')}}s_{32}(t') & \int_{t_0}^{t}dt'\frac{1}{2}s_{33}(t') & \int_{t_0}^{t}dt'\frac{1}{2}\sqrt{\frac{p_4(t')}{p_3(t')}}s_{34}(t') \\ 
\int_{t_0}^{t}dt'\frac{1}{2}\sqrt{\frac{p_1(t')}{p_4(t')}}s_{41}(t') & \int_{t_0}^{t}dt'\frac{1}{2}\sqrt{\frac{p_2(t')}{p_4(t')}}s_{42}(t') & \int_{t_0}^{t}dt'\frac{1}{2}\sqrt{\frac{p_3(t')}{p_4(t')}}s_{43}(t') & \int_{t_0}^{t}dt's_{44}(t') \\ 
\end{pmatrix} \Bigg]^T \Bigg]
\end{eqnarray}
We can compute the reduced density matrix of subsystem A and B in standard way and we can establish von-Neumann entropy relation.

By introducing $t_1=i\hbar t$ we have $\frac{t_1}{i\hbar}=t$ and it leads to
\begin{eqnarray}
\begin{pmatrix}
\frac{1}{2}s_{11}(\frac{t_1}{i\hbar}) & \frac{1}{2}\sqrt{\frac{p_2(\frac{t_1}{i\hbar})}{p_1(\frac{t_1}{i\hbar})}}s_{12}(\frac{t_1}{i\hbar}) & \frac{1}{2}\sqrt{\frac{p_3(\frac{t_1}{i\hbar})}{p_1(\frac{t_1}{i\hbar})}}s_{13}(\frac{t_1}{i\hbar}) & \frac{1}{2}\sqrt{\frac{p_4(\frac{t_1}{i\hbar})}{p_1(\frac{t_1}{i\hbar})}}s_{14}(\frac{t_1}{i\hbar}) \\ 
\frac{1}{2}\sqrt{\frac{p_1(\frac{t_1}{i\hbar})}{p_2(\frac{t_1}{i\hbar})}}s_{21}(\frac{t_1}{i\hbar}) & \frac{1}{2}s_{22}(\frac{t_1}{i\hbar}) & \frac{1}{2}\sqrt{\frac{p_3(\frac{t_1}{i\hbar})}{p_2(t)}}s_{23}(\frac{t_1}{i\hbar}) & \frac{1}{2}\sqrt{\frac{p_4(\frac{t_1}{i\hbar})}{p_2(\frac{t_1}{i\hbar})}}s_{24}(\frac{t_1}{i\hbar}) \\ 
\frac{1}{2}\sqrt{\frac{p_1(\frac{t_1}{i\hbar})}{p_3(\frac{t_1}{i\hbar})}}s_{31}(\frac{t_1}{i\hbar}) & \frac{1}{2}\sqrt{\frac{p_2(\frac{t_1}{i\hbar})}{p_3(\frac{t_1}{i\hbar})}}s_{32}(\frac{t_1}{i\hbar}) & \frac{1}{2}s_{33}(\frac{t_1}{i\hbar}) & \frac{1}{2}\sqrt{\frac{p_4(\frac{t_1}{i\hbar})}{p_3(\frac{t_1}{i\hbar})}}s_{34}(\frac{t_1}{i\hbar}) \\ 
\frac{1}{2}\sqrt{\frac{p_1(\frac{t_1}{i\hbar})}{p_4(\frac{t_1}{i\hbar})}}s_{41}(\frac{t_1}{i\hbar}) & \frac{1}{2}\sqrt{\frac{p_2(\frac{t_1}{i\hbar})}{p_4(\frac{t_1}{i\hbar})}}s_{42}(\frac{t_1}{i\hbar}) & \frac{1}{2}\sqrt{\frac{p_3(\frac{t_1}{i\hbar})}{p_4(\frac{t_1}{i\hbar})}}s_{43}(\frac{t_1}{i\hbar}) & s_{44}(\frac{t_1}{i\hbar}) \\ 
\end{pmatrix}
\begin{pmatrix}
\sqrt{p_1(\frac{t_1}{i\hbar})} \\
\sqrt{p_2(\frac{t_1}{i\hbar})} \\
\sqrt{p_3(\frac{t_1}{i\hbar})} \\
\sqrt{p_4(\frac{t_1}{i\hbar})} \\
\end{pmatrix}
=
i\hbar
\frac{d}{dt_1}
\begin{pmatrix}
\sqrt{p_1(\frac{t_1}{i\hbar})} \\
\sqrt{p_2(\frac{t_1}{i\hbar})} \\
\sqrt{p_3(\frac{t_1}{i\hbar})} \\
\sqrt{p_4(\frac{t_1}{i\hbar})} \\
\end{pmatrix}
\end{eqnarray}
It leads us towards analytical solution

\begin{eqnarray}
\begin{pmatrix}
\sqrt{p_1(\frac{t_1}{i\hbar})} \\
\sqrt{p_2(\frac{t_1}{i\hbar})} \\
\sqrt{p_3(\frac{t_1}{i\hbar})} \\
\sqrt{p_4(\frac{t_1}{i\hbar})} \\
\end{pmatrix}
= \nonumber \\ =
\exp [ \Bigg[
\begin{pmatrix}
\frac{1}{i\hbar}\int_{0}^{t_1}dt_1'\frac{1}{2}s_{11}(\frac{t_1}{i\hbar}) & \frac{1}{i\hbar}\int_{0}^{t_1}dt_1\frac{1}{2}\sqrt{\frac{p_2(\frac{t_1}{i\hbar})}{p_1(\frac{t_1}{i\hbar})}}s_{12}(\frac{t_1}{i\hbar}) & \frac{1}{i\hbar}\int_{0}^{t_1}dt_1'\frac{1}{2}\sqrt{\frac{p_3(\frac{t_1}{i\hbar})}{p_1(\frac{t_1}{i\hbar})}}s_{13}(\frac{t_1}{i\hbar}) & \frac{1}{i\hbar}\int_{0}^{t_1}dt_1'\frac{1}{2}\sqrt{\frac{p_4(\frac{t_1}{i\hbar})}{p_1(\frac{t_1}{i\hbar})}}s_{14}(\frac{t_1}{i\hbar}) \\ 
\frac{1}{i\hbar}\int_{0}^{t_1}dt_1'\frac{1}{2}\sqrt{\frac{p_1(\frac{t_1}{i\hbar})}{p_2(\frac{t_1}{i\hbar})}}s_{21}(\frac{t_1}{i\hbar}) & \frac{1}{2}s_{22}(\frac{t_1}{i\hbar}) & \frac{1}{i\hbar}\int_{0}^{t_1}dt_1'\frac{1}{2}\sqrt{\frac{p_3(\frac{t_1}{i\hbar})}{p_2(t)}}s_{23}(\frac{t_1}{i\hbar}) & \frac{1}{i\hbar}\int_{0}^{t_1}dt_1'\frac{1}{2}\sqrt{\frac{p_4(\frac{t_1}{i\hbar})}{p_2(\frac{t_1}{i\hbar})}}s_{24}(\frac{t_1}{i\hbar}) \\ 
\frac{1}{i\hbar}\int_{0}^{t_1}dt_1'\frac{1}{2}\sqrt{\frac{p_1(\frac{t_1}{i\hbar})}{p_3(\frac{t_1}{i\hbar})}}s_{31}(\frac{t_1}{i\hbar}) & \frac{1}{i\hbar}\int_{0}^{t_1}dt_1'\frac{1}{2}\sqrt{\frac{p_2(\frac{t_1}{i\hbar})}{p_3(\frac{t_1}{i\hbar})}}s_{32}(\frac{t_1}{i\hbar}) & \frac{1}{2}s_{33}(\frac{t_1}{i\hbar}) & \frac{1}{i\hbar}\int_{0}^{t_1}dt_1'\frac{1}{2}\sqrt{\frac{p_4(\frac{t_1}{i\hbar})}{p_3(\frac{t_1}{i\hbar})}}s_{34}(\frac{t_1}{i\hbar}) \\ 
\frac{1}{i\hbar}\int_{0}^{t_1}dt_1'\frac{1}{2}\sqrt{\frac{p_1(\frac{t_1}{i\hbar})}{p_4(\frac{t_1}{i\hbar})}}s_{41}(\frac{t_1}{i\hbar}) & \frac{1}{i\hbar}\int_{0}^{t_1}dt_1'\frac{1}{2}\sqrt{\frac{p_2(\frac{t_1}{i\hbar})}{p_4(\frac{t_1}{i\hbar})}}s_{42}(\frac{t_1}{i\hbar}) & \frac{1}{i\hbar}\int_{0}^{t_1}dt_1'\frac{1}{2}\sqrt{\frac{p_3(\frac{t_1}{i\hbar})}{p_4(\frac{t_1}{i\hbar})}}s_{43}(\frac{t_1}{i\hbar}) & \frac{1}{i\hbar}\int_{0}^{t_1}dt_1s_{44}(\frac{t_1}{i\hbar}) \\ 
\end{pmatrix}
\Bigg]
\begin{pmatrix}
\sqrt{p_1(\frac{0}{i\hbar})} \\
\sqrt{p_2(\frac{0}{i\hbar})} \\
\sqrt{p_3(\frac{0}{i\hbar})} \\
\sqrt{p_4(\frac{0}{i\hbar})} \\
\end{pmatrix}
\end{eqnarray}
Dealing in similar fashion we can obtain quantum density matrix given as
\begin{eqnarray}
\begin{pmatrix}
\sqrt{p_1(\frac{t_1}{i\hbar})} \\
\sqrt{p_2(\frac{t_1}{i\hbar})} \\
\sqrt{p_3(\frac{t_1}{i\hbar})} \\
\sqrt{p_4(\frac{t_1}{i\hbar})} \\
\end{pmatrix}
\begin{pmatrix}
\sqrt{p_1(\frac{t_1}{i\hbar})}^{*} & \sqrt{p_2(\frac{t_1}{i\hbar})}^{*} & \sqrt{p_3(\frac{t_1}{i\hbar})}^{*} & \sqrt{p_4(\frac{t_1}{i\hbar})}^{*} \\
\end{pmatrix}
= \nonumber \\ =
\exp [ \Bigg[
\begin{pmatrix}
\frac{1}{i\hbar}\int_{0}^{t_1}dt_1'\frac{1}{2}s_{11}(\frac{t_1}{i\hbar}) & \frac{1}{i\hbar}\int_{0}^{t_1}dt_1\frac{1}{2}\sqrt{\frac{p_2(\frac{t_1}{i\hbar})}{p_1(\frac{t_1}{i\hbar})}}s_{12}(\frac{t_1}{i\hbar}) & \frac{1}{i\hbar}\int_{0}^{t_1}dt_1'\frac{1}{2}\sqrt{\frac{p_3(\frac{t_1}{i\hbar})}{p_1(\frac{t_1}{i\hbar})}}s_{13}(\frac{t_1}{i\hbar}) & \frac{1}{i\hbar}\int_{0}^{t_1}dt_1'\frac{1}{2}\sqrt{\frac{p_4(\frac{t_1}{i\hbar})}{p_1(\frac{t_1}{i\hbar})}}s_{14}(\frac{t_1}{i\hbar}) \\ 
\frac{1}{i\hbar}\int_{0}^{t_1}dt_1'\frac{1}{2}\sqrt{\frac{p_1(\frac{t_1}{i\hbar})}{p_2(\frac{t_1}{i\hbar})}}s_{21}(\frac{t_1}{i\hbar}) & \frac{1}{2}s_{22}(\frac{t_1}{i\hbar}) & \frac{1}{i\hbar}\int_{0}^{t_1}dt_1'\frac{1}{2}\sqrt{\frac{p_3(\frac{t_1}{i\hbar})}{p_2(t)}}s_{23}(\frac{t_1}{i\hbar}) & \frac{1}{i\hbar}\int_{0}^{t_1}dt_1'\frac{1}{2}\sqrt{\frac{p_4(\frac{t_1}{i\hbar})}{p_2(\frac{t_1}{i\hbar})}}s_{24}(\frac{t_1}{i\hbar}) \\ 
\frac{1}{i\hbar}\int_{0}^{t_1}dt_1'\frac{1}{2}\sqrt{\frac{p_1(\frac{t_1}{i\hbar})}{p_3(\frac{t_1}{i\hbar})}}s_{31}(\frac{t_1}{i\hbar}) & \frac{1}{i\hbar}\int_{0}^{t_1}dt_1'\frac{1}{2}\sqrt{\frac{p_2(\frac{t_1}{i\hbar})}{p_3(\frac{t_1}{i\hbar})}}s_{32}(\frac{t_1}{i\hbar}) & \frac{1}{2}s_{33}(\frac{t_1}{i\hbar}) & \frac{1}{i\hbar}\int_{0}^{t_1}dt_1'\frac{1}{2}\sqrt{\frac{p_4(\frac{t_1}{i\hbar})}{p_3(\frac{t_1}{i\hbar})}}s_{34}(\frac{t_1}{i\hbar}) \\ 
\frac{1}{i\hbar}\int_{0}^{t_1}dt_1'\frac{1}{2}\sqrt{\frac{p_1(\frac{t_1}{i\hbar})}{p_4(\frac{t_1}{i\hbar})}}s_{41}(\frac{t_1}{i\hbar}) & \frac{1}{i\hbar}\int_{0}^{t_1}dt_1'\frac{1}{2}\sqrt{\frac{p_2(\frac{t_1}{i\hbar})}{p_4(\frac{t_1}{i\hbar})}}s_{42}(\frac{t_1}{i\hbar}) & \frac{1}{i\hbar}\int_{0}^{t_1}dt_1'\frac{1}{2}\sqrt{\frac{p_3(\frac{t_1}{i\hbar})}{p_4(\frac{t_1}{i\hbar})}}s_{43}(\frac{t_1}{i\hbar}) & \frac{1}{i\hbar}\int_{0}^{t_1}dt_1s_{44}(\frac{t_1}{i\hbar}) \\ 
\end{pmatrix}
\Bigg] \times \nonumber \\ \times
\begin{pmatrix}
\sqrt{p_1(\frac{0}{i\hbar})} \\
\sqrt{p_2(\frac{0}{i\hbar})} \\
\sqrt{p_3(\frac{0}{i\hbar})} \\
\sqrt{p_4(\frac{0}{i\hbar})} \\
\end{pmatrix}
\begin{pmatrix}
\sqrt{p_1(\frac{t_1}{i\hbar})}^{*} & \sqrt{p_2(\frac{t_1}{i\hbar})}^{*} & \sqrt{p_3(\frac{t_1}{i\hbar})}^{*} & \sqrt{p_4(\frac{t_1}{i\hbar})}^{*} \\
\end{pmatrix}
\times \nonumber \\
\times \exp [ -\Bigg[
\begin{pmatrix}
\frac{1}{i\hbar}\int_{0}^{t_1}dt_1'\frac{1}{2}s_{11}(\frac{t_1}{i\hbar}) & \frac{1}{i\hbar}\int_{0}^{t_1}dt_1\frac{1}{2}\sqrt{\frac{p_2(\frac{t_1}{i\hbar})}{p_1(\frac{t_1}{i\hbar})}}s_{12}(\frac{t_1}{i\hbar}) & \frac{1}{i\hbar}\int_{0}^{t_1}dt_1'\frac{1}{2}\sqrt{\frac{p_3(\frac{t_1}{i\hbar})}{p_1(\frac{t_1}{i\hbar})}}s_{13}(\frac{t_1}{i\hbar}) & \frac{1}{i\hbar}\int_{0}^{t_1}dt_1'\frac{1}{2}\sqrt{\frac{p_4(\frac{t_1}{i\hbar})}{p_1(\frac{t_1}{i\hbar})}}s_{14}(\frac{t_1}{i\hbar}) \\ 
\frac{1}{i\hbar}\int_{0}^{t_1}dt_1'\frac{1}{2}\sqrt{\frac{p_1(\frac{t_1}{i\hbar})}{p_2(\frac{t_1}{i\hbar})}}s_{21}(\frac{t_1}{i\hbar}) & \frac{1}{2}s_{22}(\frac{t_1}{i\hbar}) & \frac{1}{i\hbar}\int_{0}^{t_1}dt_1'\frac{1}{2}\sqrt{\frac{p_3(\frac{t_1}{i\hbar})}{p_2(t)}}s_{23}(\frac{t_1}{i\hbar}) & \frac{1}{i\hbar}\int_{0}^{t_1}dt_1'\frac{1}{2}\sqrt{\frac{p_4(\frac{t_1}{i\hbar})}{p_2(\frac{t_1}{i\hbar})}}s_{24}(\frac{t_1}{i\hbar}) \\ 
\frac{1}{i\hbar}\int_{0}^{t_1}dt_1'\frac{1}{2}\sqrt{\frac{p_1(\frac{t_1}{i\hbar})}{p_3(\frac{t_1}{i\hbar})}}s_{31}(\frac{t_1}{i\hbar}) & \frac{1}{i\hbar}\int_{0}^{t_1}dt_1'\frac{1}{2}\sqrt{\frac{p_2(\frac{t_1}{i\hbar})}{p_3(\frac{t_1}{i\hbar})}}s_{32}(\frac{t_1}{i\hbar}) & \frac{1}{2}s_{33}(\frac{t_1}{i\hbar}) & \frac{1}{i\hbar}\int_{0}^{t_1}dt_1'\frac{1}{2}\sqrt{\frac{p_4(\frac{t_1}{i\hbar})}{p_3(\frac{t_1}{i\hbar})}}s_{34}(\frac{t_1}{i\hbar}) \\ 
\frac{1}{i\hbar}\int_{0}^{t_1}dt_1'\frac{1}{2}\sqrt{\frac{p_1(\frac{t_1}{i\hbar})}{p_4(\frac{t_1}{i\hbar})}}s_{41}(\frac{t_1}{i\hbar}) & \frac{1}{i\hbar}\int_{0}^{t_1}dt_1'\frac{1}{2}\sqrt{\frac{p_2(\frac{t_1}{i\hbar})}{p_4(\frac{t_1}{i\hbar})}}s_{42}(\frac{t_1}{i\hbar}) & \frac{1}{i\hbar}\int_{0}^{t_1}dt_1'\frac{1}{2}\sqrt{\frac{p_3(\frac{t_1}{i\hbar})}{p_4(\frac{t_1}{i\hbar})}}s_{43}(\frac{t_1}{i\hbar}) & \frac{1}{i\hbar}\int_{0}^{t_1}dt_1s_{44}(\frac{t_1}{i\hbar}) \\ 
\end{pmatrix}
\Bigg]
\end{eqnarray}
It is quite straightforward to establish the reduced density matrix of system A and B and hence obtain von-Neumann entropy with time that characterizes the quantum like entropy in time in classical epidemic model in 2 coupled systems. We can extract density of matrix for system A given as
\begin{eqnarray}
\rho_A=
\begin{pmatrix}
\rho(1,1)+\rho(2,2) & \rho(1,3)+\rho(2,4) \\
\rho(3,1)+\rho(4,2) & \rho(3,3)+\rho(4,4) \\
\end{pmatrix}
\end{eqnarray}
and for system B given as
\begin{eqnarray}
\rho_B=
\begin{pmatrix}
\rho(1,1)+\rho(3,3) & \rho(1,2)+\rho(3,4) \\
\rho(2,1)+\rho(4,3) & \rho(2,2)+\rho(4,4) \\
\end{pmatrix}
\end{eqnarray}.
Those density matrices needs to be normalized so finally we obtain
\begin{eqnarray}
\rho_A=\frac{1}{\rho(1,1)+\rho(2,2)+\rho(3,3)+\rho(4,4)}
\begin{pmatrix}
\rho(1,1)+\rho(2,2) & \rho(1,3)+\rho(2,4) \\
\rho(3,1)+\rho(4,2) & \rho(3,3)+\rho(4,4) \\
\end{pmatrix}
\end{eqnarray}
and for system B given as
\begin{eqnarray}
\rho_B=\frac{1}{\rho(1,1)+\rho(2,2)+\rho(3,3)+\rho(4,4)}
\begin{pmatrix}
\rho(1,1)+\rho(3,3) & \rho(1,2)+\rho(3,4) \\
\rho(2,1)+\rho(4,3) & \rho(2,2)+\rho(4,4) \\
\end{pmatrix}
\end{eqnarray}.
In such case we can apply von-Neumman entropy measure to characterize the density matrix of classical state so we have $S_A=Tr(\rho_A log(\rho_A))$ and $S_B=Tr(\rho_B log(\rho_B))$. In our general case $S_A \neq S_B$, while in case of two quantum systems with mutual interaction, but isolated from external environment we have $S_A = S_B$.

 We conduct reverse reasoning and show that any quantum system describing by tight binding model can be modeled by stochastic finite state machine. This proves the fact that classical electronics can model quantum system made from single-electron devices as in CMOS quantum computer.
This reasoning can be conduced for N coupled quantum dots implementing CMOS quantum computer. It has therefore its meaning in quantum machine learning implemented in CMOS quantum
computer.
\section{From quantum tight-binding model to classical epidemic model}
We are considering two position-based qubits interacting by means of electrostatic Coulomb force as depicted in Fig.2.
\begin{figure}
\centering
\includegraphics[scale=0.7]{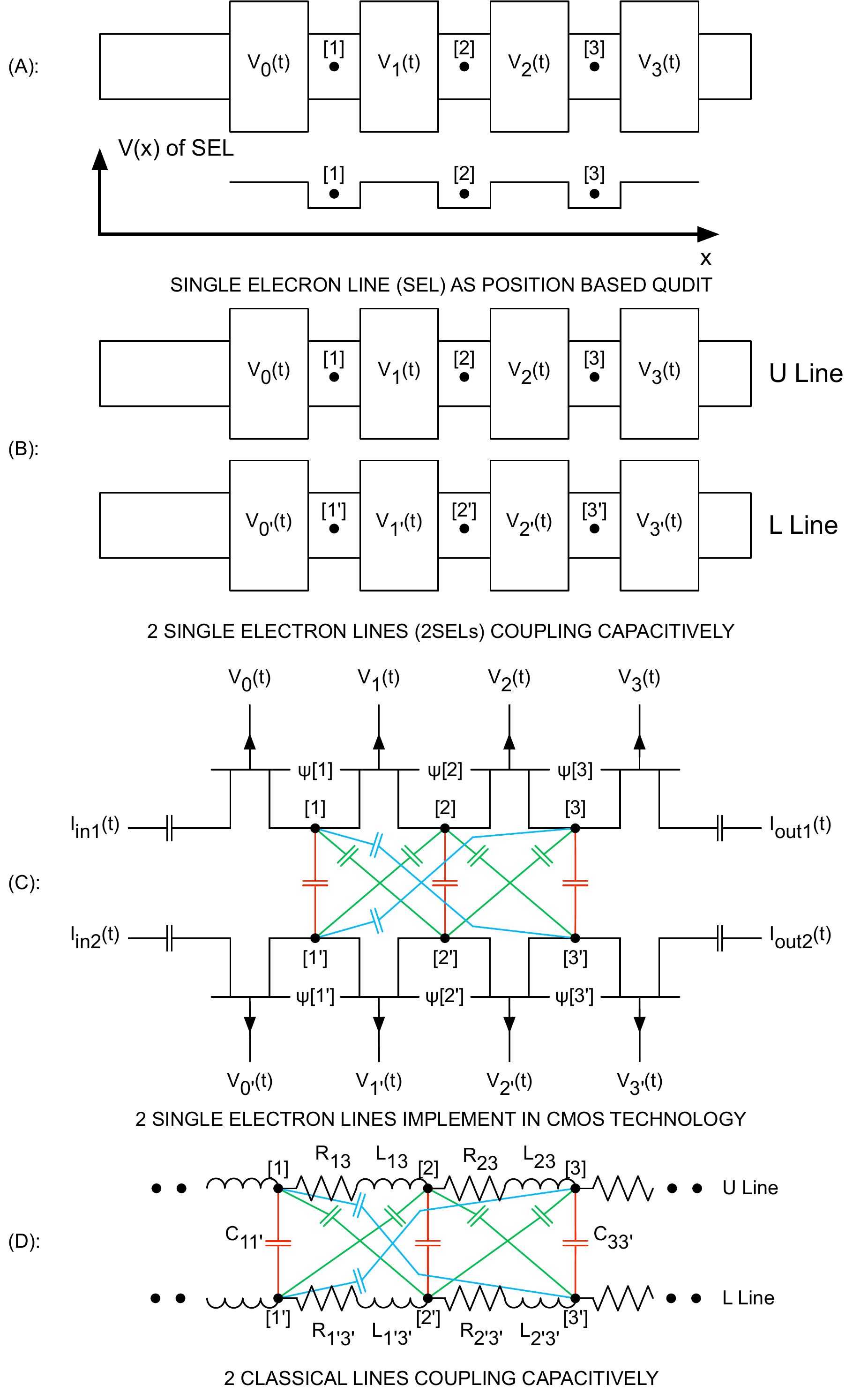} 
\caption{2 interacting position-based qubits implemented in CMOS semicondunctor technology \cite{2SEL}.}
\end{figure}
We define the Hamiltonian of 2 interacting Wannier qubits in the simplistic form given by tight-binding model. We have
\begin{eqnarray}
\hat{H}=(\hat{H}_A)_0+(\hat{H}_B)_0+\hat{H}_{A-B}= \nonumber \\
=(E_{p1A}\ket{1_A}\bra{1_A}+E_{p2A}\ket{2_A}\bra{2_A}+t_{s(1_A \rightarrow 2_A)}\ket{1_A}\bra{2_A}+t_{s(2_A \rightarrow 1_A)}\ket{2_A}\bra{1_A})_0 \times \hat{I}_B+ \nonumber \\
+\hat{I}_A\times(E_{p1B}\ket{2_B}\bra{2_B}+E_{p2B}\ket{2_B}\bra{2_B}+t_{s(1_B \rightarrow 2_B)}\ket{1_B}\bra{2_B}+t_{s(2_B \rightarrow 1_B)}\ket{2_B}\bra{1_B})_0+\nonumber \\
+\frac{q^2}{d_{1A-1B}}(\ket{1_A,1_B}\bra{1_A,1_B})+\frac{q^2}{d_{1A-2B}}\ket{1_A,2_B}\bra{1_A,2_B}+\frac{q^2}{d_{2A-1B}}\ket{2_A,1_B}\bra{2_A,1_B}+\frac{q^2}{d_{2A-2B}}\ket{2_A,2_B}\bra{2_A,2_B}, \nonumber \\
\hat{I}_A=\ket{1_A}\bra{1_A}+\ket{2_A}\bra{2_A},\hat{I}_B=\ket{1_B}\bra{1_B}+\ket{2_B}\bra{2_B}
\end{eqnarray}
and for quantum state of the form
\begin{eqnarray}
\ket{\psi(t)}=\gamma_1(t)\ket{1_A}\ket{1_B}+\gamma_2(t)\ket{1_A}\ket{2_B}+\gamma_3(t)\ket{2_A}\ket{1_B}+\gamma_4(t)\ket{2_A}\ket{2_B}= 
\begin{pmatrix}
\gamma_1(t) \\
\gamma_2(t) \\
\gamma_3(t) \\
\gamma_4(t) \\
\end{pmatrix}, 
|\gamma_1(t)|^2+|\gamma_2(t)|^2+|\gamma_3(t)|^2+|\gamma_4(t)|^2=1, \nonumber \\
.
\end{eqnarray}
and with $\gamma_1(t)=\sqrt{p_{I}(t)}e^{i\Theta_1(t)},..,\gamma_{IV}(t)=\sqrt{p_{IV}(t)}e^{i\Theta_{IV}(t)} $. We have the following evolution of non-disspative quantum state of isolated system in tight-binding model given as
\begin{eqnarray}
\begin{pmatrix}
E_{p1A}+E_{p1B}+\frac{q^2}{d_{1A-1B}} & t_{s(2_B \rightarrow 1_B)} & t_{s(2_A \rightarrow 1_A)} & 0 \\
t_{s(1_B \rightarrow 2_B)}      & E_{p1A}+E_{p2B}+\frac{q^2}{d_{1A-2B}} & 0 & t_{s(2_A \rightarrow 1_A)} \\
t_{s(1_A \rightarrow 2_A)}      & 0 & E_{p2A}+E_{p1B}+\frac{q^2}{d_{2A-1B}} & t_{s(2_B \rightarrow 1_B)} \\
0      & t_{s(1_A \rightarrow 2_A)} & t_{s(1_B \rightarrow 2_B)} & E_{p2A}+E_{p2B}+\frac{q^2}{d_{2A-2B}} \\
\end{pmatrix}
\begin{pmatrix}
\sqrt{p_{I}(t)}e^{i\Theta_{I}(t)} \\
\sqrt{p_{II}(t)}e^{i\Theta_{II}(t)} \\
\sqrt{p_{III}(t)}e^{i\Theta_{III}(t)} \\
\sqrt{p_{IV}(t)}e^{i\Theta_{IV}(t)} \\
\end{pmatrix}=\nonumber \\
=i\hbar \frac{d}{dt}
\begin{pmatrix}
\sqrt{p_{I}(t)}e^{i\Theta_{I}(t)} \\
\sqrt{p_{II}(t)}e^{i\Theta_{II}(t)} \\
\sqrt{p_{III}(t)}e^{i\Theta_{III}(t)} \\
\sqrt{p_{IV}(t)}e^{i\Theta_{IV}(t)} \\
\end{pmatrix}=\hbar
\begin{pmatrix}
(-\frac{d}{dt}\Theta_{I}(t)\sqrt{p_{I}(t)}+i\frac{1}{2 \sqrt{p_{I}(t)}}[\frac{d}{dt}p_{I}(t)])e^{i\Theta_{I}(t)} \\
(-\frac{d}{dt}\Theta_{II}(t)\sqrt{p_{II}(t)}+i\frac{1}{2 \sqrt{p_{II}(t)}}[\frac{d}{dt}p_{II}(t)])e^{i\Theta_{II}(t)} \\
(-\frac{d}{dt}\Theta_{III}(t)\sqrt{p_{III}(t)}+i\frac{1}{2 \sqrt{p_{III}(t)}}[\frac{d}{dt}p_{III}(t)])e^{i\Theta_{III}(t)} \\
(-\frac{d}{dt}\Theta_{IV}(t)\sqrt{p_{IV}(t)}+i\frac{1}{2 \sqrt{p_{IV}(t)}}[\frac{d}{dt}p_{IV}(t)])e^{i\Theta_{IV}(t)} \\
\end{pmatrix}= \nonumber \\
=\hbar
\begin{pmatrix}
(-\frac{d}{dt}\Theta_{I}(t)+i\frac{1}{2 p_{I}(t)}[\frac{d}{dt}p_{I}(t)])(\sqrt{p_{I}(t)}e^{i\Theta_{I}(t)}) \\
(-\frac{d}{dt}\Theta_{II}(t)+i\frac{1}{2 p_{II}(t)}[\frac{d}{dt}p_{II}(t)])(\sqrt{p_{II}(t)}e^{i\Theta_{II}(t)}) \\
(-\frac{d}{dt}\Theta_{III}(t)+i\frac{1}{2 p_{III}(t)}[\frac{d}{dt}p_{III}(t)])(\sqrt{p_{III}(t)}e^{i\Theta_{III}(t)}) \\
(-\frac{d}{dt}\Theta_{IV}(t)+i\frac{1}{2  p_{IV}(t)}[\frac{d}{dt}p_{IV}(t)])(\sqrt{p_{IV}(t)}e^{i\Theta_{IV}(t)}) \\
\end{pmatrix}
.
\end{eqnarray}
After reshaping we have
\tiny
\begin{eqnarray*}
\frac{1}{i\hbar}
\begin{pmatrix}
E_{p1A}+E_{p1B}+\frac{q^2}{d_{1A-1B}}+\hbar\frac{d}{dt}\Theta_{I}(t) & t_{s(2_B \rightarrow 1_B)} & t_{s(2_A \rightarrow 1_A)} & 0 \\
t_{s(1_B \rightarrow 2_B)}      & E_{p1A}+E_{p2B}+\frac{q^2}{d_{1A-2B}}+\hbar\frac{d}{dt}\Theta_{II}(t) & 0 & t_{s(2_A \rightarrow 1_A)} \\
t_{s(1_A \rightarrow 2_A)}      & 0 & E_{p2A}+E_{p1B}+\frac{q^2}{d_{2A-1B}}+\hbar\frac{d}{dt}\Theta_{III}(t) & t_{s(2_B \rightarrow 1_B)} \\
0      & t_{s(1_A \rightarrow 2_A)} & t_{s(1_B \rightarrow 2_B)} & E_{p2A}+E_{p2B}+\frac{q^2}{d_{2A-2B}}+\hbar\frac{d}{dt}\Theta_{IV}(t) \\
\end{pmatrix}
\begin{pmatrix}
\sqrt{p_{I}(t)}e^{i\Theta_{I}(t)} \\
\sqrt{p_{II}(t)}e^{i\Theta_{II}(t)} \\
\sqrt{p_{III}(t)}e^{i\Theta_{III}(t)} \\
\sqrt{p_{IV}(t)}e^{i\Theta_{IV}(t)} \\
\end{pmatrix}= 
\begin{pmatrix}
\frac{e^{i\Theta_{I}(t)}}{2\sqrt{p_{I}(t)}}\frac{d}{dt}p_{I}(t) \\
\frac{e^{i\Theta_{II}(t)}}{2\sqrt{p_{II}(t)}}\frac{d}{dt}p_{II}(t) \\
\frac{e^{i\Theta_{III}(t)}}{2\sqrt{p_{III}(t)}}\frac{d}{dt}p_{III}(t) \\
\frac{e^{i\Theta_{IV}(t)}}{2\sqrt{p_{IV}(t)}}\frac{d}{dt}p_{IV}(t) \\
\end{pmatrix}.
\end{eqnarray*}
\normalsize
Equivalently we have
\tiny
\begin{eqnarray*}
\frac{1}{i\hbar}
\begin{pmatrix}
E_{p1A}+E_{p1B}+\frac{q^2}{d_{1A-1B}}+\hbar\frac{d}{dt}\Theta_{I}(t) & t_{s(2_B \rightarrow 1_B)} & t_{s(2_A \rightarrow 1_A)} & 0 \\
t_{s(1_B \rightarrow 2_B)}      & E_{p1A}+E_{p2B}+\frac{q^2}{d_{1A-2B}}+\hbar\frac{d}{dt}\Theta_{II}(t) & 0 & t_{s(2_A \rightarrow 1_A)} \\
t_{s(1_A \rightarrow 2_A)}      & 0 & E_{p2A}+E_{p1B}+\frac{q^2}{d_{2A-1B}}+\hbar\frac{d}{dt}\Theta_{III}(t) & t_{s(2_B \rightarrow 1_B)} \\
0      & t_{s(1_A \rightarrow 2_A)} & t_{s(1_B \rightarrow 2_B)} & E_{p2A}+E_{p2B}+\frac{q^2}{d_{2A-2B}}+\hbar\frac{d}{dt}\Theta_{IV}(t) \\
\end{pmatrix} \times \nonumber \\
\times
\begin{pmatrix}
\frac{e^{i\Theta_{I}(t)}}{\sqrt{p_{I}(t)}} & 0 & 0 & 0 \\
0 & \frac{e^{i\Theta_{II}(t)}}{\sqrt{p_{II}(t)}} & 0 & 0 \\
0 & 0 & \frac{e^{i\Theta_{III}(t)}}{\sqrt{p_{III}(t)}} & 0  \\
0 & 0 & 0 & \frac{e^{i\Theta_{IV}(t)}}{\sqrt{p_{IV}(t)}}\\
\end{pmatrix}
\begin{pmatrix}
\sqrt{p_{I}(t)}e^{-i\Theta_{I}(t)} & 0 & 0 & 0 \\
0 & \sqrt{p_{II}(t)}e^{-i\Theta_{II}(t)} & 0 & 0 \\
0 & 0 & \sqrt{p_{III}(t)}e^{-i\Theta_{III}(t)} & 0  \\
0 & 0 & 0 & \sqrt{p_{IV}(t)}e^{-i\Theta_{IV}(t)}\\
\end{pmatrix}
\begin{pmatrix}
\sqrt{p_{I}(t)}e^{i\Theta_{I}(t)} \\
\sqrt{p_{II}(t)}e^{i\Theta_{II}(t)} \\
\sqrt{p_{III}(t)}e^{i\Theta_{III}(t)} \\
\sqrt{p_{IV}(t)}e^{i\Theta_{IV}(t)} \\
\end{pmatrix}= 
\begin{pmatrix}
\frac{e^{i\Theta_{I}(t)}}{2\sqrt{p_{I}(t)}}\frac{d}{dt}p_{I}(t) \\
\frac{e^{i\Theta_{II}(t)}}{2\sqrt{p_{II}(t)}}\frac{d}{dt}p_{II}(t) \\
\frac{e^{i\Theta_{III}(t)}}{2\sqrt{p_{III}(t)}}\frac{d}{dt}p_{III}(t) \\
\frac{e^{i\Theta_{IV}(t)}}{2\sqrt{p_{IV}(t)}}\frac{d}{dt}p_{IV}(t) \\
\end{pmatrix}.
\end{eqnarray*}
\normalsize
and final form of classical epidemic model mimicking the quantum tight-binding model in the form as
\small
\begin{eqnarray*}
\begin{pmatrix}
e^{-i\Theta_{I}(t)}\sqrt{p_{I}(t)} & 0 & 0 & 0 \\
0 & e^{-i\Theta_{II}(t)}\sqrt{p_{II}(t)} & 0 & 0 \\
0 & 0 & e^{-i\Theta_{III}(t)}\sqrt{p_{III}(t)} & 0 \\
0 & 0 & 0 & e^{-i\Theta_{IV}(t)}\sqrt{p_{IV}(t)} \\
\end{pmatrix}.
\frac{2}{i\hbar} \times \nonumber \\
\times
\begin{pmatrix}
E_{p1A}+E_{p1B}+\frac{q^2}{d_{1A-1B}}+\hbar\frac{d}{dt}\Theta_{I}(t) & t_{s(2_B \rightarrow 1_B)} & t_{s(2_A \rightarrow 1_A)} & 0 \\
t_{s(1_B \rightarrow 2_B)}      & E_{p1A}+E_{p2B}+\frac{q^2}{d_{1A-2B}}+\hbar\frac{d}{dt}\Theta_{II}(t) & 0 & t_{s(2_A \rightarrow 1_A)} \\
t_{s(1_A \rightarrow 2_A)}      & 0 & E_{p2A}+E_{p1B}+\frac{q^2}{d_{2A-1B}}+\hbar\frac{d}{dt}\Theta_{III}(t) & t_{s(2_B \rightarrow 1_B)} \\
0      & t_{s(1_A \rightarrow 2_A)} & t_{s(1_B \rightarrow 2_B)} & E_{p2A}+E_{p2B}+\frac{q^2}{d_{2A-2B}}+\hbar\frac{d}{dt}\Theta_{IV}(t) \\
\end{pmatrix} \times \nonumber \\
\times
\begin{pmatrix}
\frac{e^{i\Theta_{I}(t)}}{\sqrt{p_{I}(t)}} & 0 & 0 & 0 \\
0 & \frac{e^{i\Theta_{II}(t)}}{\sqrt{p_{II}(t)}} & 0 & 0 \\
0 & 0 & \frac{e^{i\Theta_{III}(t)}}{\sqrt{p_{III}(t)}} & 0  \\
0 & 0 & 0 & \frac{e^{i\Theta_{IV}(t)}}{\sqrt{p_{IV}(t)}}\\
\end{pmatrix}
\begin{pmatrix}
p_{I}(t) \\
p_{II}(t) \\
p_{III}(t) \\
p_{IV}(t) \\
\end{pmatrix}= 
\frac{d}{dt}
\begin{pmatrix}
p_{I}(t) \\
p_{II}(t) \\
p_{III}(t) \\
p_{IV}(t) \\
\end{pmatrix}.
\end{eqnarray*}
\normalsize
We obtain
\tiny
\begin{eqnarray*}
\frac{d}{dt}
\begin{pmatrix}
p_{I}(t) \\
p_{II}(t) \\
p_{III}(t) \\
p_{IV}(t) \\
\end{pmatrix}=
\begin{pmatrix}
e^{-i\Theta_{I}(t)}\sqrt{p_{I}(t)} & 0 & 0 & 0 \\
0 & e^{-i\Theta_{II}(t)}\sqrt{p_{II}(t)} & 0 & 0 \\
0 & 0 & e^{-i\Theta_{III}(t)}\sqrt{p_{III}(t)} & 0 \\
0 & 0 & 0 & e^{-i\Theta_{IV}(t)}\sqrt{p_{IV}(t)} \\
\end{pmatrix}.
\frac{2}{i\hbar} \times \nonumber \\
\times
\begin{pmatrix}
[E_{p1A}+E_{p1B}+\frac{q^2}{d_{1A-1B}}+\hbar\frac{d}{dt}\Theta_{I}(t)]\frac{e^{i\Theta_{I}(t)}}{\sqrt{p_{I}(t)}} & [t_{s(2_B \rightarrow 1_B)}]\frac{e^{i\Theta_{II}(t)}}{\sqrt{p_{II}(t)}} & [t_{s(2_A \rightarrow 1_A)}]\frac{e^{i\Theta_{III}(t)}}{\sqrt{p_{III}(t)}} & 0 \\
t_{s(1_B \rightarrow 2_B)}\frac{e^{i\Theta_{I}(t)}}{\sqrt{p_{I}(t)}}      & [E_{p1A}+E_{p2B}+\frac{q^2}{d_{1A-2B}}+\hbar\frac{d}{dt}\Theta_{II}(t)]\frac{e^{i\Theta_{II}(t)}}{\sqrt{p_{II}(t)}}  & 0 & [t_{s(2_A \rightarrow 1_A)}]\frac{e^{i\Theta_{IV}(t)}}{\sqrt{p_{IV}(t)}} \\
t_{s(1_A \rightarrow 2_A)}\frac{e^{i\Theta_{I}(t)}}{\sqrt{p_{I}(t)}}      & 0 & [E_{p2A}+E_{p1B}+\frac{q^2}{d_{2A-1B}}+\hbar\frac{d}{dt}\Theta_{III}(t)]\frac{e^{i\Theta_{III}(t)}}{\sqrt{p_{III}(t)}} & t_{s(2_B \rightarrow 1_B)}\frac{e^{i\Theta_{IV}(t)}}{\sqrt{p_{IV}(t)}} \\
0      & t_{s(1_A \rightarrow 2_A)}\frac{e^{i\Theta_{II}(t)}}{\sqrt{p_{II}(t)}} & t_{s(1_B \rightarrow 2_B)}\frac{e^{i\Theta_{III}(t)}}{\sqrt{p_{III}(t)}} & [E_{p2A}+E_{p2B}+\frac{q^2}{d_{2A-2B}}+\hbar\frac{d}{dt}\Theta_{IV}(t)]\frac{e^{i\Theta_{IV}(t)}}{\sqrt{p_{IV}(t)}} \\
\end{pmatrix} 
\begin{pmatrix}
p_{I}(t) \\
p_{II}(t) \\
p_{III}(t) \\
p_{IV}(t) \\
\end{pmatrix} 
\end{eqnarray*}
\normalsize
and in compact form we have \tiny
\begin{eqnarray*}
\frac{d}{dt}
\begin{pmatrix}
p_{I}(t) \\
p_{II}(t) \\
p_{III}(t) \\
p_{IV}(t) \\
\end{pmatrix}=
\frac{2}{i\hbar} 
\times
\begin{pmatrix}
[E_{p1A}+E_{p1B}+\frac{q^2}{d_{1A-1B}}+\hbar\frac{d}{dt}\Theta_{I}(t)] & [t_{s(2_B \rightarrow 1_B)}]\frac{e^{i(\Theta_{II}(t)-\Theta_{I}(t))}\sqrt{p_{I}(t)}}{\sqrt{p_{II}(t)}} & [t_{s(2_A \rightarrow 1_A)}]\frac{e^{i(\Theta_{III}(t)-\Theta_{I}(t))}\sqrt{p_{I}(t)}}{\sqrt{p_{III}(t)}} & 0 \\
t_{s(1_B \rightarrow 2_B)}\frac{e^{i(\Theta_{I}(t)-\Theta_{II}(t))}\sqrt{p_{II}(t)}}{\sqrt{p_{I}(t)}}      & [E_{p1A}+E_{p2B}+\frac{q^2}{d_{1A-2B}}+\hbar\frac{d}{dt}\Theta_{II}(t)]  & 0 & [t_{s(2_A \rightarrow 1_A)}]\frac{e^{i(\Theta_{IV}(t)-\Theta_{II}(t))}}{(\sqrt{p_{IV}(t)}-\sqrt{p_{II}(t)})} \\
t_{s(1_A \rightarrow 2_A)}\frac{e^{i(\Theta_{I}(t)-\Theta_{III}(t))}\sqrt{p_{III}(t)}}{\sqrt{p_{I}(t)}}      & 0 & [E_{p2A}+E_{p1B}+\frac{q^2}{d_{2A-1B}}+\hbar\frac{d}{dt}\Theta_{III}(t)]& t_{s(2_B \rightarrow 1_B)}\frac{e^{i(\Theta_{IV}(t)-\Theta_{III}(t))}\sqrt{p_{III}(t)}}{\sqrt{p_{IV}(t)}} \\
0      & t_{s(1_A \rightarrow 2_A)}\frac{e^{i(\Theta_{II}(t)-\Theta_{IV}(t))}\sqrt{p_{IV}(t)}}{\sqrt{p_{II}(t)}} & t_{s(1_B \rightarrow 2_B)}\frac{e^{i(\Theta_{III}(t)-\Theta_{IV}(t))}\sqrt{p_{IV}(t)}}{\sqrt{p_{III}(t)}} & [E_{p2A}+E_{p2B}+\frac{q^2}{d_{2A-2B}}+\hbar\frac{d}{dt}\Theta_{IV}(t)] \\
\end{pmatrix} 
\begin{pmatrix}
p_{I}(t) \\
p_{II}(t) \\
p_{III}(t) \\
p_{IV}(t) \\
\end{pmatrix}. 
\end{eqnarray*}
\normalsize
\section{Mapping quantum tight-binding model to stochastic finite state machine}
Let us start from equations of motion for 2 electrostatically coupled position based qubits as by \cite{2SEL} expressed by minimalistic tight-binding model and we have
\small
\begin{eqnarray}
\begin{pmatrix}
E_{p1A}+E_{p1B}+\frac{q^2}{d_{1A-1B}} & t_{s(2_B \rightarrow 1_B)} & t_{s(2_A \rightarrow 1_A)} & 0 \\
t_{s(1_B \rightarrow 2_B)}      & E_{p1A}+E_{p2B}+\frac{q^2}{d_{1A-2B}} & 0 & t_{s(2_A \rightarrow 1_A)} \\
t_{s(1_A \rightarrow 2_A)}      & 0 & E_{p2A}+E_{p1B}+\frac{q^2}{d_{2A-1B}} & t_{s(2_B \rightarrow 1_B)} \\
0      & t_{s(1_A \rightarrow 2_A)} & t_{s(1_B \rightarrow 2_B)} & E_{p2A}+E_{p2B}+\frac{q^2}{d_{2A-2B}} \\
\end{pmatrix}
\begin{pmatrix}
\sqrt{p_{I}(t)cos(\Theta_I(t))}+i\sqrt{p_{I}(t)sin(\Theta_I(t))} \\
\sqrt{p_{II}(t)cos(\Theta_{II}(t))}+i\sqrt{p_{II}(t)sin(\Theta_{II}(t))}  \\
\sqrt{p_{III}(t)cos(\Theta_{III}(t))}+i\sqrt{p_{III}(t)sin(\Theta_{III}(t))}  \\
\sqrt{p_{IV}(t)cos(\Theta_{IV}(t))}+i\sqrt{p_{IV}(t)sin(\Theta_{IV}(t))}  \\
\end{pmatrix}= \nonumber \\
=i \hbar \frac{d}{dt}
\begin{pmatrix}
\sqrt{p_{I}(t)cos(\Theta_I(t))}+i\sqrt{p_{I}(t)sin(\Theta_I(t))} \\
\sqrt{p_{II}(t)cos(\Theta_{II}(t))}+i\sqrt{p_{II}(t)sin(\Theta_{II}(t))}  \\
\sqrt{p_{III}(t)cos(\Theta_{III}(t))}+i\sqrt{p_{III}(t)sin(\Theta_{III}(t))}  \\
\sqrt{p_{IV}(t)cos(\Theta_{IV}(t))}+i\sqrt{p_{IV}(t)sin(\Theta_{IV}(t))}
\end{pmatrix}=
\begin{pmatrix}
i \hbar \frac{d}{dt}\sqrt{p_{I}(t)cos(\Theta_I(t))}- \hbar \frac{d}{dt}\sqrt{p_{I}(t)sin(\Theta_I(t))} \\
i \hbar \frac{d}{dt}\sqrt{p_{II}(t)cos(\Theta_{II}(t))}- \hbar \frac{d}{dt}\sqrt{p_{II}(t)sin(\Theta_{II}(t))}  \\
i \hbar \frac{d}{dt}\sqrt{p_{III}(t)cos(\Theta_{III}(t))}- \hbar \frac{d}{dt}\sqrt{p_{III}(t)sin(\Theta_{III}(t))}  \\
i \hbar \frac{d}{dt}\sqrt{p_{IV}(t)cos(\Theta_{IV}(t))}- \hbar \frac{d}{dt}\sqrt{p_{IV}(t)sin(\Theta_{IV}(t))}
\end{pmatrix}
\end{eqnarray}
and equivalently we can write
\begin{eqnarray}
(E_{p1A}+E_{p1B}+\frac{q^2}{d_{1A-1B}})\sqrt{p_{I}(t)cos(\Theta_I(t))} +( t_{sr(2_B \rightarrow 1_B)})\sqrt{p_{II}(t)cos(\Theta_{II}(t))}+ \nonumber \\
-( t_{sim(2_B \rightarrow 1_B)})\sqrt{p_{II}(t)sin(\Theta_{II}(t))}+( t_{sr(2_A \rightarrow 1_A)})\sqrt{p_{III}(t)cos(\Theta_{III}(t))}-( t_{sim(2_A \rightarrow 1_A)})\sqrt{p_{III}(t)sin(\Theta_{III}(t))}=-\hbar \frac{d}{dt}\sqrt{p_{I}(t)sin(\Theta_I(t))}
\nonumber \\
( t_{sim(2_B \rightarrow 1_B)})\sqrt{p_{II}(t)cos(\Theta_{II}(t))} + t_{sre(2_B \rightarrow 1_B)}\sqrt{p_{II}(t)sin(\Theta_{II}(t))}+( t_{sim(2_A \rightarrow 1_A)})\sqrt{p_{III}(t)cos(\Theta_{III}(t))} + t_{sre(2_A \rightarrow 1_A)}\sqrt{p_{III}(t)sin(\Theta_{III}(t))} , \nonumber \\
=\hbar \frac{d}{dt}\sqrt{p_{I}(t)cos(\Theta_I(t))} \nonumber \\
(E_{p1A}+E_{p2B}+\frac{q^2}{d_{1A-2B}})\sqrt{p_{II}(t)cos(\Theta_{II}(t))} +( t_{sr(1_B \rightarrow 2_B)})\sqrt{p_{I}(t)cos(\Theta_I(t))} - t_{sim(1_B \rightarrow 2_B)}\sqrt{p_{I}(t)sin(\Theta_I(t))}+ \nonumber \\(t_{sr(2_A \rightarrow 1_A)}\sqrt{p_{IV}(t)cos(\Theta_{IV}(t))}) - t_{sim(2_A \rightarrow 1_A)}\sqrt{p_{IV}(t)sin(\Theta_{IV}(t))}  = - \hbar \frac{d}{dt}\sqrt{p_{II}(t)sin(\Theta_{II}(t))} \\
( t_{sim(1_B \rightarrow 2_B)})\sqrt{p_{I}(t)cos(\Theta_I(t))} + t_{sre(1_B \rightarrow 2_B)}\sqrt{p_{I}(t)sin(\Theta_I(t))}+( t_{sim(2_A \rightarrow 1_A)})\sqrt{p_{IV}(t)cos(\Theta_{IV}(t))} + t_{sre(2_A \rightarrow 1_A)}\sqrt{p_{IV}(t)sin(\Theta_{IV}(t))} , \nonumber \\
=\hbar \frac{d}{dt}\sqrt{p_{II}(t)cos(\Theta_{II}(t))} \nonumber \\
(E_{p2A}+E_{p1B}+\frac{q^2}{d_{2A-1B}})\sqrt{p_{III}(t)cos(\Theta_{III}(t))} +( t_{sr(2_B \rightarrow 1_B)})\sqrt{p_{IV}(t)cos(\Theta_{IV}(t))} - t_{sim(2_B \rightarrow 1_B)}\sqrt{p_{IV}(t)sin(\Theta_{IV}(t))}+ \nonumber \\(t_{sr(1_A \rightarrow 2_A)}\sqrt{p_{I}(t)cos(\Theta_{I}(t))}) - t_{sim(1_A \rightarrow 2_A)}\sqrt{p_{I}(t)sin(\Theta_{I}(t))}  = - \hbar \frac{d}{dt}\sqrt{p_{III}(t)sin(\Theta_{III}(t))} \\
( t_{sim(1_A \rightarrow 2_A)})\sqrt{p_{I}(t)cos(\Theta_I(t))} + t_{sre(1_A \rightarrow 2_A)}\sqrt{p_{I}(t)sin(\Theta_{I}(t))}+( t_{sim(2_B \rightarrow 1_B)})\sqrt{p_{IV}(t)cos(\Theta_{IV}(t))} + t_{sre(2_B \rightarrow 1_B)}\sqrt{p_{IV}(t)sin(\Theta_{IV}(t))} , \nonumber \\
=\hbar \frac{d}{dt}\sqrt{p_{III}(t)cos(\Theta_{III}(t))} \nonumber \\
(E_{p2A}+E_{p2B}+\frac{q^2}{d_{2A-2B}})\sqrt{p_{IV}(t)cos(\Theta_{IV}(t))} +( t_{sr(1_B \rightarrow 2_B)})\sqrt{p_{III}(t)cos(\Theta_{III}(t))} - t_{sim(1_B \rightarrow 2_B)}\sqrt{p_{III}(t)sin(\Theta_{III}(t))}+ \nonumber \\(t_{sr(1_A \rightarrow 2_A)}\sqrt{p_{II}(t)cos(\Theta_{II}(t))}) - t_{sim(1_A \rightarrow 2_A)}\sqrt{p_{II}(t)sin(\Theta_{II}(t))}  = - \hbar \frac{d}{dt}\sqrt{p_{IV}(t)sin(\Theta_{IV}(t))} \\
( t_{sim(1_A \rightarrow 2_A)})\sqrt{p_{II}(t)cos(\Theta_{II}t))} + t_{sre(1_A \rightarrow 2_A)}\sqrt{p_{II}(t)sin(\Theta_{II}(t))}+( t_{sim(1_B \rightarrow 2_B)})\sqrt{p_{III}(t)cos(\Theta_{III}(t))} + t_{sre(1_B \rightarrow 2_B)}\sqrt{p_{III}(t)sin(\Theta_{III}(t))} , \nonumber \\
=\hbar \frac{d}{dt}\sqrt{p_{IV}(t)cos(\Theta_{IV}(t))} \nonumber \\
\end{eqnarray}
The established relation can be written in the compact form as
\begin{landscape}
\tiny
\begin{eqnarray}
\begin{pmatrix}
0 & (E_{p1A}+E_{p1B}+\frac{q^2}{d_{1A-1B}}) & ( t_{sim(2_B \rightarrow 1_B)}) & ( t_{sr(2_B \rightarrow 1_B)}) & ( t_{sim(2_A \rightarrow 1_A)}) & ( t_{sr(2_A \rightarrow 1_A)}) & 0 & 0 \\
(E_{p1A}+E_{p1B}+\frac{q^2}{d_{1A-1B}}) & 0 & ( t_{sr(2_B \rightarrow 1_B)}) & ( t_{im(2_B \rightarrow 1_B)}) & ( t_{sr(2_A \rightarrow 1_A)}) & ( t_{sim(2_A \rightarrow 1_A)}) & 0 & 0 \\
( t_{sim(1_B \rightarrow 2_B)}) & ( t_{sre(1_B \rightarrow 2_B)}) & 0 & (E_{p1A}+E_{p2B}+\frac{q^2}{d_{1A-2B}}) & 0 & 0 & ( t_{sim(2_A \rightarrow 1_A)}) & ( t_{sr(2_A \rightarrow 1_A)}) \\
( t_{sre(1_B \rightarrow 2_B)}) & ( t_{sim(1_B \rightarrow 2_B)}) & (E_{p1A}+E_{p2B}+\frac{q^2}{d_{1A-2B}}) & 0 & 0 & 0 & ( t_{sr(2_A \rightarrow 1_A)}) & ( t_{sim(2_A \rightarrow 1_A)}) \\
( t_{sim(1_A \rightarrow 2_A)}) & ( t_{sim(1_A \rightarrow 2_A)}) & 0 & 0 & 0 & (E_{p2A}+E_{p1B}+\frac{q^2}{d_{2A-1B}}) & ( t_{sim(2_B \rightarrow 1_B)}) & ( t_{sim(2_B \rightarrow 1_B)}) \\
( t_{sim(1_A \rightarrow 2_A)}) & ( t_{sim(1_A \rightarrow 2_A)}) & 0 & 0 & (E_{p2A}+E_{p1B}+\frac{q^2}{d_{2A-1B}}) & 0 & ( t_{sim(2_B \rightarrow 1_B)}) & ( t_{sr(2_B \rightarrow 1_B)}) \\
0 & 0 & ( t_{sr(1_A \rightarrow 2_A)}) & ( t_{im(1_A \rightarrow 2_A)}) & ( t_{sim(1_B \rightarrow 2_B)}) & ( t_{sr(1_B \rightarrow 2_B)}) & 0 & (E_{p2A}+E_{p2B}+\frac{q^2}{d_{2A-2B}}) \\
0 & 0 & ( t_{sim(1_A \rightarrow 2_A)}) & ( t_{sr(1_A \rightarrow 2_A)}) & ( t_{sim(1_B \rightarrow 2_B)}) & ( t_{sr(1_B \rightarrow 2_B)}) & (E_{p2A}+E_{p2B}+\frac{q^2}{d_{2A-2B}}) & 0 \\
\end{pmatrix} 
\begin{pmatrix}
\sqrt{p_{I}(t)cos(\Theta_{I}(t))} \\
\sqrt{p_{I}(t)sin(\Theta_{I}(t))} \\
\sqrt{p_{II}(t)cos(\Theta_{II}(t))} \\
\sqrt{p_{II}(t)sin(\Theta_{II}(t))} \\
\sqrt{p_{III}(t)cos(\Theta_{III}(t))} \\
\sqrt{p_{III}(t)sin(\Theta_{III}(t))} \\
\sqrt{p_{IV}(t)cos(\Theta_{IV}(t))} \\
\sqrt{p_{IV}(t)sin(\Theta_{IV}(t))} \\
\end{pmatrix}
=\hbar \frac{d}{dt}
\begin{pmatrix}
+\sqrt{p_{I}(t)cos(\Theta_{I}(t))} \\
-\sqrt{p_{I}(t)sin(\Theta_{I}(t))} \\
+\sqrt{p_{II}(t)cos(\Theta_{II}(t))} \\
-\sqrt{p_{II}(t)sin(\Theta_{II}(t))} \\
+\sqrt{p_{III}(t)cos(\Theta_{III}(t))} \\
-\sqrt{p_{III}(t)sin(\Theta_{III}(t))} \\
+\sqrt{p_{IV}(t)cos(\Theta_{IV}(t))} \\
-\sqrt{p_{IV}(t)sin(\Theta_{IV}(t))} \\
\end{pmatrix}
\end{eqnarray} \normalsize
and equivalently we have
\tiny
\begin{eqnarray}
2
\begin{pmatrix}
\sqrt{p_{I}(t)}cos(\Theta_{I}(t)) & 0 & 0 & 0 & 0 & 0 & 0 & 0 \\
0 & -\sqrt{p_{I}(t)}sin(\Theta_{I}(t)) & 0 & 0 & 0 & 0 & 0 & 0 \\
0 & 0 & +\sqrt{p_{II}(t)}cos(\Theta_{II}(t)) & 0 & 0 & 0 & 0 & 0 \\
0 & 0 & 0 & -\sqrt{p_{II}(t)}sin(\Theta_{II}(t)) & 0 & 0 & 0 & 0 \\
0 & 0 & 0 & 0 & +\sqrt{p_{III}(t)}cos(\Theta_{III}(t)) & 0 & 0 & 0 \\
0 & 0 & 0 & 0 & 0 & -\sqrt{p_{III}(t)}sin(\Theta_{III}(t)) & 0 & 0 \\
0 & 0 & 0 & 0 & 0 & 0 & +\sqrt{p_{IV}(t)}cos(\Theta_{IV}(t)) & 0 \\
0 & 0 & 0 & 0 & 0 & 0 & 0 & -\sqrt{p_{IV}(t)}sin(\Theta_{IV}(t)) \\
\end{pmatrix} \times \nonumber \\ \times
\begin{pmatrix}
0 & (E_{p1A}+E_{p1B}+\frac{q^2}{d_{1A-1B}}) & ( t_{sim(2_B \rightarrow 1_B)}) & ( t_{sr(2_B \rightarrow 1_B)}) & ( t_{sim(2_A \rightarrow 1_A)}) & ( t_{sr(2_A \rightarrow 1_A)}) & 0 & 0 \\
(E_{p1A}+E_{p1B}+\frac{q^2}{d_{1A-1B}}) & 0 & ( t_{sr(2_B \rightarrow 1_B)}) & ( t_{im(2_B \rightarrow 1_B)}) & ( t_{sr(2_A \rightarrow 1_A)}) & ( t_{sim(2_A \rightarrow 1_A)}) & 0 & 0 \\
( t_{sim(1_B \rightarrow 2_B)}) & ( t_{sre(1_B \rightarrow 2_B)}) & 0 & (E_{p1A}+E_{p2B}+\frac{q^2}{d_{1A-2B}}) & 0 & 0 & ( t_{sim(2_A \rightarrow 1_A)}) & ( t_{sr(2_A \rightarrow 1_A)}) \\
( t_{sre(1_B \rightarrow 2_B)}) & ( t_{sim(1_B \rightarrow 2_B)}) & (E_{p1A}+E_{p2B}+\frac{q^2}{d_{1A-2B}}) & 0 & 0 & 0 & ( t_{sr(2_A \rightarrow 1_A)}) & ( t_{sim(2_A \rightarrow 1_A)}) \\
( t_{sim(1_A \rightarrow 2_A)}) & ( t_{sim(1_A \rightarrow 2_A)}) & 0 & 0 & 0 & (E_{p2A}+E_{p1B}+\frac{q^2}{d_{2A-1B}}) & ( t_{sim(2_B \rightarrow 1_B)}) & ( t_{sim(2_B \rightarrow 1_B)}) \\
( t_{sim(1_A \rightarrow 2_A)}) & ( t_{sim(1_A \rightarrow 2_A)}) & 0 & 0 & (E_{p2A}+E_{p1B}+\frac{q^2}{d_{2A-1B}}) & 0 & ( t_{sim(2_B \rightarrow 1_B)}) & ( t_{sr(2_B \rightarrow 1_B)}) \\
0 & 0 & ( t_{sr(1_A \rightarrow 2_A)}) & ( t_{im(1_A \rightarrow 2_A)}) & ( t_{sim(1_B \rightarrow 2_B)}) & ( t_{sr(1_B \rightarrow 2_B)}) & 0 & (E_{p2A}+E_{p2B}+\frac{q^2}{d_{2A-2B}}) \\
0 & 0 & ( t_{sim(1_A \rightarrow 2_A)}) & ( t_{sr(1_A \rightarrow 2_A)}) & ( t_{sim(1_B \rightarrow 2_B)}) & ( t_{sr(1_B \rightarrow 2_B)}) & (E_{p2A}+E_{p2B}+\frac{q^2}{d_{2A-2B}}) & 0 \\
\end{pmatrix} \times \nonumber \\
\begin{pmatrix}
\frac{1}{\sqrt{p_{I}(t)}cos(\Theta_{I}(t))} & 0 & 0 & 0 & 0 & 0 & 0 & 0 \\
0 & \frac{1}{\sqrt{p_{I}(t)}sin(\Theta_{I}(t))} & 0 & 0 & 0 & 0 & 0 & 0 \\
0 & 0 & \frac{1}{\sqrt{p_{II}(t)}cos(\Theta_{II}(t))} & 0 & 0 & 0 & 0 & 0 \\
0 & 0 & 0 & \frac{1}{\sqrt{p_{II}(t)}sin(\Theta_{II}(t))} & 0 & 0 & 0 & 0 \\
0 & 0 & 0 & 0 & \frac{1}{\sqrt{p_{III}(t)}cos(\Theta_{III}(t))} & 0 & 0 & 0 \\
0 & 0 & 0 & 0 & 0 & \frac{1}{\sqrt{p_{II}(t)}sin(\Theta_{II}(t))} & 0 & 0 \\
0 & 0 & 0 & 0 & 0 & 0 & \frac{1}{\sqrt{p_{IV}(t)}cos(\Theta_{IV}(t))} & 0 \\
0 & 0 & 0 & 0 & 0 & 0 & 0 & \frac{1}{\sqrt{p_{IV}(t)}cos(\Theta_{IV}(t))} \\
\end{pmatrix} \times \nonumber \\ \times
\begin{pmatrix}
\sqrt{p_{I}(t)}cos(\Theta_{I}(t)) & 0 & 0 & 0 & 0 & 0 & 0 & 0 \\
0 & \sqrt{p_{I}(t)}sin(\Theta_{I}(t)) & 0 & 0 & 0 & 0 & 0 & 0 \\
0 & 0 & \sqrt{p_{II}(t)}cos(\Theta_{II}(t)) & 0 & 0 & 0 & 0 & 0 \\
0 & 0 & 0 & \sqrt{p_{II}(t)}sin(\Theta_{II}(t)) & 0 & 0 & 0 & 0 \\
0 & 0 & 0 & 0 & \sqrt{p_{III}(t)}cos(\Theta_{III}(t)) & 0 & 0 & 0 \\
0 & 0 & 0 & 0 & 0 & \sqrt{p_{III}(t)}sin(\Theta_{III}(t)) & 0 & 0 \\
0 & 0 & 0 & 0 & 0 & 0 & \sqrt{p_{IV}(t)}cos(\Theta_{IV}(t)) & 0 \\
0 & 0 & 0 & 0 & 0 & 0 & 0 & \sqrt{p_{IV}(t)}sin(\Theta_{IV}(t)) \\
\end{pmatrix}
\begin{pmatrix}
\sqrt{p_{I}(t)}cos(\Theta_{I}) \\
\sqrt{p_{I}(t)}sin(\Theta_{I}) \\
\sqrt{p_{II}(t)}cos(\Theta_{II}) \\
\sqrt{p_{II}(t)}sin(\Theta_{II}) \\
\sqrt{p_{III}(t)}cos(\Theta_{III}) \\
\sqrt{p_{III}(t)}sin(\Theta_{III}) \\
\sqrt{p_{IV}(t)}cos(\Theta_{IV}) \\
\sqrt{p_{IV}(t)}sin(\Theta_{IV}) \\
\end{pmatrix}= \nonumber \\
=2
\begin{pmatrix}
\sqrt{p_{I}(t)}cos(\Theta_{I}(t)) & 0 & 0 & 0 & 0 & 0 & 0 & 0 \\
0 & -\sqrt{p_{I}(t)}sin(\Theta_{I}(t)) & 0 & 0 & 0 & 0 & 0 & 0 \\
0 & 0 & +\sqrt{p_{II}(t)}cos(\Theta_{II}(t)) & 0 & 0 & 0 & 0 & 0 \\
0 & 0 & 0 & -\sqrt{p_{II}(t)}sin(\Theta_{II}(t)) & 0 & 0 & 0 & 0 \\
0 & 0 & 0 & 0 & +\sqrt{p_{III}(t)}cos(\Theta_{III}(t)) & 0 & 0 & 0 \\
0 & 0 & 0 & 0 & 0 & -\sqrt{p_{III}(t)}sin(\Theta_{III}(t)) & 0 & 0 \\
0 & 0 & 0 & 0 & 0 & 0 & +\sqrt{p_{IV}(t)}cos(\Theta_{IV}(t)) & 0 \\
0 & 0 & 0 & 0 & 0 & 0 & 0 & -\sqrt{p_{IV}(t)}sin(\Theta_{IV}(t)) \\
\end{pmatrix} \times
\hbar \frac{d}{dt}
\begin{pmatrix}
+\sqrt{p_{I}(t)}cos(\Theta_{I}(t)) \\
-\sqrt{p_{I}(t)}sin(\Theta_{I}(t)) \\
+\sqrt{p_{II}(t)}cos(\Theta_{II}(t)) \\
-\sqrt{p_{II}(t)}sin(\Theta_{II}(t)) \\
+\sqrt{p_{III}(t)}cos(\Theta_{III}(t)) \\
-\sqrt{p_{III}(t)}sin(\Theta_{III}(t)) \\
+\sqrt{p_{IV}(t)}cos(\Theta_{IV}(t)) \\
-\sqrt{p_{IV}(t)}sin(\Theta_{IV}(t)) \\
\end{pmatrix}=
\hbar \frac{d}{dt}
\begin{pmatrix}
p_{I}(t)(cos(\Theta_{I}(t)))^2 \\
p_{I}(t)(sin(\Theta_{I}(t)))^2 \\
p_{II}(t)(cos(\Theta_{II}(t)))^2 \\
p_{II}(t)(sin(\Theta_{II}(t)))^2 \\
p_{III}(t)(cos(\Theta_{III}(t)))^2 \\
p_{III}(t)(sin(\Theta_{III}(t)))^2 \\
p_{IV}(t)(cos(\Theta_{IV}(t)))^2 \\
p_{IV}(t)(sin(\Theta_{IV}(t)))^2 \\
\end{pmatrix}=
\hbar \frac{d}{dt}
\begin{pmatrix}
p_1(t) \\
p_2(t) \\
p_3(t) \\
p_4(t) \\
p_5(t) \\
p_6(t) \\
p_7(t) \\
p_8(t) \\
\end{pmatrix}
\end{eqnarray}
\normalsize
The S matrix from classical epidemic model can be written as
\begin{eqnarray}
\hat{S}(t)
\begin{pmatrix}
p_1(t) \\
p_2(t) \\
p_3(t) \\
p_4(t) \\
p_5(t) \\
p_6(t) \\
p_7(t) \\
p_8(t) \\
\end{pmatrix} =
\begin{pmatrix}
s_{1,1}(t) & s_{1,2}(t) & s_{1,3}(t) & s_{1,4}(t) & s_{1,5}(t) & s_{1,6}(t) & s_{1,7}(t) & s_{1,8}(t) \\
s_{2,1}(t) & s_{2,2}(t) & s_{2,3}(t) & s_{2,4}(t) & s_{2,5}(t) & s_{2,6}(t) & s_{2,7}(t) & s_{2,8}(t) \\
s_{3,1}(t) & s_{3,2}(t) & s_{3,3}(t) & s_{3,4}(t) & s_{3,5}(t) & s_{3,6}(t) & s_{3,7}(t) & s_{3,8}(t) \\
s_{4,1}(t) & s_{4,2}(t) & s_{4,3}(t) & s_{4,4}(t) & s_{4,5}(t) & s_{4,6}(t) & s_{4,7}(t) & s_{4,8}(t) \\
s_{5,1}(t) & s_{5,2}(t) & s_{5,3}(t) & s_{5,4}(t) & s_{5,5}(t) & s_{5,6}(t) & s_{5,7}(t) & s_{5,8}(t) \\
s_{6,1}(t) & s_{6,2}(t) & s_{6,3}(t) & s_{6,4}(t) & s_{6,5}(t) & s_{6,6}(t) & s_{6,7}(t) & s_{6,8}(t) \\
s_{7,1}(t) & s_{7,2}(t) & s_{7,3}(t) & s_{7,4}(t) & s_{7,5}(t) & s_{7,6}(t) & s_{7,7}(t) & s_{7,8}(t) \\
s_{8,1}(t) & s_{8,2}(t) & s_{8,3}(t) & s_{8,4}(t) & s_{8,5}(t) & s_{8,6}(t) & s_{8,7}(t) & s_{8,8}(t) \\
\end{pmatrix}
\begin{pmatrix}
p_1(t) \\
p_2(t) \\
p_3(t) \\
p_4(t) \\
p_5(t) \\
p_6(t) \\
p_7(t) \\
p_8(t) \\
\end{pmatrix}
=\frac{d}{dt}
\begin{pmatrix}
p_1(t) \\
p_2(t) \\
p_3(t) \\
p_4(t) \\
p_5(t) \\
p_6(t) \\
p_7(t) \\
p_8(t) \\
\end{pmatrix}.
\end{eqnarray}
and can be written in relation to 2 coupled single electron devices as
\tiny
\begin{eqnarray}
\hat{S}(t)=
\begin{pmatrix}
s_{1,1}(t) & s_{1,2}(t) & s_{1,3}(t) & s_{1,4}(t) & s_{1,5}(t) & s_{1,6}(t) & s_{1,7}(t) & s_{1,8}(t) \\
s_{2,1}(t) & s_{2,2}(t) & s_{2,3}(t) & s_{2,4}(t) & s_{2,5}(t) & s_{2,6}(t) & s_{2,7}(t) & s_{2,8}(t) \\
s_{3,1}(t) & s_{3,2}(t) & s_{3,3}(t) & s_{3,4}(t) & s_{3,5}(t) & s_{3,6}(t) & s_{3,7}(t) & s_{3,8}(t) \\
s_{4,1}(t) & s_{4,2}(t) & s_{4,3}(t) & s_{4,4}(t) & s_{4,5}(t) & s_{4,6}(t) & s_{4,7}(t) & s_{4,8}(t) \\
s_{5,1}(t) & s_{5,2}(t) & s_{5,3}(t) & s_{5,4}(t) & s_{5,5}(t) & s_{5,6}(t) & s_{5,7}(t) & s_{5,8}(t) \\
s_{6,1}(t) & s_{6,2}(t) & s_{6,3}(t) & s_{6,4}(t) & s_{6,5}(t) & s_{6,6}(t) & s_{6,7}(t) & s_{6,8}(t) \\
s_{7,1}(t) & s_{7,2}(t) & s_{7,3}(t) & s_{7,4}(t) & s_{7,5}(t) & s_{7,6}(t) & s_{7,7}(t) & s_{7,8}(t) \\
s_{8,1}(t) & s_{8,2}(t) & s_{8,3}(t) & s_{8,4}(t) & s_{8,5}(t) & s_{8,6}(t) & s_{8,7}(t) & s_{8,8}(t) \\
\end{pmatrix}= \nonumber \\
=\frac{2}{\hbar}
\begin{pmatrix}
\sqrt{p_{I}(t)}cos(\Theta_{I}(t)) & 0 & 0 & 0 & 0 & 0 & 0 & 0 \\
0 & -\sqrt{p_{I}(t)}sin(\Theta_{I}(t)) & 0 & 0 & 0 & 0 & 0 & 0 \\
0 & 0 & +\sqrt{p_{II}(t)}cos(\Theta_{II}(t)) & 0 & 0 & 0 & 0 & 0 \\
0 & 0 & 0 & -\sqrt{p_{II}(t)}sin(\Theta_{II}(t)) & 0 & 0 & 0 & 0 \\
0 & 0 & 0 & 0 & +\sqrt{p_{III}(t)}cos(\Theta_{III}(t)) & 0 & 0 & 0 \\
0 & 0 & 0 & 0 & 0 & -\sqrt{p_{III}(t)}sin(\Theta_{III}(t)) & 0 & 0 \\
0 & 0 & 0 & 0 & 0 & 0 & +\sqrt{p_{IV}(t)}cos(\Theta_{IV}(t)) & 0 \\
0 & 0 & 0 & 0 & 0 & 0 & 0 & -\sqrt{p_{IV}(t)}sin(\Theta_{IV}(t)) \\
\end{pmatrix} \times \nonumber \\ \times
\begin{pmatrix}
Im(E_{p1A}+E_{p1B} & +Re(E_{p1A}+E_{p1B}+\frac{q^2}{d_{1A-1B}}) & ( t_{sim(2_B \rightarrow 1_B)}) & +( t_{sr(2_B \rightarrow 1_B)}) & ( t_{sim(2_A \rightarrow 1_A)}) & ( t_{sr(2_A \rightarrow 1_A)}) & 0 & 0 \\
-Re(E_{p1A}+E_{p1B}+\frac{q^2}{d_{1A-1B}}) & Im(E_{p1A}+E_{p1B} & -( t_{sr(2_B \rightarrow 1_B)}) & ( t_{sim(2_B \rightarrow 1_B)}) & ( -t_{sr(2_A \rightarrow 1_A)}) & ( t_{sim(2_A \rightarrow 1_A)}) & 0 & 0 \\
( -t_{sim(1_B \rightarrow 2_B)}) & ( t_{sre(1_B \rightarrow 2_B)}) & Im(E_{p1A}+E_{p2B}) & +Re(E_{p1A}+E_{p2B}+\frac{q^2}{d_{1A-2B}}) & 0 & 0 & ( t_{sim(2_A \rightarrow 1_A)}) & ( t_{sr(2_A \rightarrow 1_A)}) \\
( -t_{sr(1_B \rightarrow 2_B)}) & ( -t_{sim(1_B \rightarrow 2_B)}) & -Re(E_{p1A}+E_{p2B}+\frac{q^2}{d_{1A-2B}}) & Im(E_{p1A}+E_{p2B}) & 0 & 0 & ( -t_{sr(2_A \rightarrow 1_A)}) & ( t_{sim(2_A \rightarrow 1_A)}) \\
( -t_{sim(1_A \rightarrow 2_A)}) & ( t_{sr(1_A \rightarrow 2_A)}) & 0 & 0 & Im(E_{p2A}+E_{p1B}) & +Re(E_{p2A}+E_{p1B}+\frac{q^2}{d_{2A-1B}}) & ( t_{sim(2_B \rightarrow 1_B)}) & ( t_{sr(2_B \rightarrow 1_B)}) \\
( -t_{sr(1_A \rightarrow 2_A)}) & ( -t_{sim(1_A \rightarrow 2_A)}) & 0 & 0 & -Re(E_{p2A}+E_{p1B}+\frac{q^2}{d_{2A-1B}}) & Im(E_{p2A}+E_{p1B}) & ( -t_{sr(2_B \rightarrow 1_B)}) & ( t_{sim(2_B \rightarrow 1_B)}) \\
0 & 0 & ( -t_{sim(1_A \rightarrow 2_A)}) & ( t_{sr(1_A \rightarrow 2_A)}) & ( -t_{im(1_B \rightarrow 2_B)}) & ( t_{sr(1_B \rightarrow 2_B)}) & Im(E_{p2A}+E_{p2B}) & +Re(E_{p2A}+E_{p2B}+\frac{q^2}{d_{2A-2B}}) \\
0 & 0 & ( -t_{sr(1_A \rightarrow 2_A)}) & ( -t_{sim(1_A \rightarrow 2_A)}) & ( -t_{sr(1_B \rightarrow 2_B)}) & ( -t_{im(1_B \rightarrow 2_B)}) & -Re(E_{p2A}+E_{p2B}+\frac{q^2}{d_{2A-2B}}) & Im(E_{p2A}+E_{p2B}) \\
\end{pmatrix} \times \nonumber \\
\begin{pmatrix}
\frac{1}{\sqrt{p_{I}(t)}cos(\Theta_{I}(t))} & 0 & 0 & 0 & 0 & 0 & 0 & 0 \\
0 & \frac{1}{\sqrt{p_{I}(t)}sin(\Theta_{I}(t))} & 0 & 0 & 0 & 0 & 0 & 0 \\
0 & 0 & \frac{1}{\sqrt{p_{II}(t)}cos(\Theta_{II}(t))} & 0 & 0 & 0 & 0 & 0 \\
0 & 0 & 0 & \frac{1}{\sqrt{p_{II}(t)}sin(\Theta_{II}(t))} & 0 & 0 & 0 & 0 \\
0 & 0 & 0 & 0 & \frac{1}{\sqrt{p_{III}(t)}cos(\Theta_{III}(t))} & 0 & 0 & 0 \\
0 & 0 & 0 & 0 & 0 & \frac{1}{\sqrt{p_{II}(t)}sin(\Theta_{II}(t))} & 0 & 0 \\
0 & 0 & 0 & 0 & 0 & 0 & \frac{1}{\sqrt{p_{IV}(t)}cos(\Theta_{IV}(t))} & 0 \\
0 & 0 & 0 & 0 & 0 & 0 & 0 & \frac{1}{\sqrt{p_{IV}(t)}cos(\Theta_{IV}(t))} \\
\end{pmatrix}=\Bigg[ \hat{S}_1, \hat{S}_2 \Bigg].
\end{eqnarray}
\normalsize
Matrix $\hat{S}_1$ is given as
\tiny
$\hat{S}_1= \\
\begin{bmatrix}
Im(E_{p1A}+E_{p1B}) & +Re(E_{p1A}+E_{p1B}+\frac{q^2}{d_{1A-1B}})\frac{\sqrt{Re(p_{I}(t))}}{\sqrt{Im(p_{I}(t))}} & ( t_{sim(2_B \rightarrow 1_B)})\frac{\sqrt{Re(p_{I}(t))}}{\sqrt{Re(p_{II}(t))}} & +( t_{sr(2_B \rightarrow 1_B)})\frac{\sqrt{Re(p_{I}(t))}}{\sqrt{Im(p_{II}(t))}} & ( t_{sim(2_A \rightarrow 1_A)})\frac{\sqrt{Re(p_{I}(t))}}{\sqrt{Re(p_{III}(t))}}
& ( t_{sr(2_A \rightarrow 1_A)})\frac{\sqrt{Re(p_{I}(t))}}{\sqrt{Im(p_{III}(t))}}
& 0
& 0 \\
-Re(E_{p1A}+E_{p1B}+\frac{q^2}{d_{1A-1B}})\frac{\sqrt{Im(p_{I}(t))}}{\sqrt{Re(p_{I}(t))}}
& Im(E_{p1A}+E_{p1B})
& -( t_{sr(2_B \rightarrow 1_B)})\frac{\sqrt{Im(p_{I}(t))}}{\sqrt{Re(p_{II}(t))}}
& ( t_{sim(2_B \rightarrow 1_B)})\frac{\sqrt{Im(p_{I}(t))}}{\sqrt{Im(p_{II}(t))}}
& ( -t_{sr(2_A \rightarrow 1_A)})\frac{\sqrt{Im(p_{I}(t))}}{\sqrt{Re(p_{III}(t))}}
& ( t_{sim(2_A \rightarrow 1_A)})\frac{\sqrt{Im(p_{I}(t))}}{\sqrt{Im(p_{III}(t))}}
& 0
& 0 \\
  ( -t_{sim(1_B \rightarrow 2_B)})\frac{\sqrt{Re(p_{II}(t))}}{\sqrt{Re(p_{I}(t))}}
& ( t_{sre(1_B \rightarrow 2_B)})\frac{\sqrt{Re(p_{II}(t))}}{\sqrt{Im(p_{I}(t))}}
&              Im(E_{p1A}+E_{p2B}) &
+Re(E_{p1A}+E_{p2B}+\frac{q^2}{d_{1A-2B}})\frac{\sqrt{Re(p_{II}(t))}}{\sqrt{Im(p_{II}(t))}}
& 0
& 0
& ( t_{sim(2_A \rightarrow 1_A)})\frac{\sqrt{Re(p_{II}(t))}}{\sqrt{Im(p_{IV}(t))}}
& ( t_{sr(2_A \rightarrow 1_A)}) \frac{\sqrt{Re(p_{II}(t))}}{\sqrt{Re(p_{IV}(t))}} \\
   ( -t_{sr(1_B \rightarrow 2_B)})\frac{\sqrt{Im(p_{II}(t))}}{\sqrt{Re(p_{I}(t))}}
& ( -t_{sim(1_B \rightarrow 2_B)})\frac{\sqrt{Im(p_{II}(t))}}{\sqrt{Im(p_{I}(t))}}
& -Re(E_{p1A}+E_{p2B}+\frac{q^2}{d_{1A-2B}})\frac{\sqrt{Im(p_{II}(t))}}{\sqrt{Re(p_{II}(t))}}
& Im(E_{p1A}+E_{p2B})
& 0
& 0
& ( -t_{sr(2_A \rightarrow 1_A)})\frac{\sqrt{Im(p_{II}(t))}}{\sqrt{Re(p_{IV}(t))}}
& ( t_{sim(2_A \rightarrow 1_A)})\frac{\sqrt{Im(p_{II}(t))}}{\sqrt{Im(p_{IV}(t))}} \\
( -t_{sim(1_A \rightarrow 2_A)})\frac{\sqrt{Re(p_{III}(t))}}{\sqrt{Re(p_{I}(t))}}
& ( t_{sr(1_A \rightarrow 2_A)})\frac{\sqrt{Re(p_{III}(t))}}{\sqrt{Im(p_{I}(t))}}
& 0
& 0
& Im(E_{p2A}+E_{p1B})
& +Re(E_{p2A}+E_{p1B}+\frac{q^2}{d_{2A-1B}})\frac{\sqrt{Re(p_{III}(t))}}{\sqrt{Im(p_{III}(t))}}
& ( t_{sim(2_B \rightarrow 1_B)})
& ( t_{sr(2_B \rightarrow 1_B)})\frac{\sqrt{Re(p_{III}(t))}}{\sqrt{Im(p_{IV}(t))}} \\
   ( -t_{sr(1_A \rightarrow 2_A)})\frac{\sqrt{Im(p_{III}(t))}}{\sqrt{Re(p_{I}(t))}}
& ( -t_{sim(1_A \rightarrow 2_A)})\frac{\sqrt{Im(p_{III}(t))}}{\sqrt{Im(p_{I}(t))}}
& 0
& 0
& -Re(E_{p2A}+E_{p1B}+\frac{q^2}{d_{2A-1B}})\frac{\sqrt{Im(p_{III}(t))}}{\sqrt{Re(p_{III}(t))}}
& Im(E_{p2A}+E_{p1B})
& ( -t_{sr(2_B \rightarrow 1_B)})           \frac{\sqrt{Im(p_{III}(t))}}{\sqrt{Re(p_{IV}(t))}}
& ( t_{sim(2_B \rightarrow 1_B)})           \frac{\sqrt{Im(p_{III}(t))}}{\sqrt{Im(p_{IV}(t))}} \\
0
& 0
& (-t_{sim(1_A \rightarrow 2_A)})\frac{\sqrt{Re(p_{IV}(t))}}{\sqrt{Re(p_{II}(t))}}
& ( t_{sr(1_A \rightarrow 2_A)})\frac{\sqrt{Re(p_{IV}(t))}}{\sqrt{Im(p_{II}(t))}}
& (-t_{im(1_B \rightarrow 2_B)})\frac{\sqrt{Re(p_{IV}(t))}}{\sqrt{Re(p_{III}(t))}}
& ( t_{sr(1_B \rightarrow 2_B)})\frac{\sqrt{Re(p_{IV}(t))}}{\sqrt{Im(p_{III(t}}}
&            Im(E_{p2A}+E_{p2B})
& +Re(E_{p2A}+E_{p2B}+\frac{q^2}{d_{2A-2B}}) \frac{\sqrt{Re(p_{IV}(t))}}{\sqrt{Im(p_{IV}(t))}} \\
0 &
0 &
( -t_{sr(1_A \rightarrow 2_A)})               \frac{\sqrt{Im(p_{IV}(t))}}{\sqrt{Re(p_{II}(t))}}
& ( -t_{sim(1_A \rightarrow 2_A)})            \frac{\sqrt{Im(p_{IV}(t))}}{\sqrt{Im(p_{II}(t))}}
& ( -t_{sr(1_B \rightarrow 2_B)})             \frac{\sqrt{Im(p_{IV}(t))}}{\sqrt{Re(p_{III}(t))}}
& ( -t_{im(1_B \rightarrow 2_B)})             \frac{\sqrt{Im(p_{IV}(t))}}{\sqrt{Im(p_{III}(t))}}
& \frac{-Re(E_{p2A}+E_{p2B}+\frac{q^2}{d_{2A-2B}})}{\sqrt{Im(p_{IV}(t))}/\sqrt{Re(p_{IV}(t))}}
& Im(E_{p2A}+E_{p2B})                          \\
\end{bmatrix},$
\normalsize
and matrix $\hat{S}_2$ is given as   \\
\tiny
$\hat{S}_2=
\begin{bmatrix}
0 \\
0 \\
( t_{sr(2_A \rightarrow 1_A)})\frac{\sqrt{Re(p_{II}(t))}}{\sqrt{Im(p_{IV}(t))}} \\
( t_{sim(2_A \rightarrow 1_A)})\frac{\sqrt{Im(p_{II}(t))}}{\sqrt{Im(p_{IV}(t))}} \\
( t_{sim(2_B \rightarrow 1_B)})\frac{\sqrt{Re(p_{III}(t))}}{\sqrt{Im(p_{IV}(t))}} \\
 ( t_{sim(2_B \rightarrow 1_B)})\frac{\sqrt{Im(p_{III}(t))}}{\sqrt{Im(p_{IV}(t))}} \\
+Re(E_{p2A}+E_{p2B}+\frac{q^2}{d_{2A-2B}})\frac{\sqrt{Re(p_{IV}(t))}}{\sqrt{Im(p_{IV}(t))}} \\
Im(E_{p2A}+E_{p2B}) \\
\end{bmatrix}$ \\ \normalsize
and with
\begin{eqnarray}
\frac{2}{\hbar}\Bigg[ \hat{S}_1(t), \hat{S}_2(t) \Bigg]
\begin{pmatrix}
Re(p_{I}(t))\\
Im(p_{I}(t))\\
Re(p_{II}(t))\\
Im(p_{II}(t))\\
Re(p_{III}(t))\\
Im(p_{III}(t))\\
Re(p_{IV}(t))\\
Im(p_{IV}(t))\\
\end{pmatrix}=
\frac{d}{dt}
\begin{pmatrix}
Re(p_{I}(t))\\
Im(p_{I}(t))\\
Re(p_{II}(t))\\
Im(p_{II}(t))\\
Re(p_{III}(t))\\
Im(p_{III}(t))\\
Re(p_{IV}(t))\\
Im(p_{IV}(t))\\
\end{pmatrix},
\end{eqnarray}
with
\begin{eqnarray}
Re(p_{I}(t))=\cos(\Theta_I(t))^2p_{I}(t),  \\
Im(p_{I}(t))=\sin(\Theta_I(t))^2p_{I}(t),  \\
Re(p_{II}(t))=\cos(\Theta_{II}(t))^2p_{II}(t), \\
Im(p_{II}(t))=\sin(\Theta_{II}(t))^2p_{II}(t), \\
Re(p_{III}(t))=\cos(\Theta_{III}(t))^2p_{III}(t), \\
Im(p_{III}(t))=\sin(\Theta_{III}(t))^2p_{III}(t), \\
Re(p_{IV}(t))=\cos(\Theta_{IV}(t))^2p_{IV}(t), \\
Im(p_{IV}(t))=\sin(\Theta_{IV}(t))^2p_{IV}(t).  \\
\end{eqnarray}
After conducted considerations it becomes explictly visible that usage of magnetic Aharonov-Bohm effect will have impact on the occupancy of nodes $(1_A,1_B)$,$(1_A,2_B)$,$(2_A,1_B)$,$(2_A,2_B)$. Let us assume that two interacting qubits are symmetric and aligned along x axes. In such case we need to renormalize phases dynamics with
time so $\Theta_{I}(t) \rightarrow \Theta_{I}(t)+(A_x(1_A,t)+A_x(1_B,t))\Delta L\frac{e}{\hbar}$, $\Theta_{II}(t) \rightarrow \Theta_{II}(t)+(A_x(1_A,t)+A_x(2_B,t))\Delta L\frac{e}{\hbar}$,
$\Theta_{III}(t) \rightarrow \Theta_{III}(t)+(A_x(2_A,t)+A_x(1_B,t))\Delta L\frac{e}{\hbar}$, $\Theta_{IV}(t) \rightarrow \Theta_{IV}(t)+(A_x(2_A,t)+A_x(2_B,t))\Delta L \frac{e}{\hbar}$, where $\delta L$ is the diameter each of 4 quantum dots and e is elementary charge of electron. Such situation was described by means of Schroedinger formalism in \cite{Spie}. Furthermore the noninvasive detection of charge movement by single electron devices
 \cite{Noninvasive} can also be also mapped to epidemic model. Various analytical solutions of tight-binding model that can be used in epidemic model can be found in \cite{Decoherence}, \cite{Photonic}, \cite{Nbodies}, \cite{Cryogenics}.

\end{landscape}
\normalsize
Here we have assumed that there is possible escape of electron(s) from the system of quantum dots to outside environment by tunneling from the system of 2 coupled quantum dots what can be reflected in imaginary values of $E_{p1A}$, $E_{p2A}$, $E_{p1B}$, $E_{p2B}$. Furthermore we can also assume that electron is being injected to the structure of
2 coupled quantum dots what also corresponds to non-zero imaginary value of $E_{p1A}$, $E_{p2A}$, $E_{p1B}$, $E_{p2B}$ and is denoted as $Im(E_{p1A})$, $Im(E_{p2A})$, $Im(E_{p1B})$, $Im(E_{p2B})$. Presence of electrons at positions (1A,1B),(1A,2B), (2A,1B), (2A,2B) is reflected by probabilities $p_1(t)+p_2(t)$, $p_3(t)+p_4(t)$, $p_5(t)+p_6(t)$, $p_7(t)+p_8(t)$. Dependence of evolution of quantum phases $\Theta_{I}(t)$,$\Theta_{II}(t)$,$\Theta_{III}(t)$,$\Theta_{IV}(t)$ is encoded by square of tangents of phases given by $tan(\Theta_{I}(t))^2=\frac{p_2(t)}{p_1(t)}$, $tan(\Theta_{II}(t))^2=\frac{p_4(t)}{p_3(t)}$, $tan(\Theta_{III}(t))^2=\frac{p_6(t)}{p_5(t)}$, $tan(\Theta_{IV}(t))^2=\frac{p_8(t)}{p_7(t)}$. If tight-binding model is time-independent the evolution of phase can be tracked in unique way by classical epidemic model.
Furthermore matrix $\hat{S}$(t) of classical epidemic model can be determined in unique way, since all its eigenstates and eigenvectors can be determined analytically.
This is not the case of bigger matrices than 4 by 4. Furthermore one can incorporate the noise in case of 2 coupled position based qubits with use of delta functions and such situation can also be mapped to classical epidemic model. We notice that for the case of coupled position based qubits system expressed by matrix N by N and N possible states we have mapping to classical epidemic model with obtained matrix $\hat{S}$ that has dimension 2N by 2N and it described evolution of 2N classical states in the framework of stochastic finite state machine.
\section{Conclusions}
The obtained results shows that quantum mechanical phenomena might be almost entirely simulated by classical statistical model. It includes the quantum like entanglement and superposition of states. Therefore coupled epidemic models expressed by classical systems in terms of classical physics can be the base for possible incorporation of quantum technologies and in particular
for quantum like computation and quantum like communication. In the conduced computations Wolfram software was used \cite{Mathematica}. All work presented at \cite{QHSchannel},
\cite{Nbodies} can be expressed by classical epidemic model. It is expected that time crystals can be also described in the given framework \cite{Wilczek}, \cite{Sacha}.
It is open issue to what extent we can parameterize various condensed matter phenomena by stochastic finite state machine \cite{Spalek}.

\vspace{12pt}

\end{document}